\documentclass[a4paper]{article}
\usepackage{subfig}

\newsavebox{\measurebox}

\usepackage[T1]{fontenc}
\usepackage[utf8]{inputenc}
\usepackage{amsmath}
\usepackage{amsfonts}
\usepackage{amssymb}
\usepackage{graphicx}
\usepackage{tikz}
\usepackage{mathtools}
\usepackage[top=3cm, bottom=3cm, left=3cm, right=3cm]{geometry}
\usepackage{calrsfs}
\usepackage{bbm}
\usepackage{algorithm}
\usepackage[noend]{algpseudocode}
\usepackage{float}
\DeclareMathAlphabet{\pazocal}{OMS}{zplm}{m}{n}

\makeatletter
\newcommand*\bigcdot{\mathpalette\bigcdot@{.5}}
\newcommand*\bigcdot@[2]{\mathbin{\vcenter{\hbox{\scalebox{#2}{$\m@th#1\bullet$}}}}}
\makeatother

\newtheorem{definition-theorem}[theorem]{\indent Definition-Theorem}

\def \F{\mathcal{F}}

\title{\bf Boundary solution based on rescaling method: recoup the first and second-order statistics of neuron network dynamics}
\author{C. Romaro, A.C. Roque, J.R.C. Piqueira \\ \textit{Physics department and Escola Politécnica} \\ \textit{University of São Paulo.}}
\date{October 2019}

\date{\today}

\begin{document}

\maketitle

\section*{Abstract}
There is a strong nexus between the network size and the computational resources available, which may impede a neuroscience study. In the meantime, rescaling the network while maintaining its behavior is not a trivial mission. Additionally, modeling patterns of connections under topographic organization presents an extra challenge: to solve the network boundaries or mingled with an unwished behavior. This behavior, for example, could be an inset oscillation due to the torus solution; or a blend with/of unbalanced neurons due to a lack (or overdose) of connections.
We detail the network rescaling method able to sustain behavior statistical utilized in \cite{romaro2018implementation} and present a boundary solution method based on the previous statistics recoup idea.
\vspace{0.6 cm}

\section{Introduction}

\vspace{0.4 cm}

Understanding the brain is challenging, given both its complex mechanisms and its inaccessibility. Modeling in neuroscience  has been typically used to understand the neurons \cite{lapicque1907recherches, HH-1952} and neuronal system \cite{tsodyks1998neural, millman2010self, kriener2014pattern}. Computational resources \cite{carnevale2006neuron, lytton2016simulation, gewaltig2007nest, goodman2009brian} continually contribute to the study and understanding of neuronal pathways \cite{hines2004modeldb}, channels \cite{ranjan2011channelpedia}, proteins and other discovered mechanisms \cite{mcdougal2013reaction} however computational resources \cite{jordan2018extremely} still pose a challenge to network dynamics studies even in neuroscience \cite{markram2015reconstruction, potjans2012cell, wagatsuma2013spatial,lee2017computational, tsodyks1998neural}. 
Consequently there seems to be a compromise between the increase in detail or the size of network models and the computational resource available. To make more detailed simulations computationally feasible we could therefore reduce the size of the network. 
\vspace{0.4 cm}

Rescaling the network to decrease or increase its size is, however, a challenging process. For example, as we reduce the number of neurons, an increase in the number of connections or the synaptic weight is needed to balance the external inputs. However, this can lead to an undesired spiking synchrony and regularity \cite{brunel2000dynamics, van2015scalability, iyer2013influence}.
\vspace{0.4 cm}

An additional challenge arises in the modeling of somatotopic regions or networks with boundary conditions. The topographic pattern of connection is interrupted in the network edges, changing the activity in the network boundary \cite{markram2015reconstruction, mazza2004dynamical}. A classic solution adopted to this problem is the torus connection, which introduces undesired oscillations to the network \cite{senk2018conditions, senk2018reconciliation, veltz2015periodic}.  
\vspace{0.4 cm}

The purpose of this work is to explain a method to rescaling the network recouping the first and second order statistics and, , present a method to boundary solution of topographic network based on the rescaling model previous presented.
\vspace{0.4 cm}

This paper is organized as follows. In Section \ref{Rescaling Method} we present the Rescaling method. In Subsection \ref{Rescaling Method algorithm} we give a description of the algorithm and, than, in Subsection \ref{Rescaling Applied} we present applied examples and their results. Finally in Subsection \ref{Model requirements, math explication and method limitations} we present the mathematical explanation and discuss model requirements as well as method limitations. 
\vspace{0.4 cm}

Following this pattern, in Section \ref{Boundary Method} we present the Rescaling method. In Subsection \ref{Boundary Algorithm} we give a description of the algorithm and, than, in Subsection \ref{Boundary correction applied} we present applied examples. And, finally, in Subsection \ref{Boundary:Model requirements, mathematical explication and method limitations}, we discuss the sufficient conditions of the model for the method application.
\vspace{0.6 cm}

\section{Rescaling Method}
\label{Rescaling Method}
\vspace{0.4 cm}

A neuron network structure can be defined by the number of neurons N, a function of connection ${\F(o_{pre},o_{post})}$ between neuron pre-synaptic $o_{pre}$ and post-synaptic $o_{post}$, and the synaptic strength ${w_{pre,post}}$. This network can be a slice in inner and inter connected subsets on neurons (populations or layers). Other neuron-model-dependent parameters such as firing threshold ${V_{th}}$, reset potential after spike ${V_{res}}$, absolute refractory period  ${\tau_{ref}}$; or synapse-model-dependent parameters such as synapse time constant ${\tau_{syn}}$, synaptic transmission delays ${\Delta_{syn}}$; can integrate the model. Those parameters, definitely bias the neuron network activity and behavior but do not compose the network structure parameters. 
\vspace{0.4 cm}

Our rescaling is dependent on a single parameter $k$ positive in the interval ]$0, \infty $ [, which is used to resize down ( ${k \in ]0,1[} $ ) or up (${k \in ]1,\infty[}$ ) the numbers of network neurons, connections, external inputs, and synaptic weights, while maintaining fixed the function of connection ${\F(o_{pre},o_{post})}$ and the proportions of cells per subset of neurons.
\vspace{0.4 cm}

This method is able to maintain the first and second-order statistics, and, therefore, the layer-specific average firing rates, the synchrony, the irregularity features and the network behavior similar to the ones observed in the full version. That happens essentially because it holds fixed the probability and the pattern of connections, it keeps the average random input \cite{van2015scalability}, and the fixed proportion between the firing threshold and the square root of the number of connections \cite{vreeswijk1998chaotic}.
\vspace{0.6 cm}

\subsection{Rescaling Method algorithm}
\label{Rescaling Method algorithm}
\vspace{0.4 cm}

The algorithm of the rescaling method can be found in any example-application on
Section \ref{Rescaling Applied}, also available on GitHub (https://github.com/ceciliaromaro/recoup-the-first-and-second-order-statistics-of-neuron-network-dynamics) and it is informally described as previous in \cite{romaro2018implementation} as follows:
\vspace{0.4 cm}

\begin{itemize}
\item{\textbf{Step 1:}} Decreasing the number of neurons and external input per neuron  by multiplying them by the scale factor while keeping the proportions of cells per population fixed;
\vspace{0.4 cm}

\item{\textbf{Step 2:}}  Decreasing the number of connections per population  by multiplying them by the square of the scale factor  while keeping the functions of connections (probabilities) between populations unchanged;
\vspace{0.4 cm}

\item{\textbf{Step 3:}}  Increasing the synaptic weights  by dividing them by the square root of the scale factor;
\vspace{0.4 cm}

\item{\textbf{Step 4:}}   Providing each cell with a DC input current with a value corresponding to the total input lost due to rescaling.
\vspace{0.6 cm}

\end{itemize}
\vspace{0.4 cm}

The first three steps keep the proportional balance of network through neurons, external inputs and layers. The fourth step changes the threshold to guarantee the neuron/layer activity.
\vspace{0.4 cm}

\subsubsection{Rescaling method for 1 layer model}

\begin{algorithm}[H]
\caption{Rescaling method for model with 1 set of neurons}\label{alg:sim1}
\begin{algorithmic}[1]

\State $N$ the number of neurons.
\State $C$ the probability of connection (${\F(o_{pre},o_{post})=C}$).
\State $X$ the total number of connections ($x = X/N$ the average number of connections per neuron)
\State $X_{ext}$ the number on average of external neurons connected to each neuron in N.
\State $w$ (pA or mV) the weight of synaptic strength.
\State $k$ the factor of rescaling.
\State $f_{ext}$ (Hz) the average firing rate of the external input.
\State $f$ (Hz) the average firing rate of the set of neurons.
\State $\tau _{syn}$ (ms) synapse time constant.

\vspace{0.3 cm}

1.NUMBER OF NEURONS

\vspace{0.1 cm}

\State $N' \gets k*N$
\State $X_{ext}' \gets k*X_{ext}$

\vspace{0.3 cm}

2.NUMBER OF CONNECTIONS
\vspace{0.1 cm}

\State   $C' \gets   C$
\State  $X' \gets  k^2 *X$  \Comment{ corolario of $X$ = $C*N_{pre}*N_{pos}$} 

\vspace{0.3 cm}

3. SYNAPTIC STRENGHT
\vspace{0.1 cm}

\State $w' \gets w/ \sqrt{k}$

\vspace{0.3 cm}

4. THRESHOLD ADJUSTMENT

\vspace{0.1 cm}

\State $q_{sum} = w * f * x $
\State $q_{ext} = w * f_{ext}* X_{ext} $
\State $I_{DC}' = \tau _{syn} * (  
		        (1 - \sqrt{k}) * (q_{sum} + q_{ext}))$ \Comment{Extra DC (pA  or mV) input to compensate resize}
		        \label{Calcule IDC}

\vspace{0.3 cm}

\State Done!   \Comment{Notice that step 4 uses parameters without resizing.}
\end{algorithmic}
\end{algorithm}

Notice that if $w$ is given by $mV$, it is not necessary to multiply the DC input by  $\tau _{syn}$ in step \ref{Calcule IDC}. Instead, \ref{Calcule IDC}: $V_{DC}' =   
		        (1 - \sqrt{k}) * (q_{sum} + q_{ext})$ \Comment{Extra DC (mV) input to compensate resizing.}  Notice that $\tau_m$ and $C_m$ are neurons parameters, not network parameters.

\vspace{0.6 cm}

\subsubsection{Rescaling method for n layers model}

\vspace{0.4 cm}

The same idea applied for 1 layer is recurrently apply for all layers. The one attention is to calculate the compensation threshold current correctly: a weighted average connections number-frequency-weight of each presynaptic layer.
\vspace{0.4 cm}

\begin{algorithm}[H]
\caption{Rescaling method for model with n set of neurons}\label{alg:sim2}
\begin{algorithmic}[1]

\State $n$ the number of layers/sets of neuros. 
\State $N_i$ the number of presynaptic neurons in layer $i$. (${i \in n}$)
\State $N_j$ the number of possynaptic neurons in layer $j$. (${j \in n}$)
\State $C_{ij}$ the probability of connection from layer $i$ to layer $j$ (${\F(i,j)=C_{ij}}$).
\State $X_{ij}$ the total number of connections between layer $i$ and layer $j$ 
\vspace{0.0 cm}

($x_j = X_{ij}/N_j$ the average number of received connections per neuron).
\State $X_{ext,j}$ the number on average of external neurons connected to each neurons of layer $j$.
\State $w_{ij}$ (pA or mV) the average weight of synaptic strength in $i$ target $j$.
\State $w_{ext,j}$ (pA or mV) the average weight of synaptic strength in $X_{ext,j}$ to layer $j$.
\State $k$ the factor of rescaling.
\State $f_{ext,j}$ (Hz) the average firing rate of the external input target set $j$.
\State $f_i$ (Hz) the average firing rate of the presynaptic neurons.
\State $\tau _{syn}$ (ms) synapse time constant.

\vspace{0.3 cm}

1.NUMBER OF NEURONS

\vspace{0.2 cm}

\For{each $j$ in $n$}
    \State $N'_j \gets k*N_j$
    \State $X'_{ext,j} \gets k*X_{ext,j}$
\EndFor

\vspace{0.4 cm}

2.NUMBER OF CONNECTIONS
\vspace{0.2 cm}

\For{each $j$ in $n$}
\For{each $i$ in $n$}
    \State   $C_{ij}' \gets   C_{ij}$
    \State  $X_{ij}' \gets  k^2 *X_{ij} $  \Comment{ corolario of $X_{ij}$ = $C_{pre,pos}$*$N_i$*$N_j$} 
\EndFor
\EndFor

\vspace{0.4 cm}

3. SYNAPTIC STRENGHT
\vspace{0.2 cm}

\For{each $j$ in $n$}
\For{each $i$ in $n$}
    \State $w_{ij}' \gets w_{ij}/ \sqrt{k}$
\EndFor
\State $w_{ext,j}' \gets w_{ext,j}/ \sqrt{k}$
\EndFor

\vspace{0.4 cm}

4. THRESHOLD ADJUSTMENT

\vspace{0.2 cm}

\For{each $j$ in $n$}
    \State $q_{sum_j} = \sum_{i=1}^n w_{ij}  * f_i * X_{ij}/N_j$
    \State $q_{ext,j} = w_{ext,j}  * f_{ext,j}* X_{ext,j}$
\State $I_{DC_j}' = \tau _{syn} * (  
		        (1 - \sqrt{k}) * (q_{sum_j} + q_{ext,j}))$ \Comment{DC (pA  or mV) input to compensate resize}
\EndFor
\vspace{0.4 cm}

\State Done!
\end{algorithmic}
\end{algorithm}

\newpage

\subsection{Rescaling Applied}
\label{Rescaling Applied}
\vspace{0.4 cm}

All applications of this method presented in this publication were implemented in Python (with Brian2 or NetPyNE) and can be found on GitHub (https://github.com/ceciliaromaro/recoup-the-first-and-second-order-statistics-of-neuron-network-dynamics). The neuron model is the Leaky Integrate and Fire (LIF).

\vspace{0.6 cm}

\subsubsection{Sparse random connected network: Inhibitory neurons}
\label{sec:Sparse random connected network}
\vspace{0.4 cm}

For any pre-synaptic neuron i and any post-synaptic neuron j in the network, a fixed probability p of connection i-j is called random inner connection. For a probability p lower than 0.1 we can say it is sparse \cite{vreeswijk1998chaotic}.
The first illustrative application of this method is a network of inhibitory neurons with a sparse random inner connection (p<<1) and a Poisson external input. 
\vspace{0.4 cm}

\begin{figure}[H]
\centering

\begin{minipage}[b]{.49\textwidth}
\subfloat
  []
  {\label{fig:rescaling_graf_01a}\includegraphics[width=\textwidth,height=8.65cm]{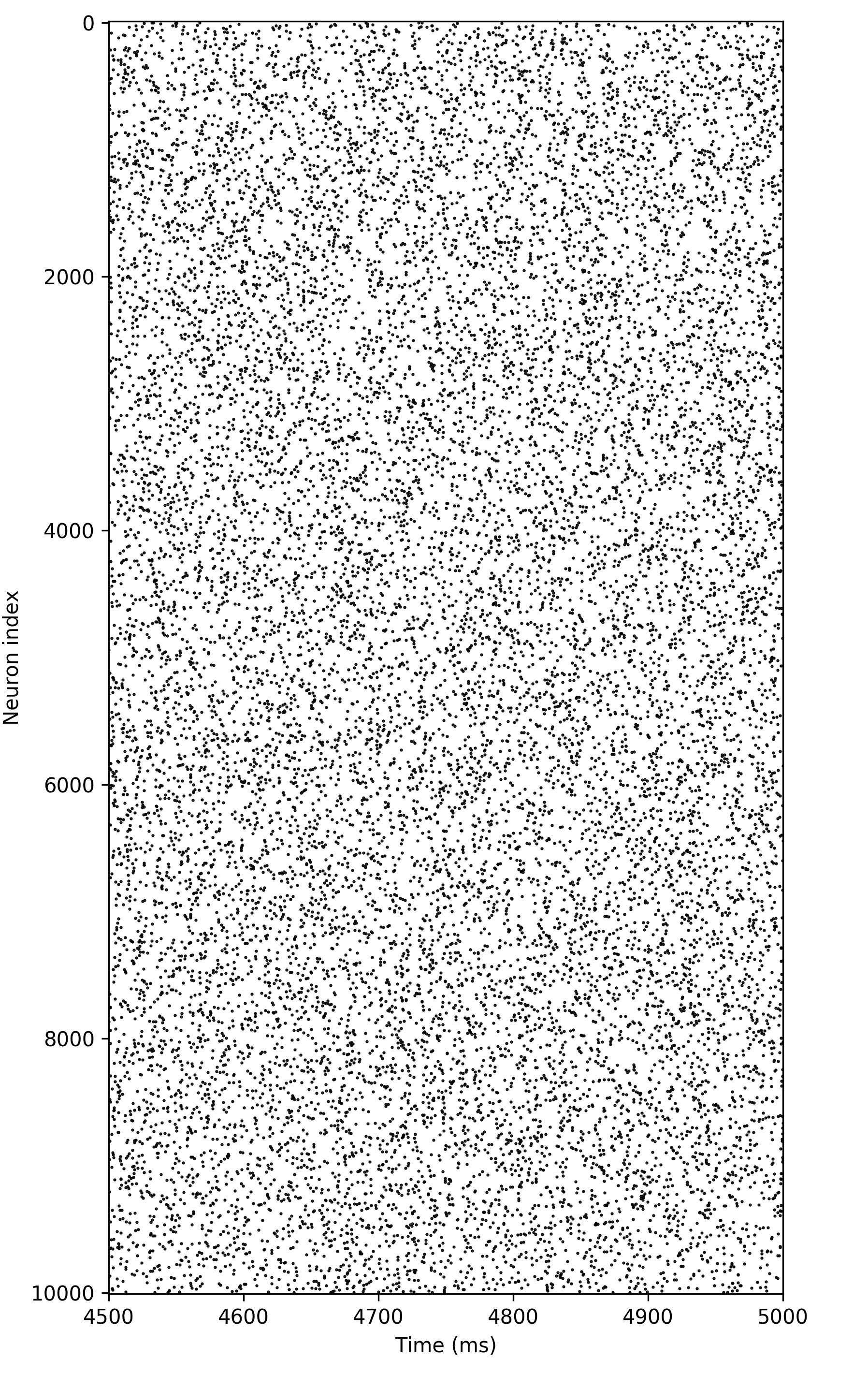}}
  
\subfloat
  []
  {\label{fig:rescaling_graf_01b}\includegraphics[width=\textwidth,height=0.35cm]{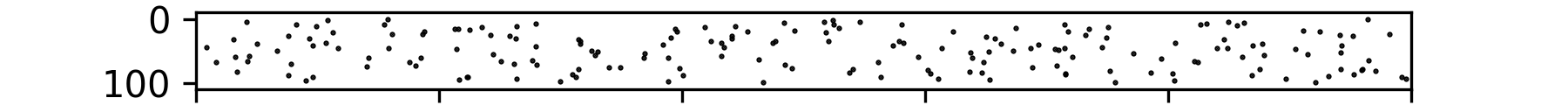}}

\end{minipage}
\begin{minipage}[b][\ht\measurebox]{.49\textwidth}
\centering
\vfill

\subfloat
  []
  {\label{fig:rescaling_graf_01c}\includegraphics[width=\textwidth,height=4.5cm]{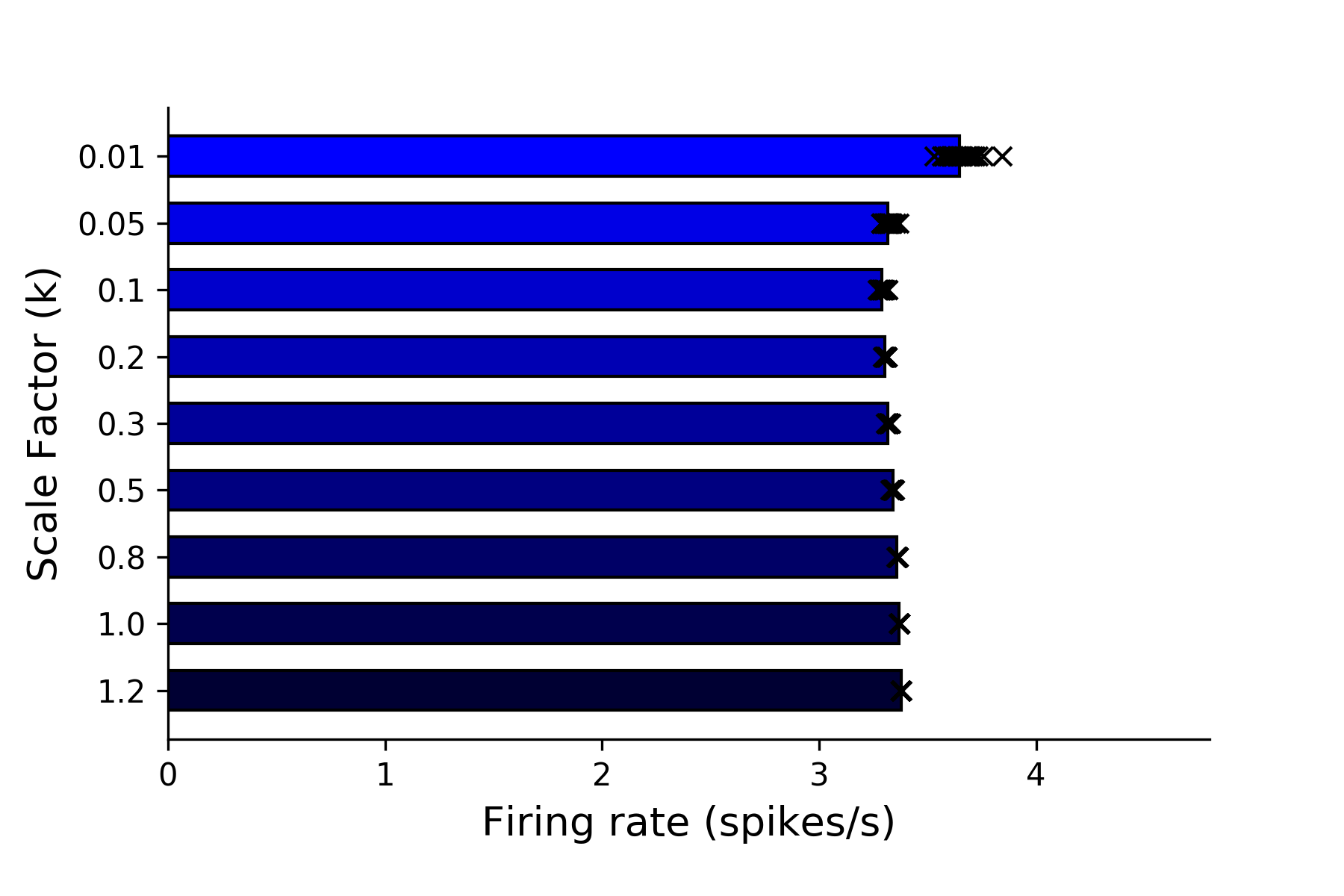}}
  
\subfloat
  []
  {\label{fig:rescaling_graf_01d}\includegraphics[width=\textwidth,height=4.5cm]{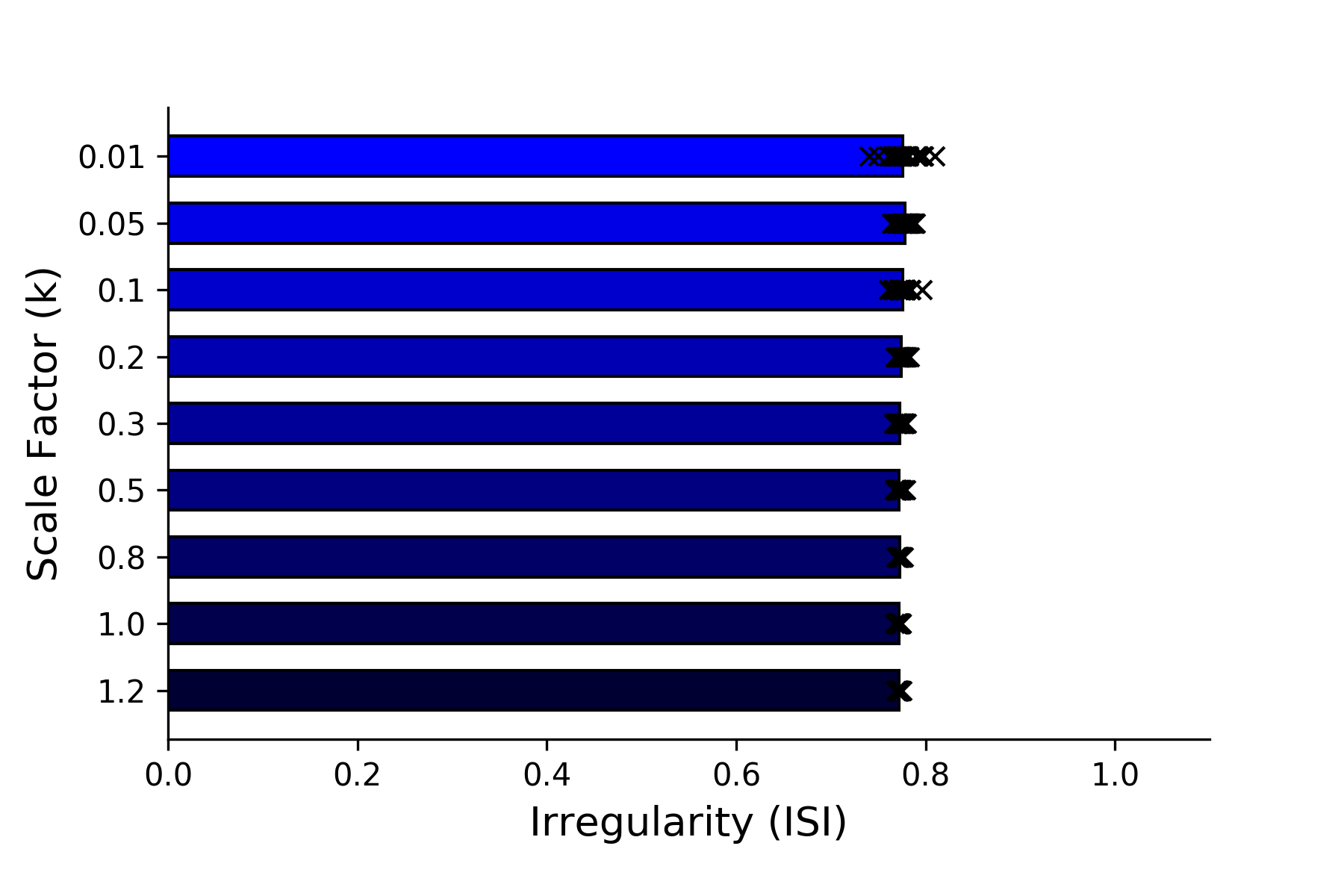}}

\end{minipage} 
\caption{Sparse random connected inhibitory neurons network model. Raster plot for (A) full network and (B) network rescaled to 1\% of the number of total neurons. (C) The average firing rate and (D) Inter Spike Interval (ISI) for different scale factors k (120\%, 100\%, 80\%, 50\%, 30\%, 20\%, 10\%, 5\%, 1\%). The 'x' is value for each simulation run and the bar is the average of the set run.}
\label{fig:rescaling_graf_01}
\end{figure}

In this appliance we rescale the network up to 1\%. The network parameters before and after rescaling are available on Table \ref{tab:rescaling_parameters_01}. Figure \ref{fig:rescaling_graf_01} presents the raster plots for the full network (\ref{fig:rescaling_graf_01a}) and for the rescaled network to 1\% (\ref{fig:rescaling_graf_01b}) and presents, for the network rescaled in different sizes (120\%, 100\%, 80\%, 50\%, 30\%, 20\%, 10\%, 5\%, 1\%): the average firing rate, a first order statistic (\ref{fig:rescaling_graf_01c}); and the inter spike interval (ISI): a second order statistic (\ref{fig:rescaling_graf_01d}). The difference between average  frequency is less the 8.5\%, and the ISI is less than 1\%. 
\vspace{0.4 cm}

\begin{table}[H]
\center{\begin{tabular}{llll}
\hline
Parameter description                  &  Variable   & Full scale  & Rescaling  \\ \hline
Factor of rescaling  & $k$ &  - & 0.01 \\
Number of inhibitory neurons  & $N_-$  & $10^4$  & $10^2$\\
Number of external input to each neuron & $X_{ext}$ &  2300 & 23 \\
Total number of inner connection & $X$  & $1,000,000$ & $100$ \\
Weight of excitatory synaptic strength & $w \pm \delta w$  (pA)  & 30$\pm$3 & 300 $\pm$30 \\ \\ \hline
Probability of connection              & $p$         &  $0.05$  & $0.05$ \\
Absolute refractory period & $\tau _{ref}$ (ms)  & 2 & 2 \\
Synapse time constant &  $\tau _{syn}$ (ms) & 0.5 & 0.5 \\
Membrane time constant & $\tau _{m}$ (ms) & 10 & 10 \\
Synaptic transmission delays  & $\Delta_{t} \pm \delta \Delta_{t} $ (ms)  & 1.5 $\pm$ 0.75 & 1.5 $\pm$ 0.75 \\
Membrane capacitance &  $C _{m}$ (pF) & 250 & 250 \\
Inhibitory/excitatory synaptic strength & $g$   & -2 & -2 \\ 
Reset potential (mV)& $V_{ret}$   & -65 & -65 \\ 
Fixed firing threshold (mV) & $V_{th}$   & -45 & -45 \\ 
the average firing rate of the external input & $f_{ext}$ (Hz) & 8 & 8\\  \\ \hline
\end{tabular}
\caption{Sparse random connected inhibitory neurons network model specification before and after of the rescaling: parameters and metrics.} \label{tab:rescaling_parameters_01}}
\end{table}
\vspace{0.4 cm}

\vspace{0.4 cm}

\vspace{0.6 cm}

\subsubsection{Avalanche network: Excitatory-inhibitory interconnected}
\vspace{0.4 cm}

The avalanche can be defined as the rise of activity above some basal or threshold level \cite{beggs2003neuronal}. This rise is triggered by the activation of a few or a single neuron, producing a cascade of firings that returns below threshold after some time. This process can have particular statistical properties like power law distribution of size and duration. 

In other works, the avalanche is a quick rise in the network activity, locally or systemic, followed by a sudden downgrade in the activity back to the previous equilibrium. This rise in activity is triggered by the activation of a few neurons with a feedback connection able to change the average activity.
\vspace{0.4 cm}

The second application of this method is a 2-layer excitatory-inhibitory neurons network with a sparse random inner connection (p<<1) and a Poisson external input. This network is similar to the first one however using excitatory neurons with inner connection able to produce avalanche. 
\vspace{0.4 cm}

\begin{figure}[H]
\centering

\begin{minipage}[b]{.49\textwidth}
\subfloat
  []
  {\label{fig:rescaling_graf_02A}\includegraphics[width=\textwidth,height=6cm]{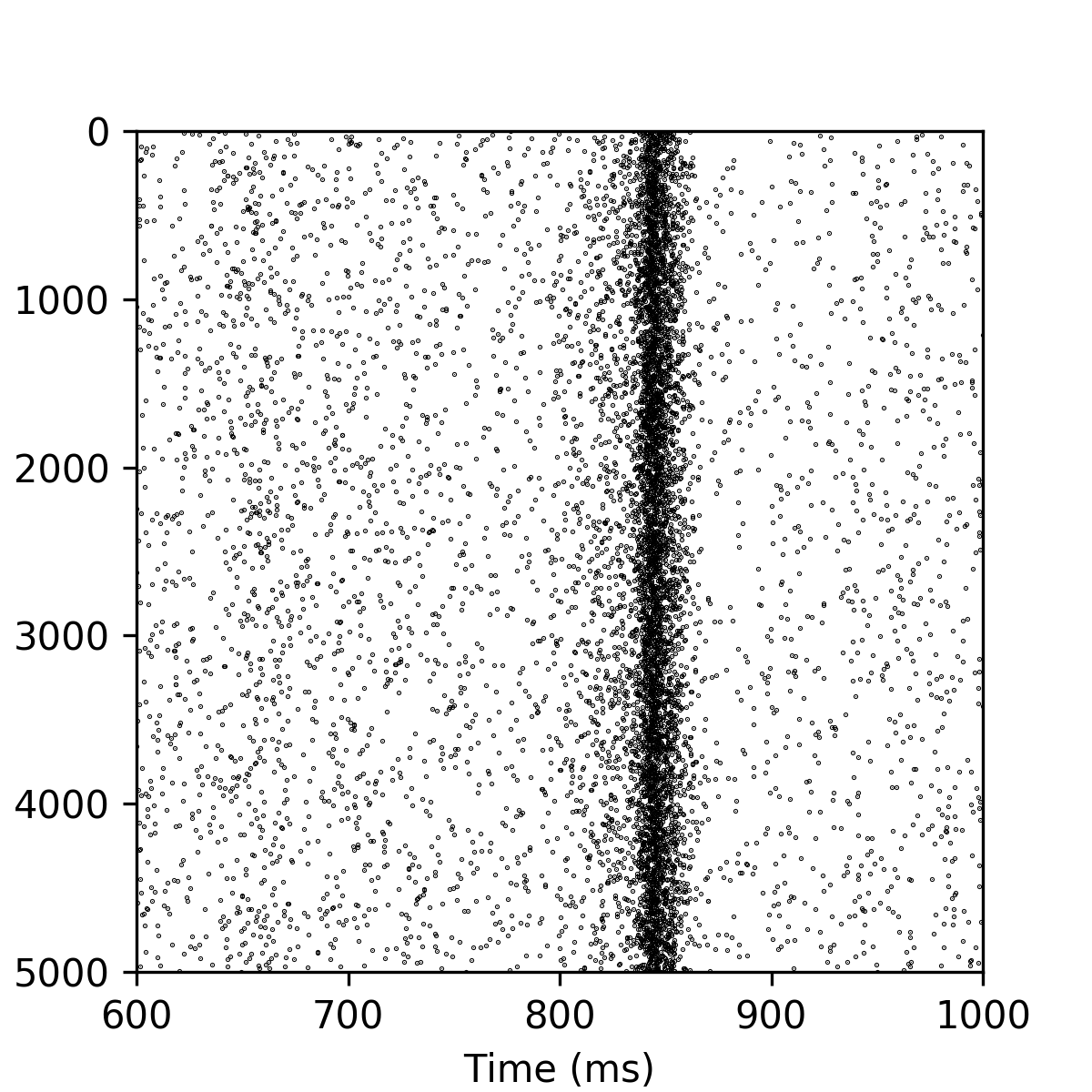}}
  
\subfloat
  []
  {\label{fig:rescaling_graf_02B}\includegraphics[width=\textwidth,height=3cm]{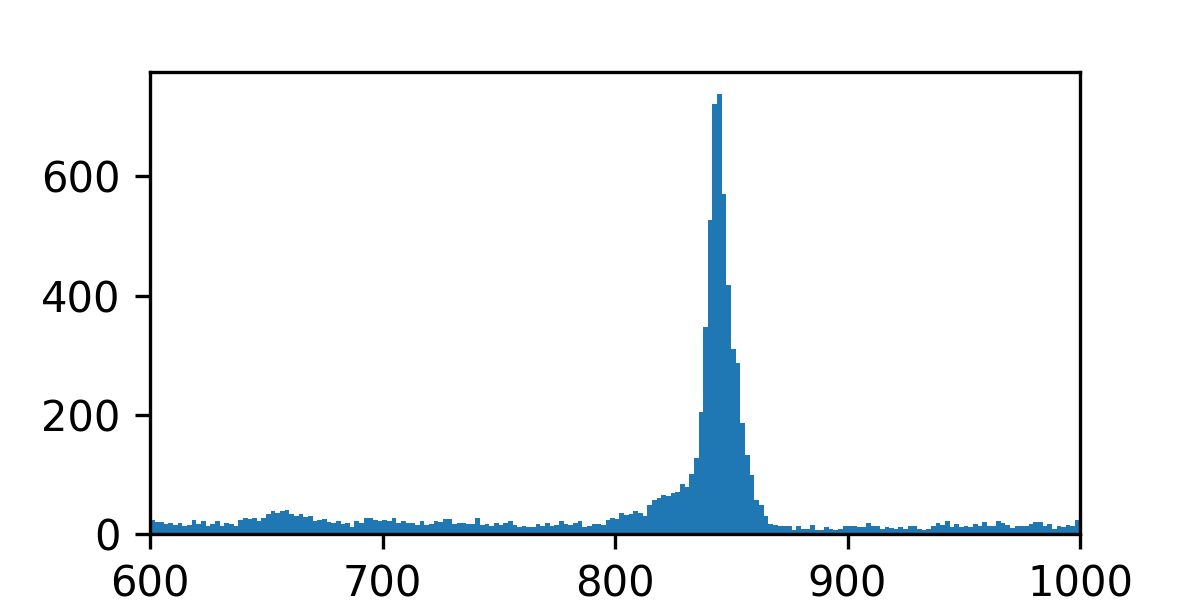}}

\end{minipage}
\begin{minipage}[b][\ht\measurebox]{.49\textwidth}
\centering
\vfill

\subfloat
  []
  {\label{fig:rescaling_graf_02C}\includegraphics[width=\textwidth,height=6cm]{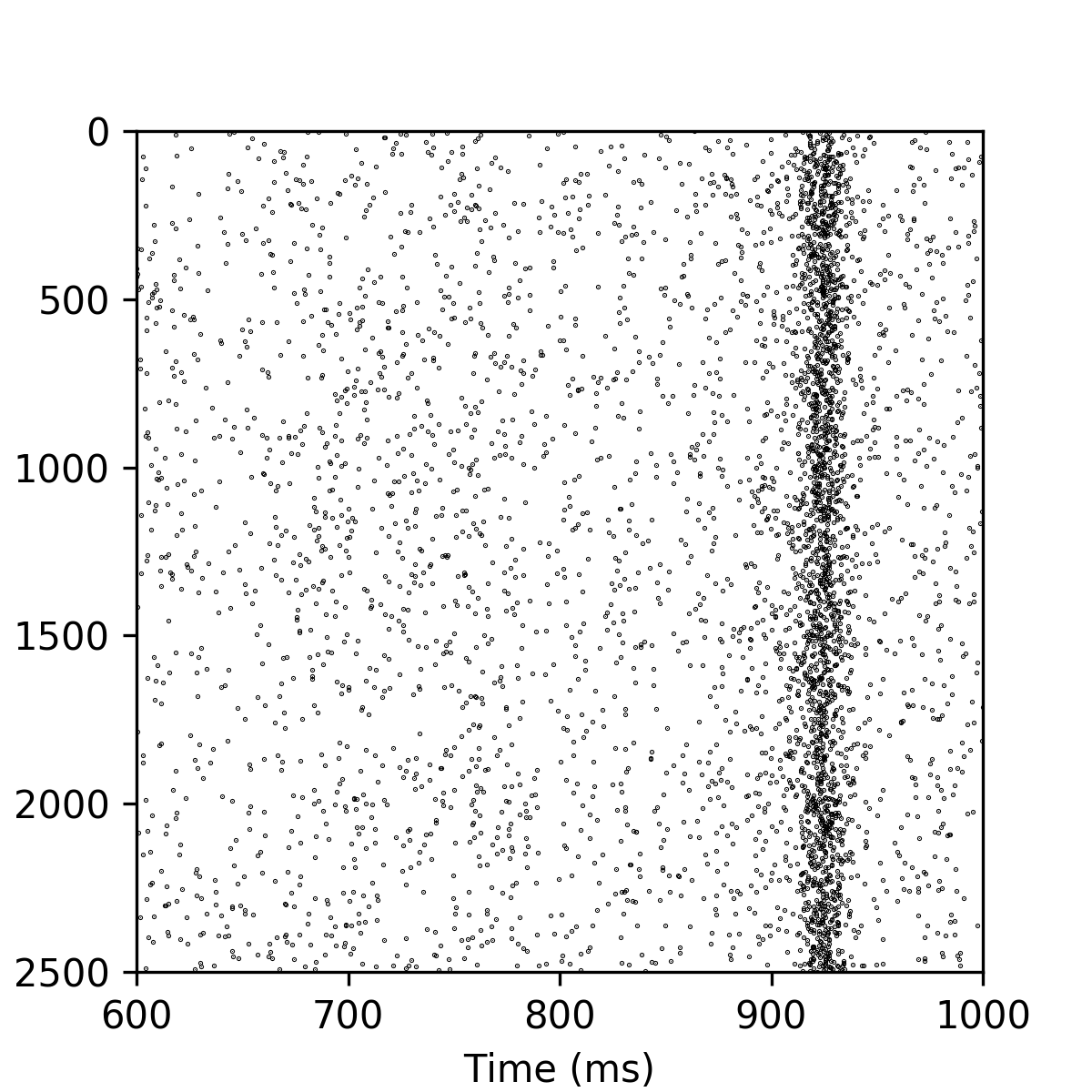}}
  
\subfloat
  []
  {\label{fig:rescaling_graf_02D}\includegraphics[width=\textwidth,height=3cm]{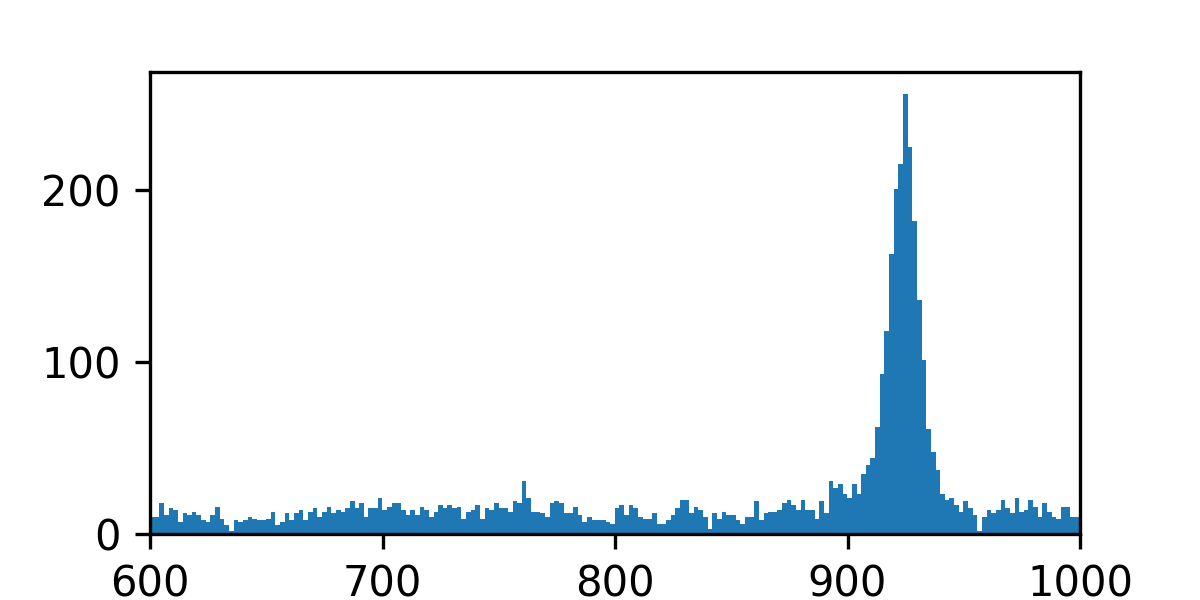}}

\end{minipage} 
\caption{Excitatory and Inhibitory interconnected neurons network with avalanches. (A) Raster plot and (B) spikes histogram for full network and (C) raster plot and (D) spikes histogram for Network rescaled to 50\% of the number of total neurons.}
\label{fig:rescaling_graf_02}
\end{figure}

In this appliance we rescale the network up to 50\%. The network parameters before and after rescaling are available on Table \ref{tab:rescaling_parameters_02}. The Figure \ref{fig:rescaling_graf_02} presents the raster plots and spike histogram for the full network and for the rescaled network to 50\%. It is possible to see the avalanche in both cases. 
\vspace{0.4 cm}

\begin{table}[H]
\center{\begin{tabular}{llll}
\hline
Parameter description                  &  Variable   & Full scale  & Rescaling  \\ \hline
Factor of rescaling  & $k$ &  - & 0.5 ou colocar a 20\%? \\
Number of excitatory neurons  & $N_+$  & 4000 & 2000  \\
Number of inhibitory neurons  & $N_-$  & 1000  & 500\\
Number of external input to each neuron & $X_{ext}$ &  50 & 25 \\
Total number of inner connection & $X$  & 250k & 62.5k \\
Weight of excitatory synaptic strength & $w \pm \delta w$  (pA)  & 400$\pm$40 & 566 $\pm$57 \\ \\ \hline
Probability of connection              & $p$         &  0.01  & 0.01 \\
Absolute refractory period & $\tau _{ref}$ (ms)  & 2 & 2 \\
Synapse time constant &  $\tau _{syn}$ (ms) & 0.5 & 0.5 \\
Membrane time constant & $\tau _{m}$ (ms) & 10 & 10 \\
Synaptic transmission delays  & $\Delta_{t} \pm \delta \Delta_{t} $ (ms)  & 1.5 $\pm$ 0.75 & 1.5 $\pm$ 0.75 \\
Membrane capacitance &  $C _{m}$ (pF) & 250 & 250 \\
Inhibitory/excitatory synaptic strength & $g$   & -4 & -4 \\ 
Reset potential (mV)& $V_{ret}$   & -65 & -65 \\ 
Fixed firing threshold (mV) & $V_{th}$   & -50 & -50 \\ 
the average firing rate of neurons & $f$ (Hz) &  & \\
the average firing rate of the external input & $f_{ext}$ (Hz) & 8 & 8\\  \\ \hline
\end{tabular}
\caption{Inhibitory-excitatory neurons network model specification before and after of the rescaling: parameters and metrics.} \label{tab:rescaling_parameters_02}}
\end{table}
\vspace{0.4 cm}

\vspace{0.6 cm}

\subsubsection{PD \cite{potjans2012cell} network: Eight layers excitatory-inhibitory interconnected Network}
\label{sec:PD network: Eight layers excitatory-inhibitory interconnected Network}

\vspace{0.4 cm}

The PD \cite{potjans2012cell} model is a four excitatory-inhibitory interconnected layers (eight sets of neurons) network with external Poisson input and some parameters based on biological data. This model is able to reproduce the average firing rate of the somatosensory cortex observed in \textit{vivo}.
\vspace{0.4 cm}

\begin{figure}[H]
\centering

\begin{minipage}[b]{.24\textwidth}
\subfloat
  []
  {\label{fig:rescaling_graf_03A}\includegraphics[width=\textwidth,height=9cm]{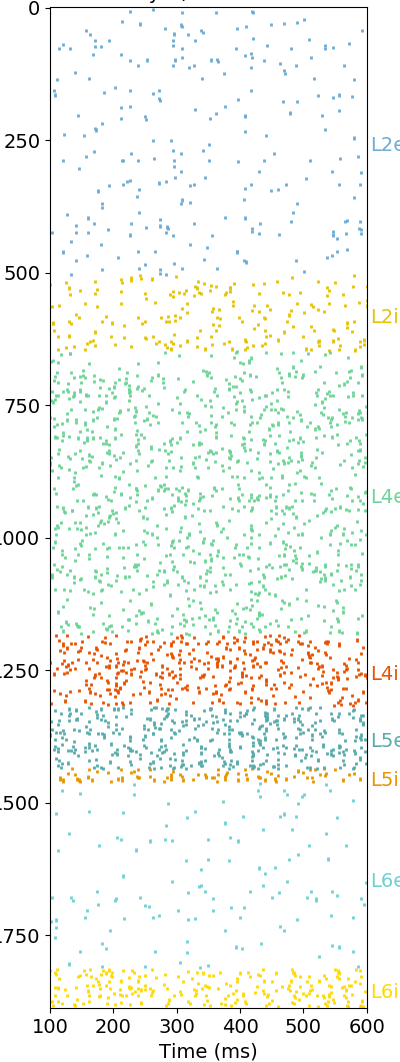}}
  \end{minipage}
\begin{minipage}[b]{.24\textwidth}
\subfloat
  []
  {\label{fig:rescaling_graf_03B}\includegraphics[width=\textwidth,height=4.2cm]{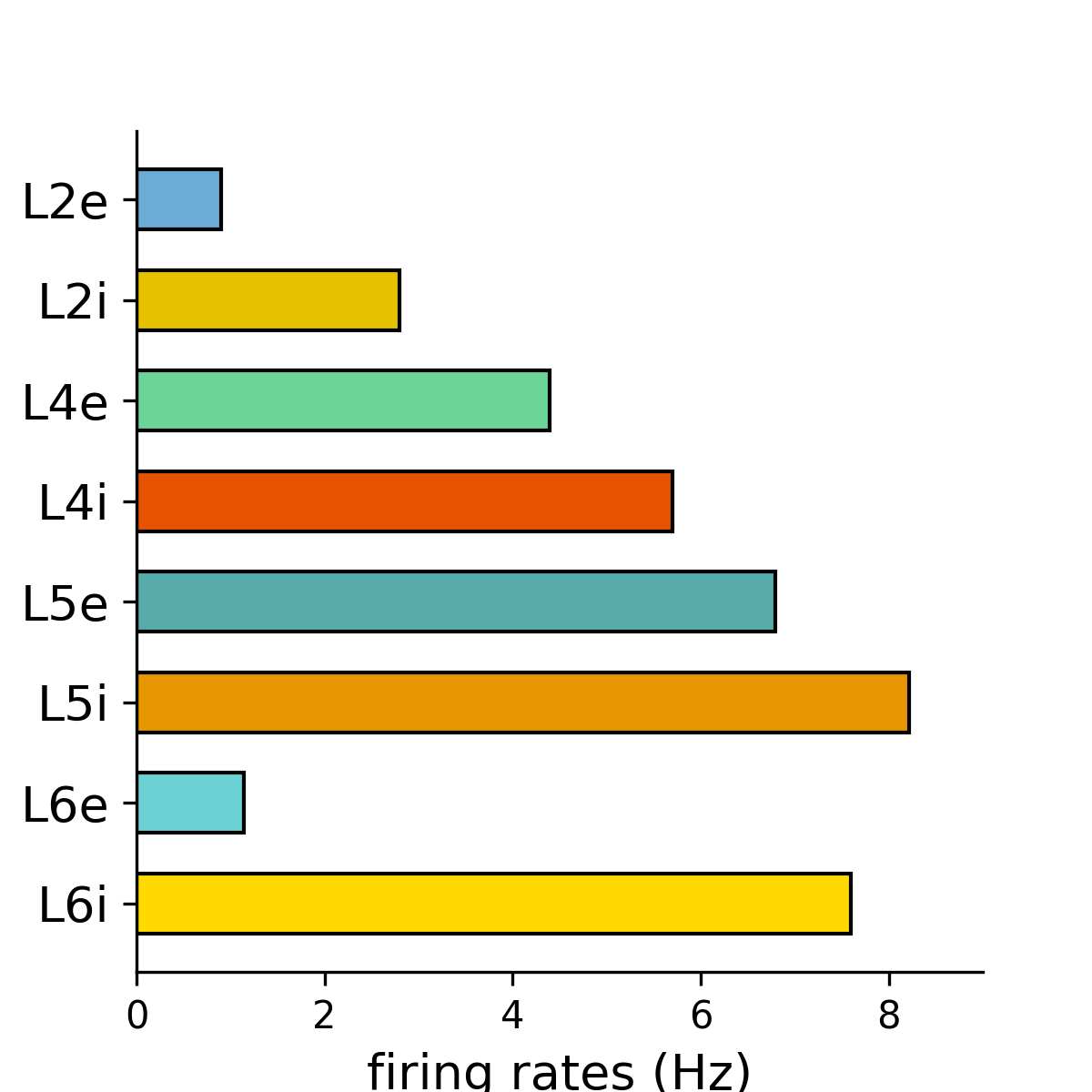}}
  
\subfloat
  []
  {\label{fig:rescaling_graf_03C}\includegraphics[width=\textwidth,height=4.2cm]{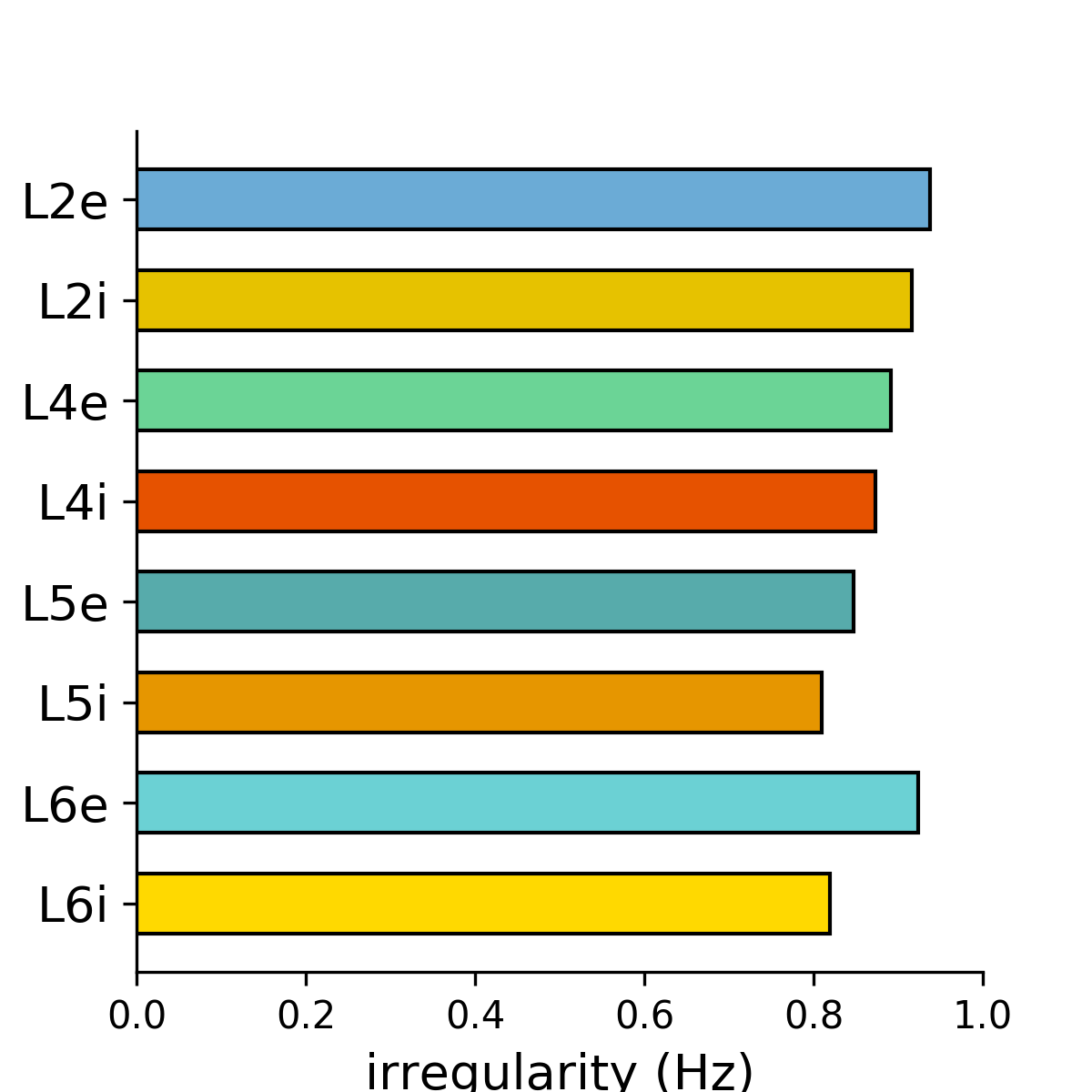}}

\end{minipage}
\begin{minipage}[b]{.24\textwidth}
  \subfloat
    []
    {\label{fig:rescaling_graf_03D}\includegraphics[width=\textwidth,height=9cm]{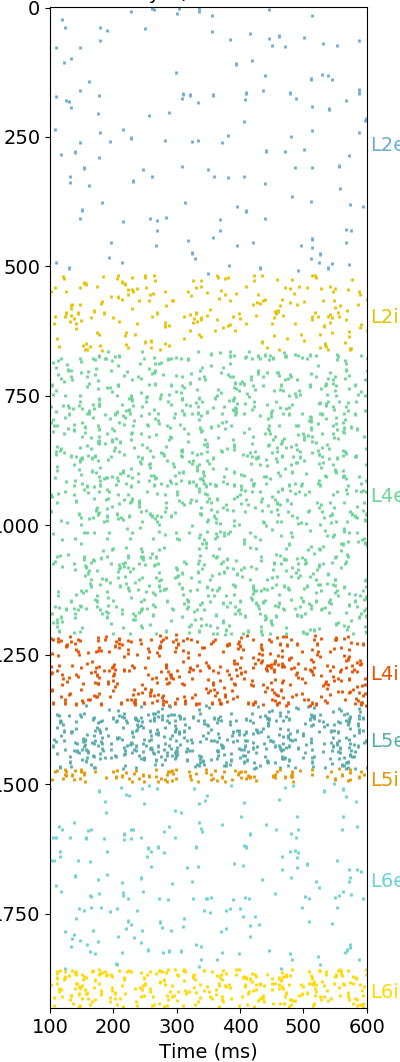}}
  \end{minipage}
\begin{minipage}[b][\ht\measurebox]{.24\textwidth}
\centering
\vfill

\subfloat
  []
  {\label{fig:rescaling_graf_03E}\includegraphics[width=\textwidth,height=4.2cm]{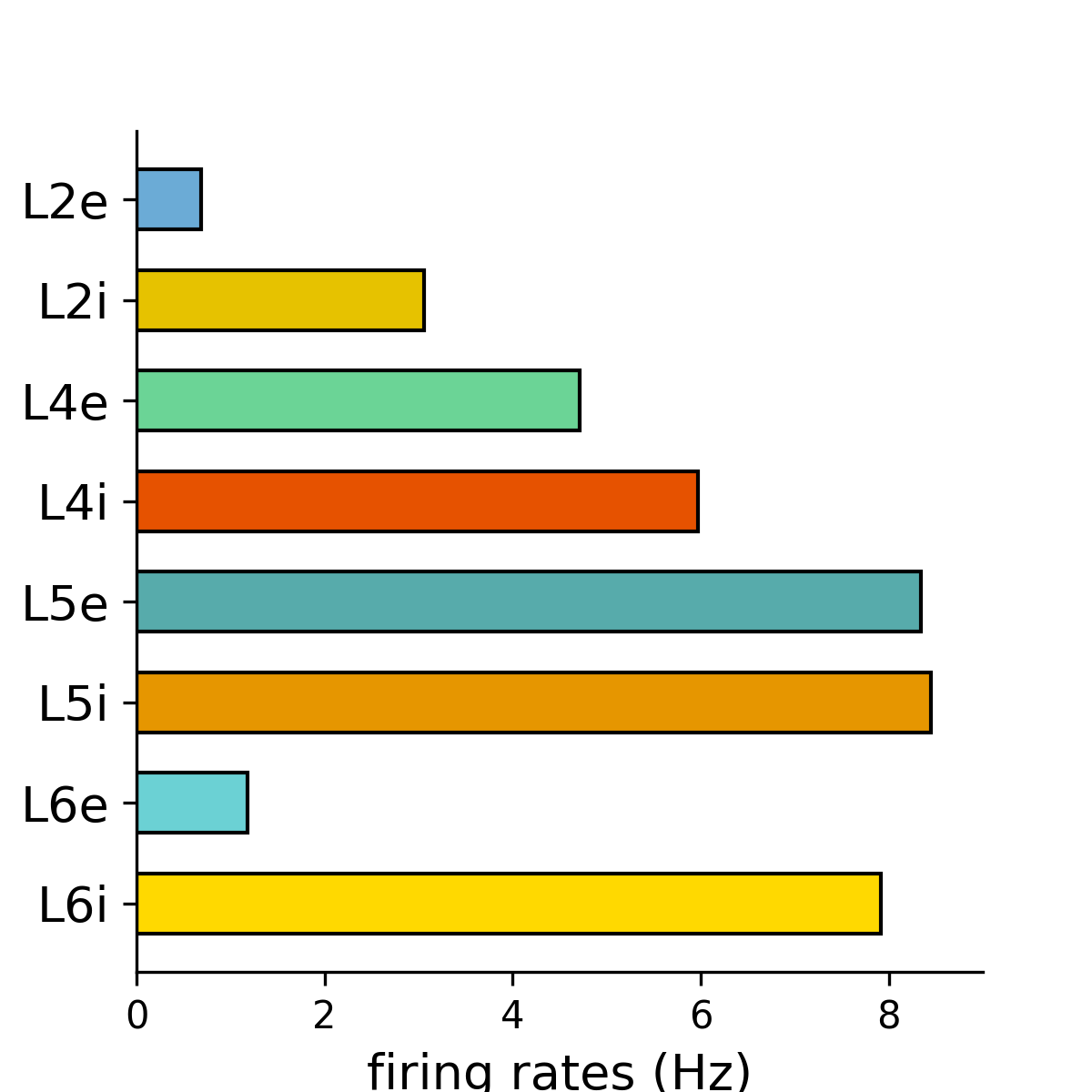}}
  
\subfloat
  []
  {\label{fig:rescaling_graf_03F}\includegraphics[width=\textwidth,height=4.2cm]{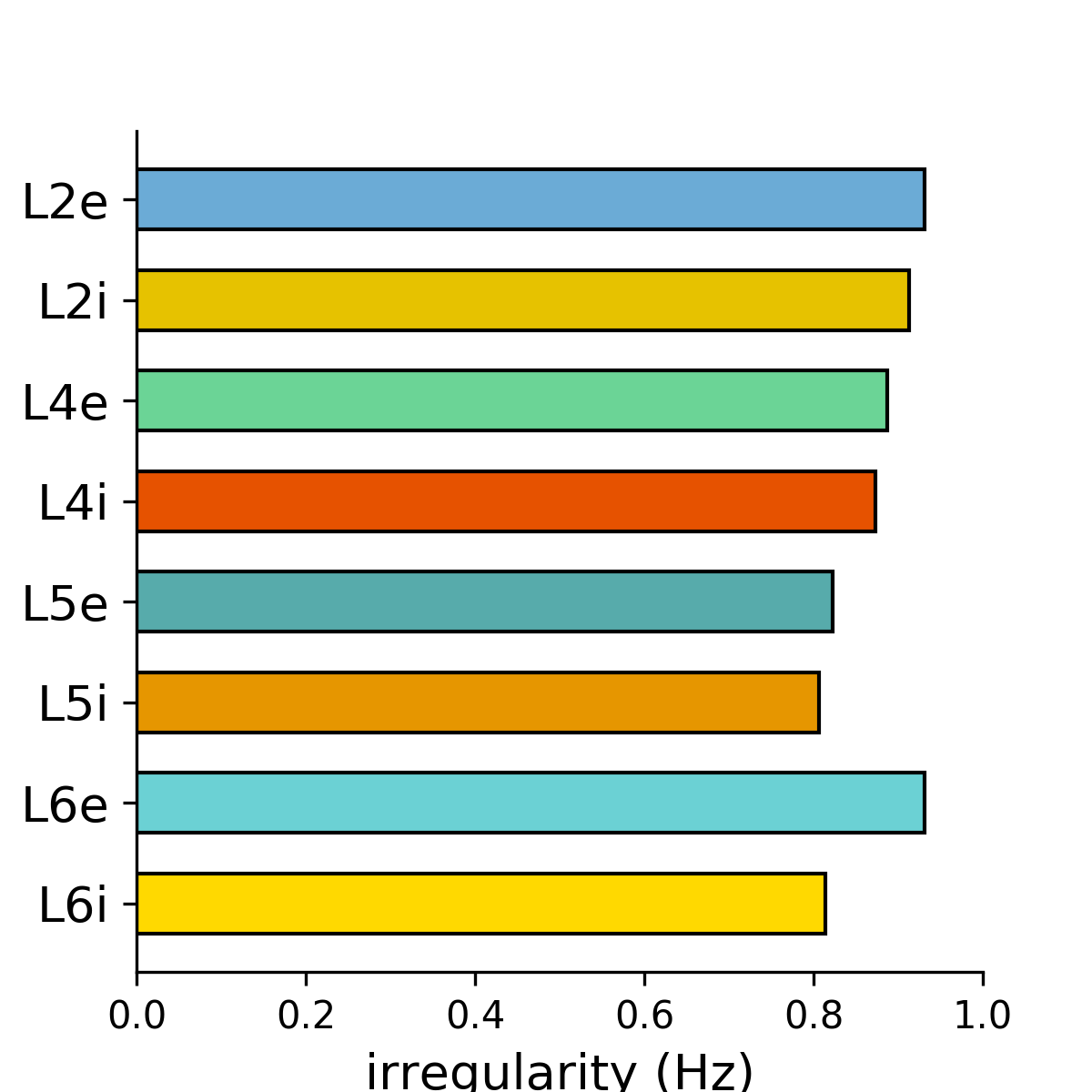}}

\end{minipage} 
\caption{Reproduction of figure 6 of \cite{potjans2012cell} (A,B,C) and Network rescaled to 30\% the number of total neurons (D,E, F):  (A) Raster plot of firing rate of the eight neuron population: 2/3 , 4, 5 and 6 for excitatory and inhibitory neurons.  The number of neurons per layer shown is proportional to the full scale of the network, resulting in a total number of approximately 1850 neurons plotted.  (B) Boxplot of 60 seconds of single unit firing rate for each population.  (C) Irregularity estimated by coefficient of variance of the interspike interval of a 60 seconds simulation. (D) Raster plot and (E,F) statistics as (A,B,C). The simulation times and number of neurons plotted were chosen as in the full scale. Adapted from \cite{romaro2018implementation}}
\label{fig:rescaling_graf_03}
\end{figure}

The rescaling of this network was implemented, discussed in detail and published at [minhaPDpublicacao]. The following shows (Table \ref{tab:rescaling_parameters_03} and Figure \ref{fig:rescaling_graf_03}) an overview of the PD rescaling network up to 30\% of the full version (k=0.3), which means less than 10\% of the total number of connections ($k^2*X$) remained and all network behavior, firing-rate specific per layer, and irregularity metrics were maintained. In \cite{romaro2018implementation} this reduction reaches 1\% of total number of neurons (k=0.01), 10 neurons for layer 5i and 0.01\% of total number of inner connections ($k^2*X$) with its limitations discussed.
\vspace{0.4 cm}

Table \ref{tab:rescaling_parameters_03} presents an overview of the network dimensions and parameters before and after rescaling to 30\%. Figure \ref{fig:rescaling_graf_03} presents the raster plots, the average firing rate per layer: a first order statistic; and the inter spike interval (ISI) per layer: a second order statistic for the full scale network and the rescaled network to 30\%.
\vspace{0.4 cm}

\begin{table}[H]
\center{\begin{tabular}{llll}
\hline
Parameter description                  &  Variable   & Full scale  & Rescaling  \\ \hline
Factor of rescaling  & $k$ &  - & 0.3 \\
Number of excitatory neurons  & $N_+$  & $\approx$ 62k* & $\approx$18.5k  \\
Number of inhibitory neurons  & $N_-$  & $\approx$ 15k*  & $\approx$4.5k\\
Number of external input to each neuron & $X_{ext}$ &  $\approx$ 2k* & $\approx$600 \\
Total number of inner connection & $X$  &  $\approx$300M &  $\approx$27M \\
Weight of excitatory synaptic strength & $w \pm \delta w$  (pA)  & 87.8$\pm$8.8 & 160 $\pm$16 \\ \\ \hline
Probability of connection              & $p$         &  $C$*  & $C$ \\
Absolute refractory period & $\tau _{ref}$ (ms)  & 2 & 2 \\
Synapse time constant &  $\tau _{syn}$ (ms) & 0.5 & 0.5 \\
Membrane time constant & $\tau _{m}$ (ms) & 10 & 10 \\
Synaptic transmission delays  & $\Delta_{t} \pm \delta \Delta_{t} $ (ms)  & $\Delta_{t}$* & $\Delta_{t}$ \\
Membrane capacitance &  $C _{m}$ (pF) & 250 & 250 \\
Inhibitory/excitatory synaptic strength & $g$   & -4 & -4 \\ 
Reset potential (mV)& $V_{ret}$   & -65 & -65 \\ 
Fixed firing threshold (mV) & $V_{th}$   & -50 & -50 \\ 
the average firing rate of neurons & $f$ (Hz) &  $f$* & $f$ \\
the average firing rate of the external input & $f_{ext}$ (Hz) & 8 & 8\\  \\ \hline
\end{tabular}
\caption{PD \cite{potjans2012cell} eight layers excitatory-inhibitory interconnected Network model specification before and after of the rescaling: parameters and metrics. (*)The value vary for each layer or type of neuron: see the specification in Table 5 from original article \cite{potjans2012cell}.} \label{tab:rescaling_parameters_03}}
\end{table}
\vspace{0.6 cm}

\vspace{0.4 cm}

\subsubsection{Brunel \cite{brunel2000dynamics} network: Excitatory-inhibitory interconnected}
\label{sec:Brunel network: Excitatory-inhibitory interconnected}
\vspace{0.4 cm}

The Brunel \cite{brunel2000dynamics} network model is an excitatory-inhibitory 2-layers neurons network
with a sparse random inner connection (p=0.1) and a DC external input. The network behavior depends on the proportion of inhibitory/excitatory weight of synaptic strength, g, the proportion of the DC external input $\Theta$, and the fixed firing threshold $V_{th}$. The variation of those proportions is able to change the average firing rate frequency, synchrony and irregularity of network.
\vspace{0.4 cm}

\begin{figure}[H]
\centering

\begin{minipage}[b]{.24\textwidth}
\subfloat
  []
  {\label{fig:Ex04A1}\includegraphics[width=\textwidth,height=3cm]{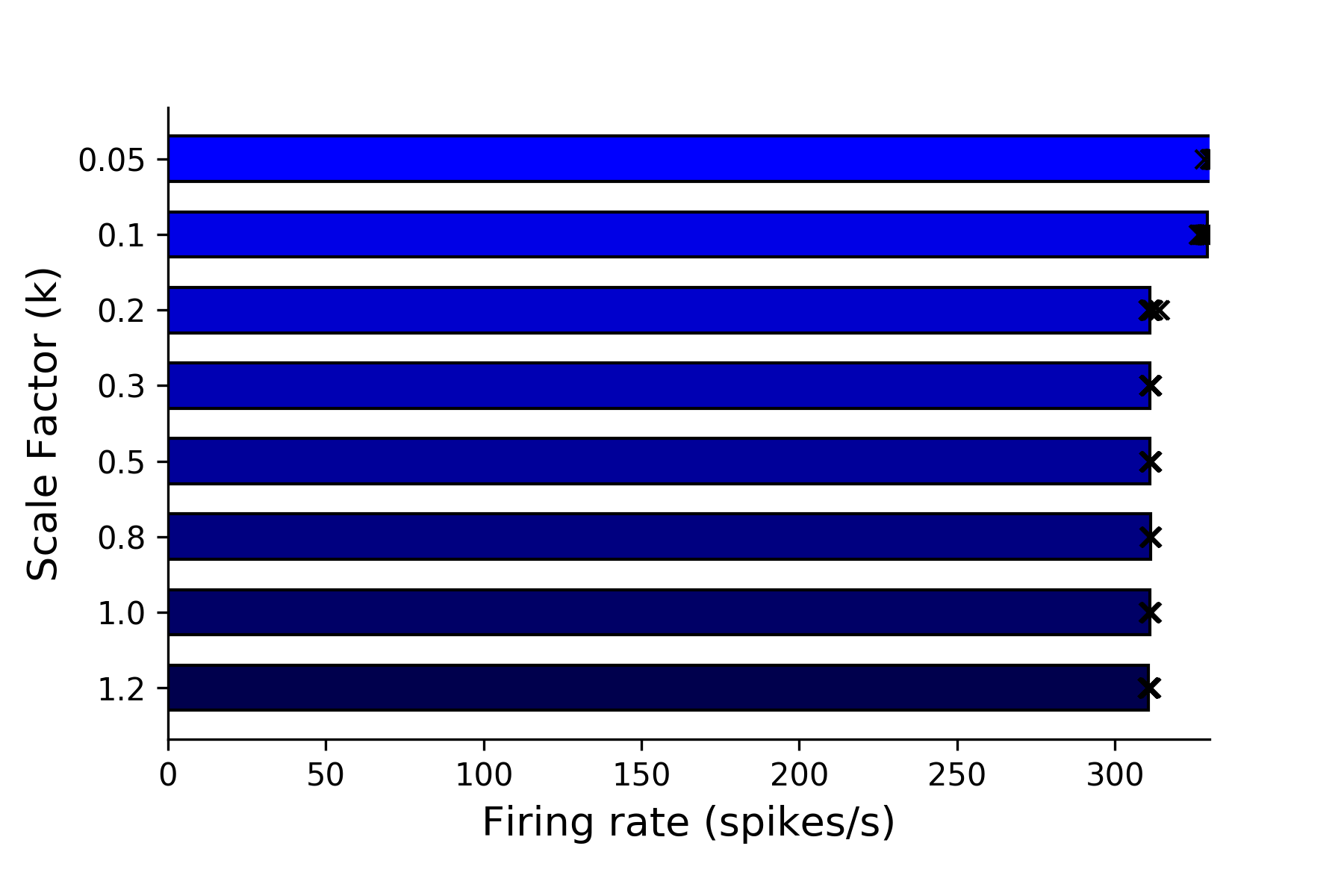}}

  \end{minipage}
\begin{minipage}[b]{.24\textwidth}
\subfloat
  []
  {\label{fig:Ex04B1}\includegraphics[width=\textwidth,height=3cm]{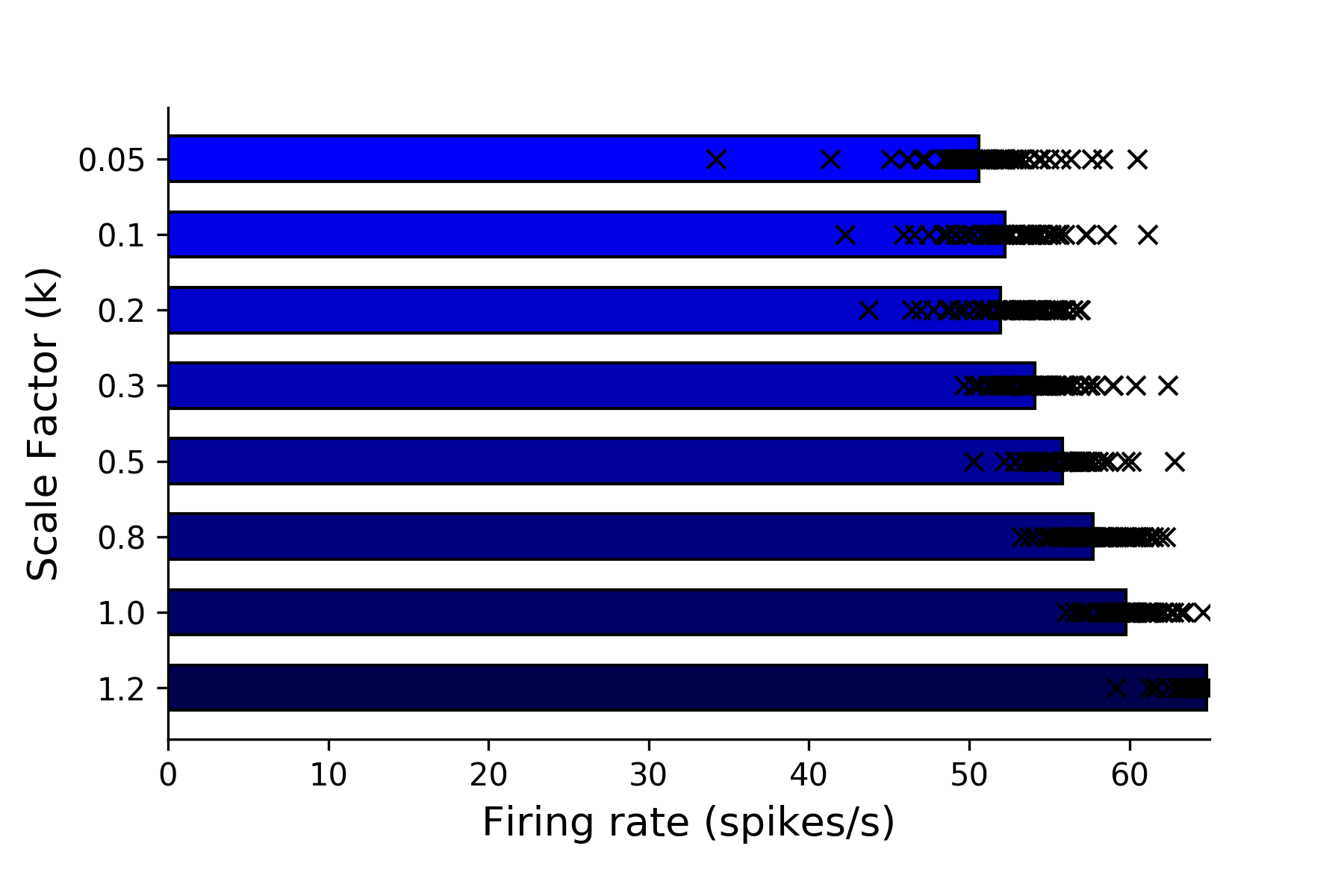}}
  
\end{minipage}
\begin{minipage}[b]{.24\textwidth}
\subfloat
  []
  {\label{fig:Ex04C1}\includegraphics[width=\textwidth,height=3cm]{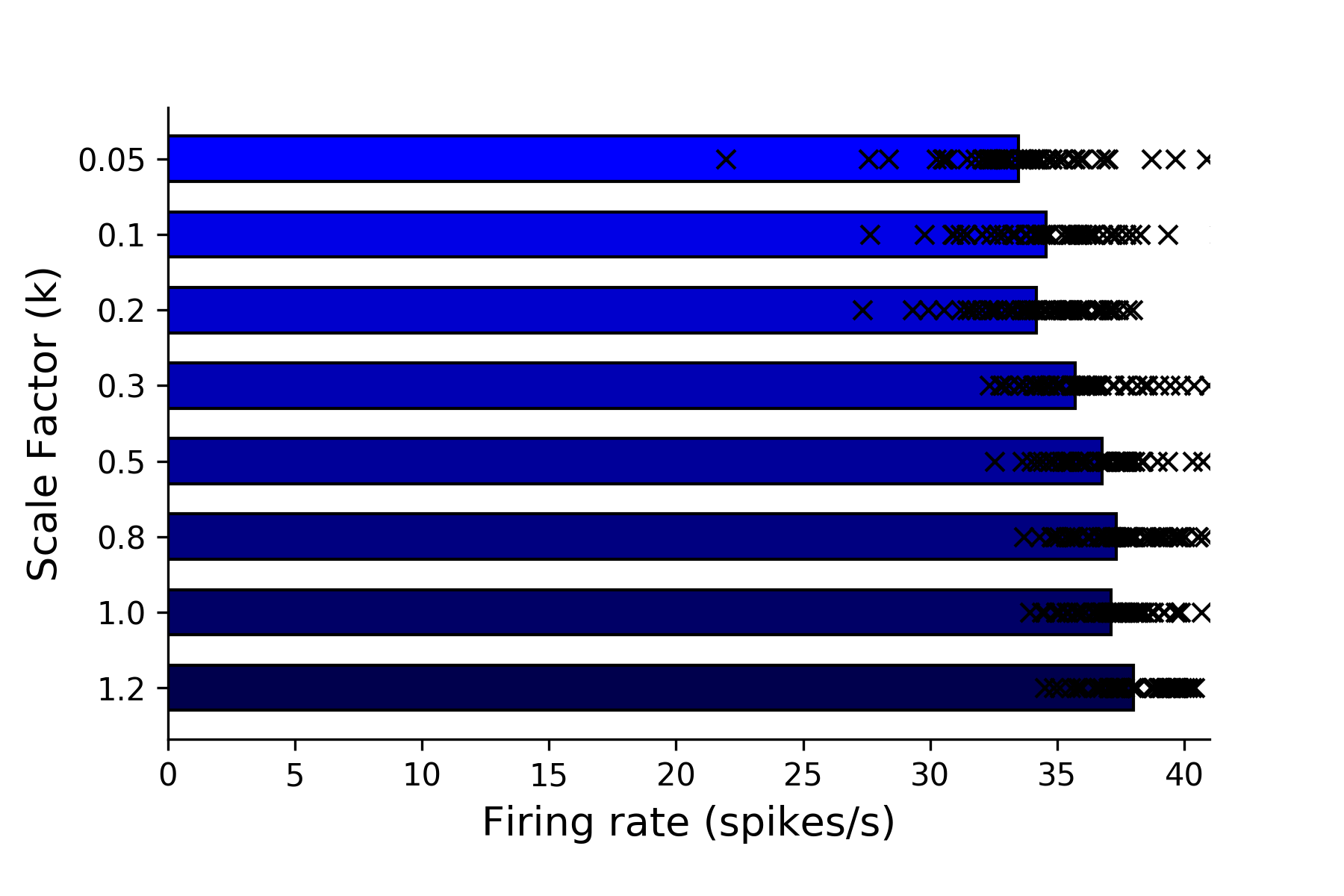}}
  
\end{minipage}
\begin{minipage}[b]{.24\textwidth}
\subfloat
  []
  {\label{fig:Ex04D1}\includegraphics[width=\textwidth,height=3cm]{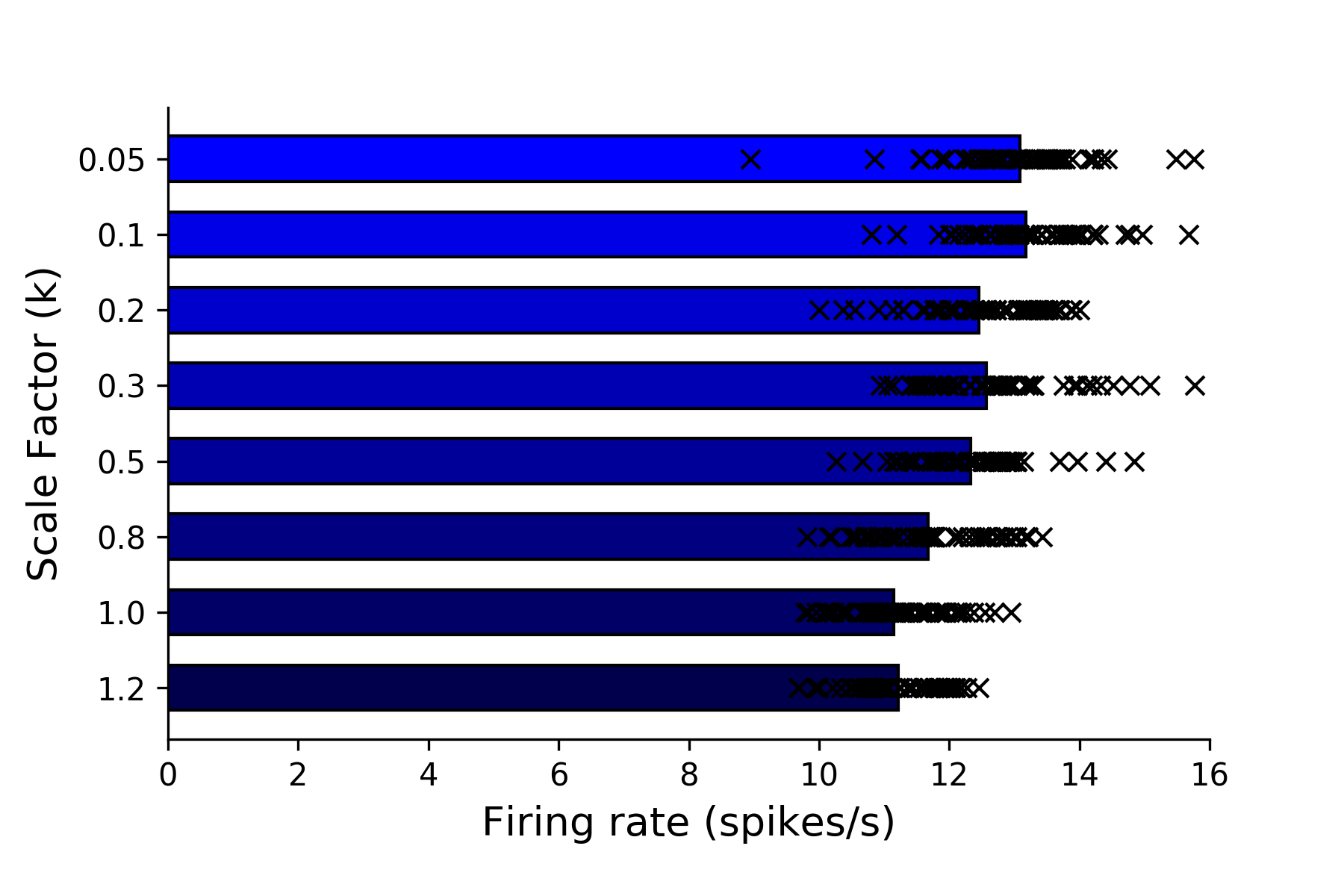}}

\end{minipage}
\caption{Average firing rate of Brunel \cite{brunel2000dynamics} network for different rescaling size (120\%, 100\%, 80\%, 50\%, 30\%, , 20\%, 10\%, 5\%) runned 30 turns during 10s each one. (A) $g=3$ and $\Theta = 2.V_{th}$; (B) $g=6$ and $\Theta=4.V_{th}$, (C) $g=5$ and $\Theta = 2.V_{th}$,  and (D) $g=4.5$ and $\Theta= 1.001.V_{th}$. The 'x' is value for each simulation run and the bar is the average of the set run. }
\label{fig:rescaling_graf_04a}
\end{figure}

\begin{figure}[H]
\centering

\begin{minipage}[b]{.24\textwidth}
\subfloat
  []
  {\label{fig:Ex04bA1}\includegraphics[width=\textwidth,height=3cm]{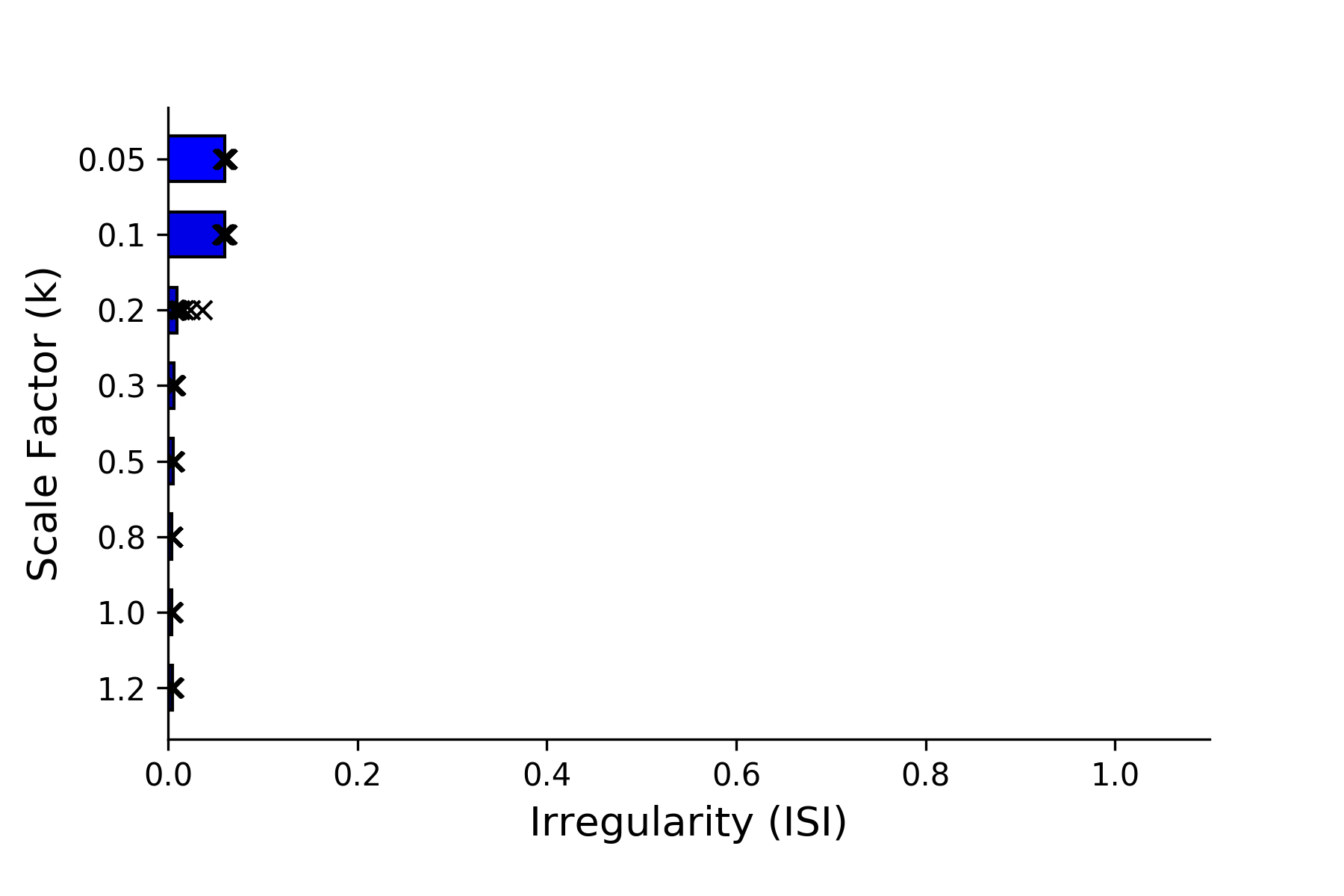}}

  \end{minipage}
\begin{minipage}[b]{.24\textwidth}
\subfloat
  []
  {\label{fig:Ex04bB1}\includegraphics[width=\textwidth,height=3cm]{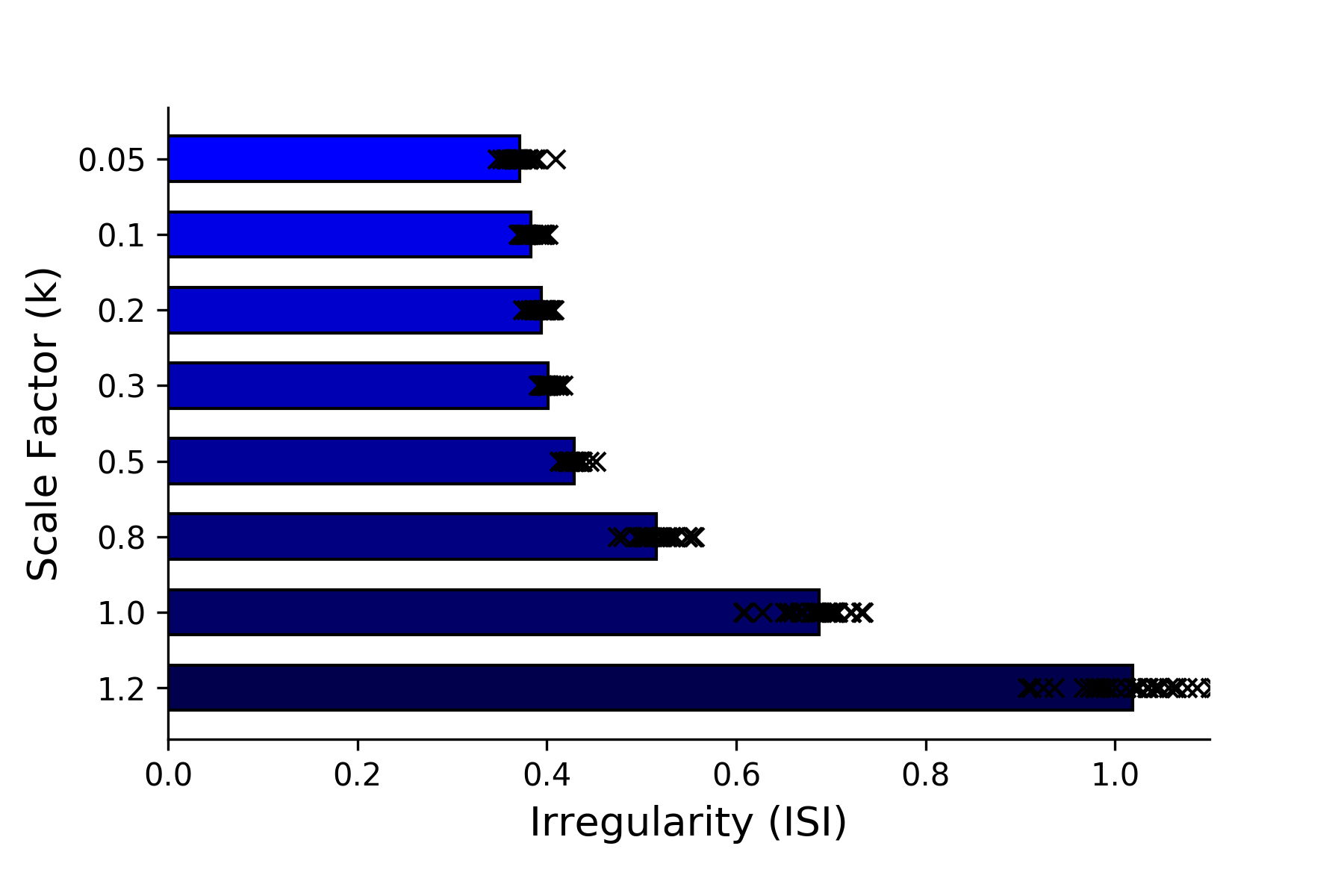}}

\end{minipage}
\begin{minipage}[b]{.24\textwidth}
\subfloat
  []
  {\label{fig:Ex04bC1}\includegraphics[width=\textwidth,height=3cm]{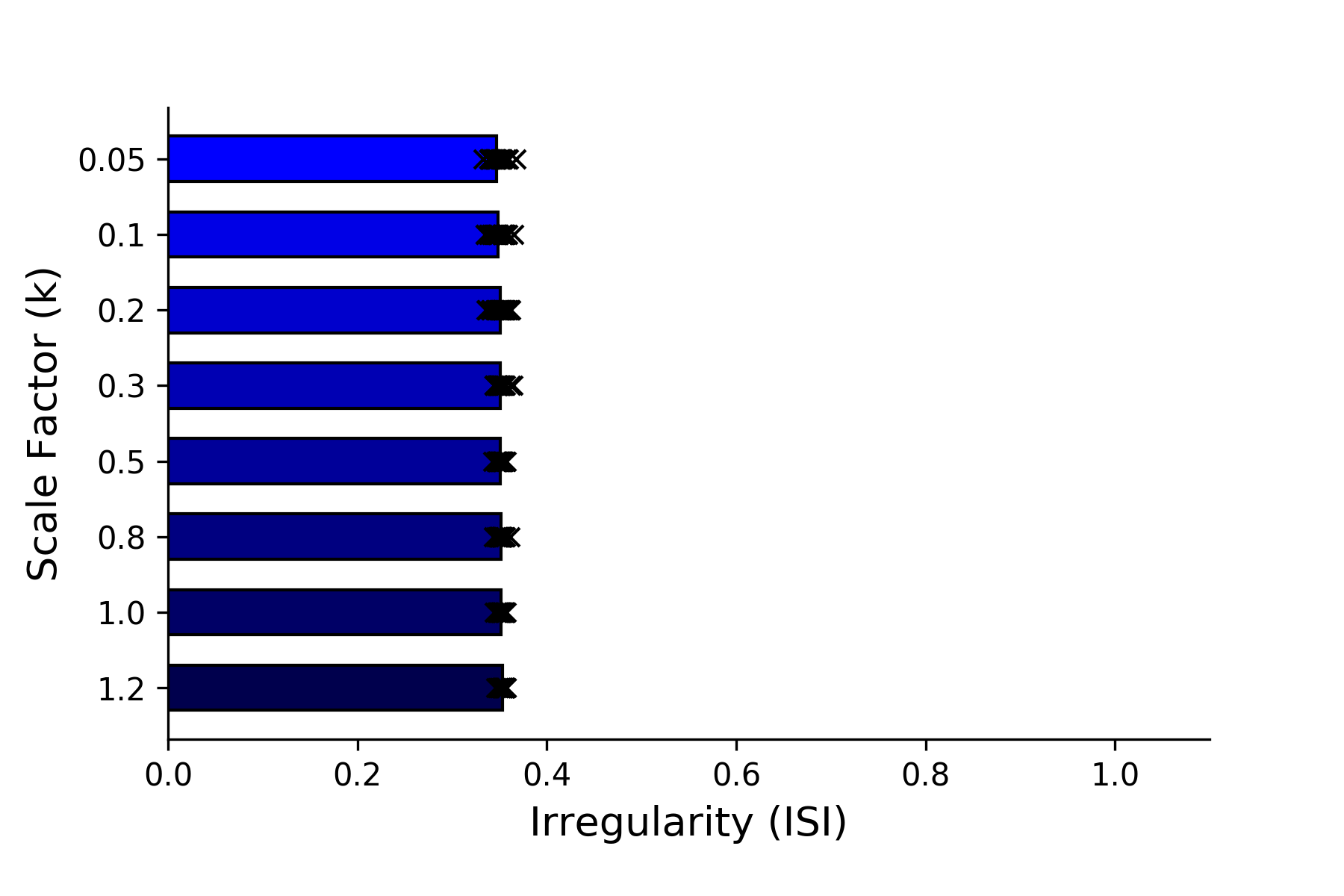}}

\end{minipage}
\begin{minipage}[b]{.24\textwidth}
\subfloat
  []
  {\label{fig:Ex04bD1}\includegraphics[width=\textwidth,height=3cm]{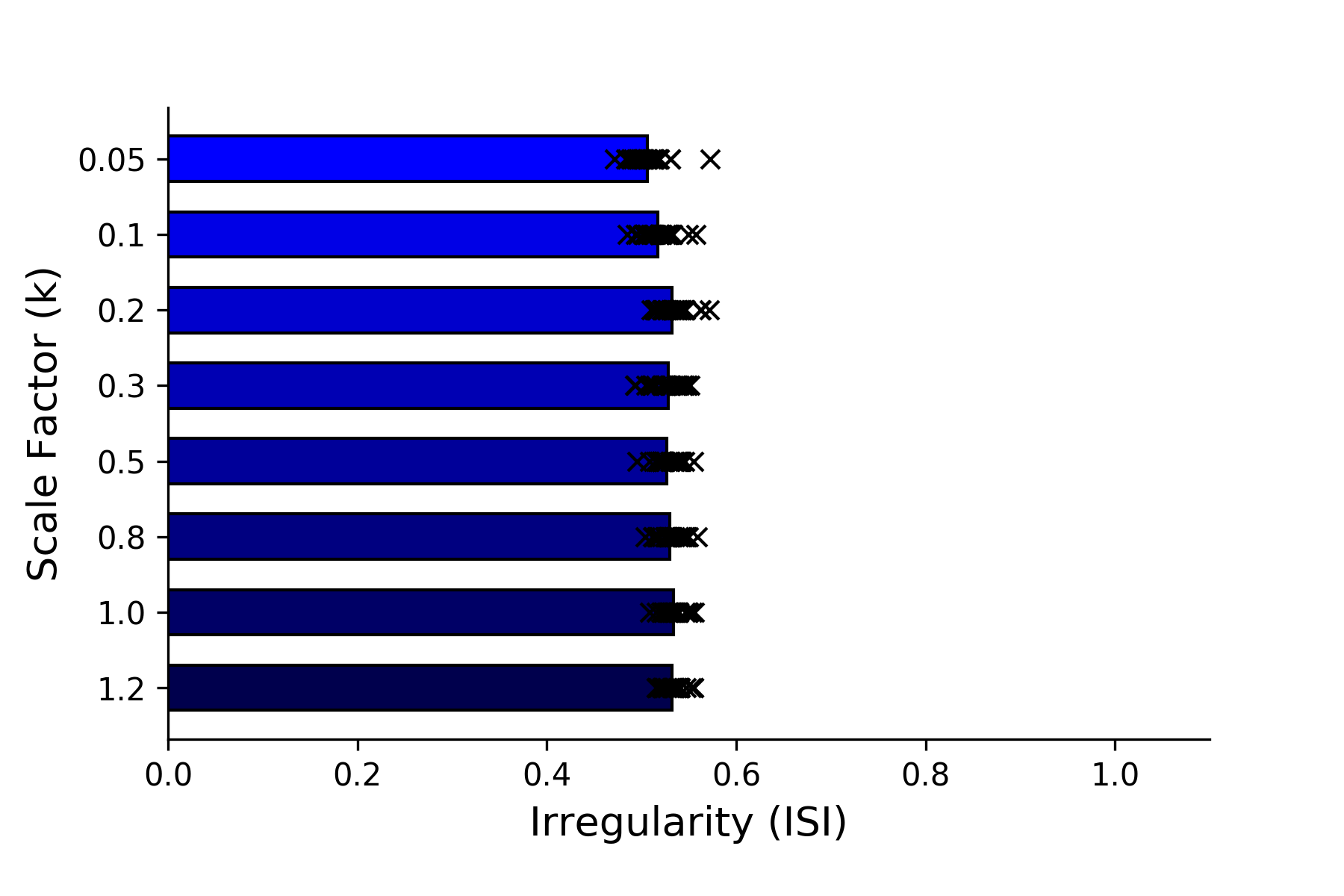}}
  
\end{minipage}
\caption{Average of irregularity of single-unit spikes calculated by the coefficient of variation of the Brunel \cite{brunel2000dynamics} network neurons ISI for different rescaling size(120\%, 100\%, 80\%, 50\%, 30\%, , 20\%, 10\%, 5\%) runned 30 turns during 10s each one. (A) $g=3$ and $\Theta = 2.V_{th}$; (B) $g=6$ and $\Theta=4.V_{th}$, (C) $g=5$ and $\Theta = 2.V_{th}$,  and (D) $g=4.5$ and $\Theta= 1.001.V_{th}$. The 'x' is value for each simulation run and the bar is the average of the set run. }
\label{fig:rescaling_graf_04c}
\end{figure}

Table \ref{tab:rescaling_parameters_04} presents general network parameters of Figure 8 in original article \cite{brunel2000dynamics}. Figure \ref{fig:rescaling_graf_04a} presents the firing rate and Figure \ref{fig:rescaling_graf_04c} presents the ISI for the rescaled network to different sizes (120\%, 100\%, 80\%, 50\%, 30\%, , 20\%, 10\%, 5\%) and four parameters combination from the Figure 8 of original article \cite{brunel2000dynamics}: (A) $g=3$ and $\Theta = 2.V_{th}$; (B) $g=6$ and $\Theta=4.V_{th}$, (C) $g=5$ and $\Theta = 2.V_{th}$,  and (D) $g=4.5$ and $\Theta= 1.001.V_{th}$. For the (D) simulations ($g=4.5$ and $\Theta= 1.001.V_{th}$), the $\Theta$ was replaced by an equivalent Poisson input with the same $w$, weight of excitatory synapses strength.
\vspace{0.4 cm}

\begin{figure}[H]
\centering

\begin{minipage}[b]{.49\textwidth}
\subfloat
  []
  {\label{fig:rescaling_graf_04bA}\includegraphics[width=\textwidth,height=6cm]{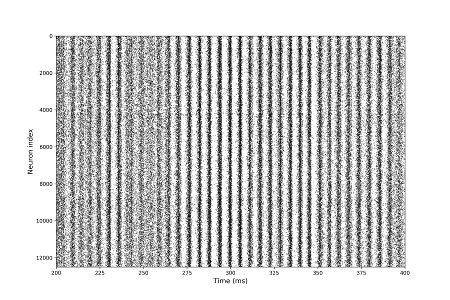}}
  
\subfloat
  []
  {\label{fig:rescaling_graf_04bB}\includegraphics[width=\textwidth,height=2cm]{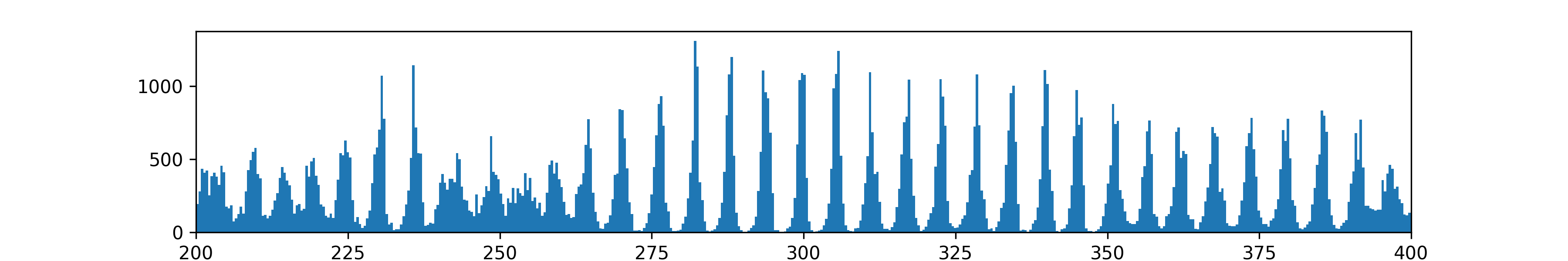}}
  
\subfloat
  []
  {\label{fig:rescaling_graf_04bC}\includegraphics[width=\textwidth,height=1.5cm]{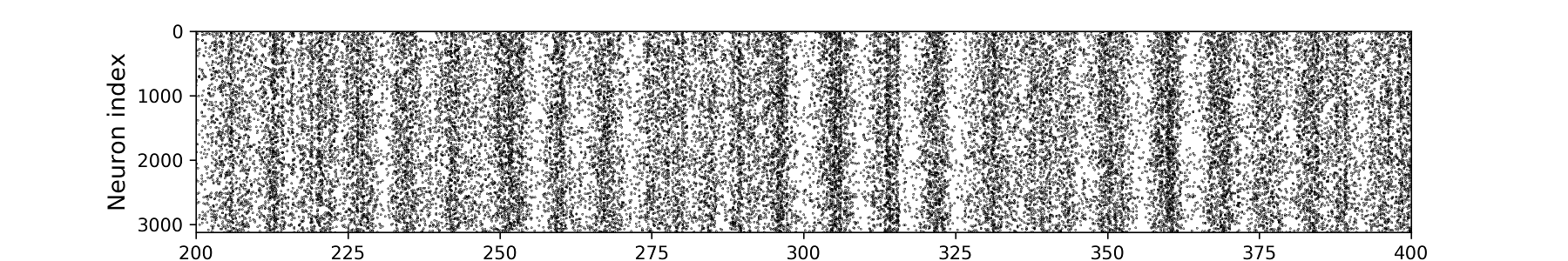}}
  
\subfloat
  []
  {\label{fig:rescaling_graf_04bD}\includegraphics[width=\textwidth,height=2cm]{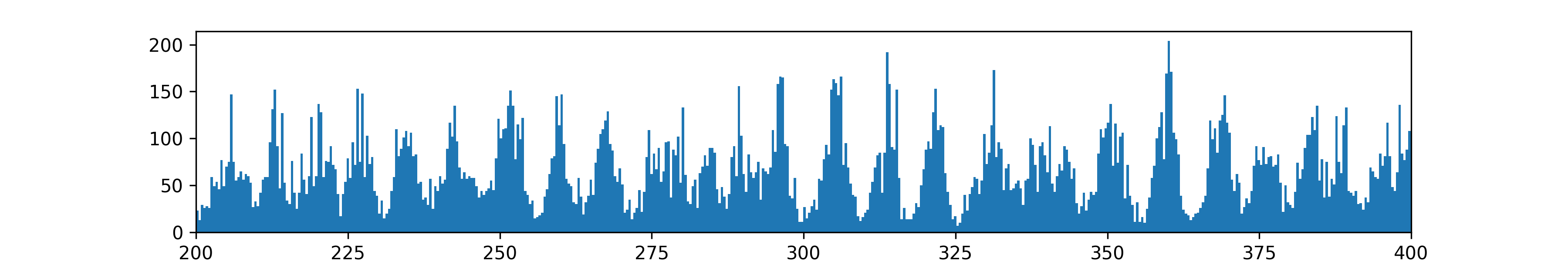}}

\end{minipage}
\begin{minipage}[b][\ht\measurebox]{.49\textwidth}
\centering
\vfill

\subfloat
  []
  {\label{fig:rescaling_graf_04bE}\includegraphics[width=\textwidth,height=6cm]{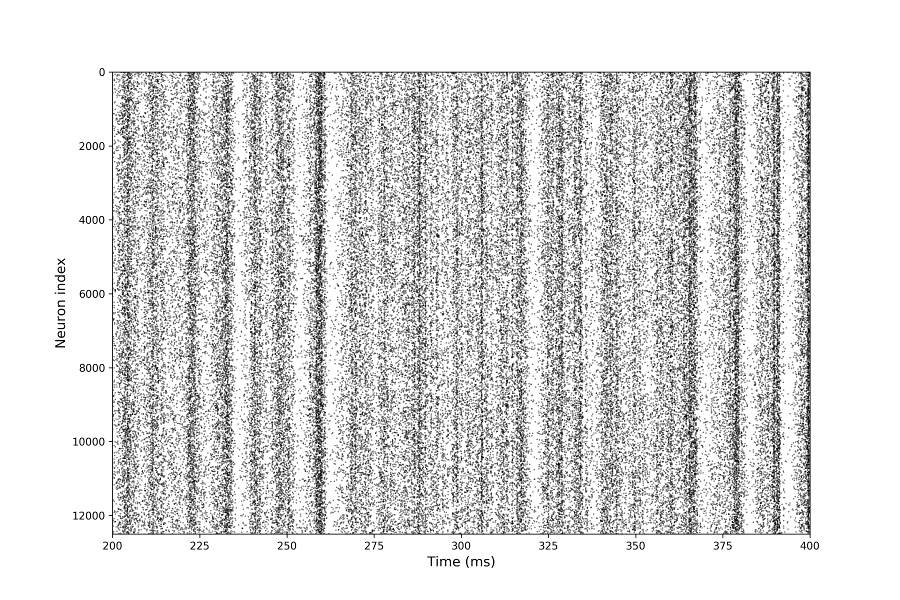}}
  
\subfloat
  []
  {\label{fig:rescaling_graf_04bF}\includegraphics[width=\textwidth,height=2cm]{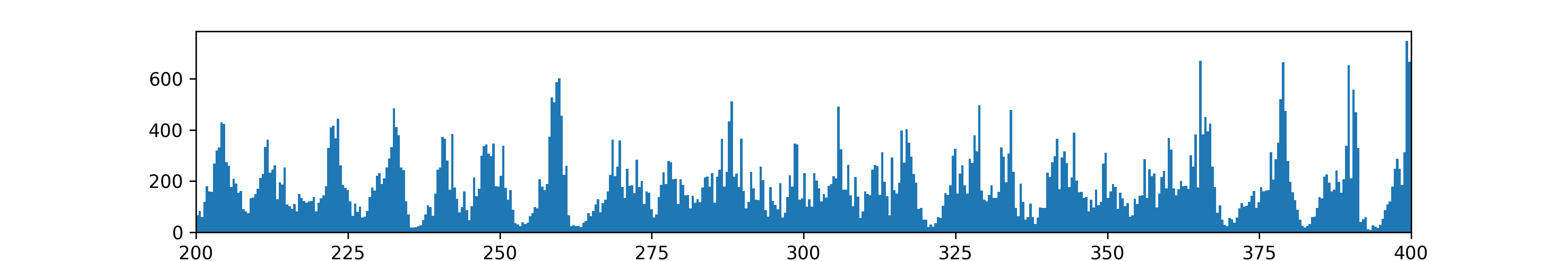}}
  
 \subfloat
  []
  {\label{fig:rescaling_graf_04bG}\includegraphics[width=\textwidth,height=1.5cm]{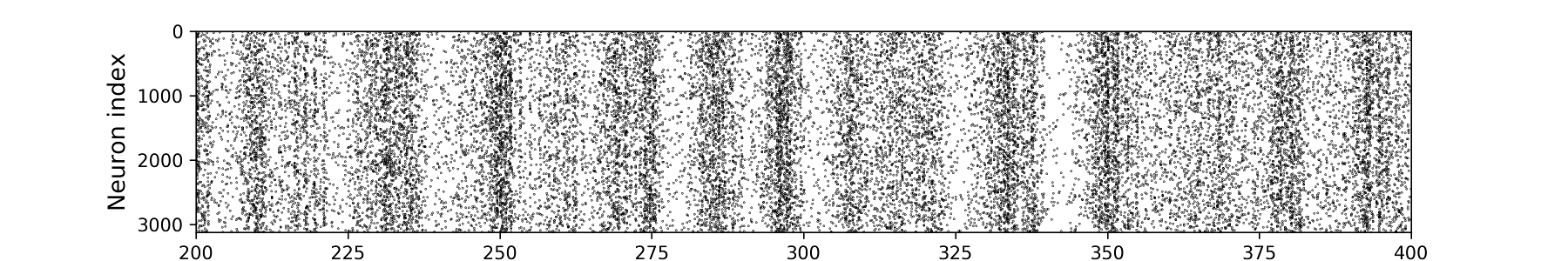}}
  
\subfloat
  []
  {\label{fig:rescaling_graf_04bH}\includegraphics[width=\textwidth,height=2cm]{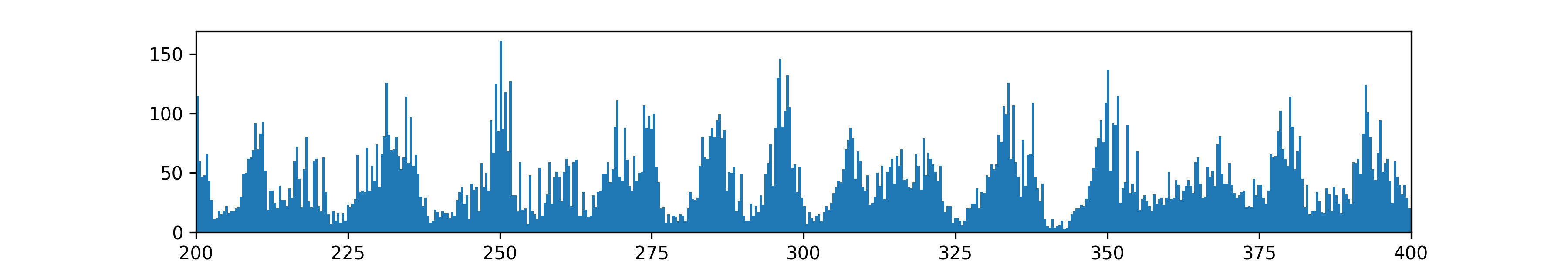}}

\end{minipage} 
\caption{Right side: Raster plots (A-full scale and C-resize to 25\%) and histogram (B-full scale and D-resize to 25\%) for the Brunel \cite{brunel2000dynamics} network with $g=6$ and $\Theta=4.V_{th}$ configuration. Left side: Raster plots (E-full scale and G-resize to 25\%) and histogram (F-full scale and H-resize to 25\%) for the Brunel \cite{brunel2000dynamics} network with $g=5$ and $\Theta = 2.V_{th}$ configuration. }
\label{fig:rescaling_graf_04b}
\end{figure}

Figure \ref{fig:rescaling_graf_04b} presents the raster plots and spike histogram for the full
network and for the rescaled network to 25\%. Those are the configuration of Figure 8B (right side of Figure \ref{fig:rescaling_graf_04b}) and Figure 8C (left side of Figure \ref{fig:rescaling_graf_04b}) in the Brunel original article \cite{brunel2000dynamics}.  It seems that the oscillation present in Figure \ref{fig:rescaling_graf_04bB} is not as strongly marked as in \ref{fig:rescaling_graf_04bD}, at least not visually. This network configuration ($g=6$ and $\Theta=4.V_{th}$) presents a large variation in irregularity with rescaling (see Figure \ref{fig:Ex04bB1}) and the highest  (15\%) average firing rate variation at rescaling, resembling the firing rate for the (D - 14.5\%) configuration but higher, even the $\theta$ replaced by a Poisson external input (in D). Those feature are explained in the end of the Section \ref{Model requirements, math explication and method limitations}: Model requirements, mathematical explication and method limitations.
\vspace{0.4 cm}

\begin{table}[H]
\center{\begin{tabular}{llll}
\hline
Parameter description                  &  Variable   & Full scale  & Rescaling  \\ \hline
Factor of rescaling  & $k$ &  - & 0.05 \\
Number of excitatory neurons  & $N_+$  & 10000 & 500  \\
Number of inhibitory neurons  & $N_-$  & 2500  & 125 \\
Number of external input to each neuron & $X_{ext}$ &  0 & 0 \\
Total number of inner connection & $X$  & 15.6M &  39k\\
Weight of excitatory synaptic strength & $w \pm \delta w$  (mV)  & 0.1 & 0.45 \\ \\ \hline
Probability of connection              & $p$         &  0.1  & 0.1 \\
Absolute refractory period & $\tau _{ref}$ (ms)  & 2 & 2 \\
Membrane time constant & $\tau _{m}$ (ms) & 20 &  20 \\
Synaptic transmission delays  & $\Delta_{t} \pm \delta \Delta_{t} $ (ms)  & 1.5 & 1.5 \\
Reset potential (mV)& $V_{ret}$   & 10 & 10 \\ Fixed firing threshold (mV) & $V_{th}$   & 20 & 20 \\   \\ \hline
\end{tabular}
\caption{Brunel \cite{brunel2000dynamics} model specification before and after of the rescaling: parameters and metrics.} \label{tab:rescaling_parameters_04}}
\end{table}

\vspace{0.4 cm}

\vspace{0.4 cm}

\subsection{Model requirements, mathematical explication and method limitations}
\label{Model requirements, math explication and method limitations}
\vspace{0.4 cm}

This method works in any network where: 

\begin{itemize}

\item{ }  the weight of synaptic strength makes a small contribution compared to the firing threshold ($w$ << $V_{th} - V_{rt}$);
\vspace{0.4 cm}

\item{} there is a low probability of connection ($p$ << 1).
\vspace{0.4 cm}

\end{itemize}

The mathematical reason is that, in those networks, the second order statistics is dependently of the number of received connections, $x$, and the square of synaptic strength, $w^2$.
\vspace{0.4 cm}

The rescaling method gives: 

\begin{equation}
\label{mfmean4}
w' =  w/ \sqrt(k)   ,
\end{equation}
\begin{equation}
\label{mfmean5}
x' =  k . x = k^2 . X / (k . N). 
\end{equation}

This maintains $x.w^2$ and, therefore, the second order statistics. 
\vspace{0.4 cm}

The first order statistics depends on the number of received connections, $x = X/N$, and the synaptic strength, $w$. So, the forth step of the method provides a DC to supply the loss of $(1-\sqrt(k))$ in the first order statistic. 
\vspace{0.4 cm}

More formally, this method works in any model that can be approximated by a sparse random connected network where the neuron activity can be approximated by an average part plus a fluctuating Gaussian part:

\begin{equation}
\label{IR}
V (t) = \mu (t)+ \sigma . \eta (t), 
\end{equation}

$\eta $ is a gaussian white noise, $\mu (t)$ is the average part,  $\sigma$ is the standard deviation  and, therefore,  $\sigma . \eta (t)$ is the fluctuating part,
where:

\begin{equation}
\label{mfmean}
\mu (t) = \mu_{int} (t)+\mu_{ext} (t),
\end{equation}

\begin{equation}
\label{mfmean2}
\sigma^2 (t) = \sigma^2_{int} (t)+\sigma^2{ext} (t),
\end{equation}
\vspace{0.4 cm}

\subsubsection{Any network: Excitatory-inhibitory interconnected}

For any network with n set of neurons, which each set may be inhibitory ($w_{pre} <0$) or excitatory ($w_{pre} >0$) neurons:

\begin{equation}
\label{mfmean1o}
\mu_{post} (t) =  \sum_{pre=1}^n ( x_{pre, post}. w_{pre,post} . f_{pre, post}. \tau_{pre, post}) + X_{ext, post} .w_{ext, post} .f_{ext, post}. \tau_{ext, post}  ,
\end{equation}

\begin{equation}
\label{mfmean3o}
\sigma^2_{post} (t) = \sum_{pre=1}^n ( x_{pre, post} . w_{pre, post}^2 . f_{pre, post}  . \tau_{pre, post} ) + X_{ext, post} .w_{ext, post}^2 .f_{ext, post}. \tau_{ext, post}.
\end{equation}
\vspace{0.4 cm}

Hence, replacing \ref{mfmean4} and \ref{mfmean5} in \ref{mfmean3o},
\begin{equation}
\label{mfmean3bo}
     \begin{aligned}
    \sigma'^2_{post} (t) =  \\
    \sum_{pre=1}^n ( (k.x_{pre, post}) . (w_{pre, post}/\sqrt k)^2 . \tau_{pre, post} ) \\
    + (k.X_{ext, post}) .(w_{ext, post}/\sqrt k)^2 .f_{ext, post}. \tau_{ext, post} \\
    =\sigma^2_{post} (t),
    \end{aligned}
\end{equation}
given that, the second order statistics is granted. Going back to the first order of statistics:
\vspace{0.4 cm}

The forth step of method grant a DC where:

\begin{equation}
\label{mfmean1bo}
DC_{post} =  (1-\sqrt(k)) . ( \sum_{pre=1}^n ( x_{pre, post}. w_{pre, post} . f_{pre, post} . \tau_{pre, post} ) + X_{ext, post} .w_{ext, post} .f_{ext, post}. \tau_{ext, post} ).
\end{equation}

Thus, the new $\mu' (t)$  is given by:

\begin{equation}
\label{mfmeanDCo}
\mu' (t) = \mu'_{int} (t)+\mu_{ext}' (t) + DC,
\end{equation}

and, replacing \ref{mfmean4}, \ref{mfmean5} and \ref{mfmean1o}   in \ref{mfmeanDCo}:

\begin{equation}
\label{mfmean10o}
    \begin{aligned}
    \mu'_{post} (t) =   \sum_{pre=1}^n ((k.x_{pre, post}). (w_{pre, post}/ \sqrt{k}) . f_{pre, post} . \tau_{pre, post}) \\
    + (k.X_{ext,post}) .(w_{ext,post}/ \sqrt{k}).f_{ext,post}. \tau_{ext, post} + DC_{post} ,
    \end{aligned}
\end{equation}

and replacing DC:

\begin{equation}
\label{mfmean11o}
    \begin{aligned}
        \mu'_{post} (t) =  \\
         \sum_{pre=1}^n ((k.x_{pre, post}). (w_{pre, post}/ \sqrt{k}) . f_{pre, post} . \tau_{pre, post} )  + k.X_{ext, post} .(w_{ext, post}/\sqrt{k} ) .f_{ext, ext} . \tau_{ext, post} ) \\
         + (1-\sqrt{k}) . \sum_{pre=1}^n ((x_{pre, post}). (w_{pre, post}) . f_{pre, post} . \tau_{pre, post} ) + (1-\sqrt{k}) (X_{ext post} .w_{ext, post} .f_{ext, post}. \tau_{ext, post}) \\
         = \mu_{post} (t),
    \end{aligned}
\end{equation}

given that, the first order statistics is granted too.

\vspace{0.6 cm}

\subsubsection{Brunel \cite{brunel2000dynamics} network: Excitatory-inhibitory interconnected}

For Brunel network, $x$, the average number of received connections per neuron, $x_e$, the average number of received excitatory connections per neuron, and $x_i$, the average number of received inhibitory connections per neuron, we have:

\begin{equation}
\label{nx}
X/N =x =   x_e+  x_i,
\end{equation}
\vspace{0.4 cm}

the mean $\mu (t)$ and the deviation $\sigma^2(t)$ can be detailed as:

\begin{equation}
\label{mfmean1}
\mu (t) =  x. w . (1-g x_i/x_e) . f . (t- \Delta_t) . \tau + X_{ext} .w .f_{ext}. \tau  ,
\end{equation}

\begin{equation}
\label{mfmean3}
\sigma^2 (t) =  x . w^2 . (1 + g^2 x_i/x_e) . f . (t-\Delta_t) . \tau + X_{ext} .w^2 .f_{ext}. \tau.
\end{equation}
\vspace{0.4 cm}

Hence, replacing \ref{mfmean4} and \ref{mfmean5} in \ref{mfmean3},
\begin{equation}
\label{mfmean3b}
\sigma'^2 (t) =  (k.x) . (w/\sqrt k)^2 . (1 + g^2 x_i/x_e) . f . (t-\Delta_t) . \tau + (k.X_{ext}) .(w/\sqrt k)^2 .f_{ext}. \tau =\sigma^2 (t),
\end{equation}
given that, the second order statistics is granted. Going back to the first order of statistics:
\vspace{0.4 cm}

The forth step of method grant a DC where:

\begin{equation}
\label{mfmean1b}
DC =  (1-\sqrt(k)) [x. w . (1-g x_i/x_e) . f . (t- \Delta_t) . \tau + X_{ext} .w .f_{ext}. \tau] .
\end{equation}

Thus, the new $\mu' (t)$  is given by:

\begin{equation}
\label{mfmeanDC}
\mu' (t) = \mu'_{int} (t)+\mu_{ext}' (t) + DC,
\end{equation}

and, replacing \ref{mfmean4}, \ref{mfmean5} and \ref{mfmean1}   in \ref{mfmeanDC}:

\begin{equation}
\label{mfmean10}
\mu' (t) =  (k.x). (w/ \sqrt{k}) . (1-g x_i/x_e) . f . (t- \Delta_t) . \tau + (k.X_{ext}) .(w/ \sqrt{k}).f_{ext}. \tau + DC ,
\end{equation}

and replacing DC:

\begin{equation}
\label{mfmean11}
    \begin{aligned}
        \mu' (t) =  \\
         (k.x). (w/ \sqrt{k}) . (1-g x_i/x_e) . f . (t- \Delta_t) . \tau + k.X_{ext} .(w/\sqrt{k} ) .f_{ext} . \tau ) \\
         + (1-\sqrt(k) [(x). (w) . (1-g x_i/x_e) . f . (t- \Delta_t) . \tau + X_{ext} .w .f_{ext}. \tau] \\
         = \mu (t),
    \end{aligned}
\end{equation}

given that, the first order statistics is granted too.
\vspace{0.6 cm}

\subsubsection{Rescaling limit and oscillation}
\label{sec:Rescaling Limitation}
\vspace{0.4 cm}

The size limit of rescaling happens when $w$ become so large that the fist model requirement ($w$ << $V_{th} - V_{rt}$) stops to be satisfied. 
\vspace{0.4 cm}

In case that the model stops working on a smaller scale, one solution is to increase the random input, which means, to artificially add an external random input and compensate it on the threshold (see Figures \ref{fig:Ex04D1} and \ref{fig:Ex04bD1}).
A massive random external input guarantees the network operation on a stable point because it reduces the perturbation point caused by under or over inter connection spike activity. It reduces the ratio between the inter connection mean $\mu_{int}$ or standard deviation $\sigma_{int}^2$ and the total mean $\mu$ or total standard deviation $\sigma^2$. It avoids changing the previous balance point of the network activity.
\vspace{0.4 cm}

Additionally, this method does not introduce any resonance or oscillation. Instead, it tends to prevent oscillations such as the application \ref{sec:Sparse random connected network}. It is due to the reduce of the ratio between the inter connection mean $\mu_{int}$ and the total standard deviation $\sigma^2$. In other words, and more formally, by equation 30 from \cite{brunel2000dynamics}

\begin{equation}
\label{eq:BrunelG}
G = \frac{ x . w . \tau . f . (gx_i/x_e - 1)}  {\sigma} =  \frac {- \mu_{int}}  {\sigma},
\end{equation}

\begin{equation}
\label{BrunelH}
H = \frac{ x . w^2 . \tau . f . (g^2x_i/x_i + 1)}  {\sigma^2} =  \frac {\sigma_{int}^2}  {\sigma^2},
\end{equation}

where $\mu_{int}$ is the mean and $\sigma_{int}^2$ is the standard deviation due to internal connections probability. 
\vspace{0.4 cm}

The oscillation on Figure \ref{fig:rescaling_graf_04bB} (and on Figure 8B in \cite{brunel2000dynamics}) is due to the $H \rightarrow {1}$ and $G ~ \sqrt( \tau / \Delta_t)$. 
Once this method does not change the ratio H, nor $\sigma$ but decreases $ \mu_{int}$, the ratio G is not satisfied anymore and, consequently, neither is the oscillation sustained in Figure \ref{fig:rescaling_graf_04bD}.

\vspace{0.6 cm}

\newpage
\section{Boundary Correction Method}
\label{Boundary Method}

\vspace{0.4 cm}

The key to this method is to apply the rescaling idea to each neuron $o$ in the network or boundary region. The rescaling factor $k_{o}$ for each neuron $o$ aims to compensate the number of connections lost due to the cut, i.e. in case the boundary was inexistent and the network had an infinite number of neurons. The key is the normalized connection density.
\vspace{0.4 cm}

Normalized connection density: For a given neuron $o$, the new number of received connections normalized by the total number of received connections if the network weas infinite (no boundary). For example, for a square $i \times j$ network, the neurons $o_{ij}$ on the corner receives at least 0.25 of the connections if the network was infinite 
- was not end in $i \times j$.
\vspace{0.4 cm}


\vspace{0.6 cm}

\subsection{Boundary correction method algorithm}
\label{Boundary Algorithm}
\vspace{0.4 cm}

The boundary correction method essentially numerically estimates the normalized density function of connection on the first step, then weights each neuron connection based on this density and finally balances the threshold to grant the neuron/layer activity.
\vspace{0.4 cm}

The laborious part of this method is to bring up the normalized density function of connection, once it depends on of the pattern of connection in each model. Below is an easy way that will work out for any pattern of connection. However, if the normalized density function of connection is analytically known, one can use it and start the boundary correction algorithm by step 2.
\vspace{0.4 cm}

The algorithmic of rescaling method can be found in any one of example-application on Section
\ref{Boundary correction applied} those are also available in GitHub (https://github.com/ceciliaromaro/recoup-the-first-and-second-order-statistics-of-neuron-network-dynamics)
as follows:
\vspace{0.4 cm}

\begin{itemize}

\item{\textbf{Step 1:}} Calculate the scale factor for any neuron $o$ in network based in the normalized connection density;
\vspace{0.4 cm}

\item{\textbf{Step 2:}} Increase the synaptic weights by dividing them by the square root of the scale factor;
\vspace{0.4 cm}

\item{\textbf{Step 3:}} Provide each cell with a DC input current with a value corresponding to the total input lost due to network edge (boundary cut).
\vspace{0.6 cm}

\end{itemize}

\subsubsection{Boundary correction method for n layers network}
\vspace{0.4 cm}

More formally, our method algorithm can be described by the following pseudo-algorithm:

\vspace{0.4 cm}

\begin{algorithm}[H]
\caption{Rescaling method for boundary correction for model with n layers}\label{alg:sim3}
\begin{algorithmic}[1]

\State $n$ number of sets in model 
\State $N_{j}$ the (finite) set of possynaptic neurons. ${j \in n}$
\State $N_i$ the (finite) set of presynaptic neurons. ${i \in n}$
\State $\overline{X_{ij}}$ the average number of connection between $N_i$ and one neuron in $N_j$ if the model was no boundary.
\State $x_{oj}$ the number of synapse connected to each neurons of $N_j$.
\State $w_{oij}$ (pA or mV) the average weight of synaptic strength the set $N_i$ target the neuron o in $N_j$.
\State $k'_{oj}$ the factor of rescaling of the neuron o from the set $N_j$. (will be calculate).
\State $\overline{f_i}$ (Hz) the average firing rate of set of neurons  $N_i$.
\State $\tau _{syn}$ (ms) synapse time constant.

\vspace{0.4 cm}

1. CALCULATE OF NORMALIZED CONNECTION DENSITY

\vspace{0.2 cm}
\For{each layer $j$ in $n$}
\For{each neuron $o$ in $N_j$}
    \State   $k_{oj}' \gets   x_{oj}/\sum_{i=1}^n (\overline{X_{ij}}$)
\EndFor
\EndFor
\vspace{0.4 cm}

2. SYNAPTIC STRENGHT
\vspace{0.2 cm}

\For{each layer $i$ in $n$}
\For{each layer $j$ in $n$}
\For{each neuron $o$ in $N_j$}
    \State $w_{oij}' \gets w_{oj}/ \sqrt{k_{oij}}$
\EndFor
\EndFor
\EndFor

\vspace{0.4 cm}

3. THRESHOLD ADJUSTMENT

\vspace{0.2 cm}

\For{each layer $j$ in $n$}
\For{each neuron $o$ in $N_j$}
    \State $c_{sum_{oij}} = w_{oij}  *  \sum_{i=1}^n \overline{f_i} * \overline{X_{ij}}$
\State $I_{DC_j}' = \tau _{syn} *  
		        (1 - \sqrt{k_{oj}'}) * c_{sum_{oij}} $ \Comment{DC (pA  or mV) input to compensate resize}
\EndFor
\EndFor

\vspace{0.4 cm}

\State Done!
\end{algorithmic}
\end{algorithm}

\newpage

\subsection{Boundary correction applied}
\label{Boundary correction applied}
\vspace{0.6 cm}

We applied the boundary solution for the models presented in Sections \ref{sec:Sparse random connected network}, \ref{sec:PD network: Eight layers excitatory-inhibitory interconnected Network} and \ref{sec:Brunel network: Excitatory-inhibitory interconnected}. 
In order to rise the boundary problem, first a model with topographic pattern of connection is needed. Therefore, we assigned a spatial position for each neuron, then applied a Gaussian with $\sigma_{g} $ as a pattern of connection and than ran the network with and without the boundary solution.
\vspace{0.4 cm}

All the applications of this method presented in this publication were implemented in Python (with
Brian2) and they can be found on GitHub (https://github.com/ceciliaromaro/recoup-the-first-and-second-order-statistics-of-neuron-network-dynamics).
\vspace{0.4 cm}

\subsubsection{Sparse random connected network: Inhibitory neurons}
\vspace{0.4 cm}

\begin{figure}[H]
\centering

\begin{minipage}[b]{.45\textwidth}
\subfloat
  []
  {\label{fig:boundary_graf_01A}\includegraphics[width=\textwidth,height=5cm]{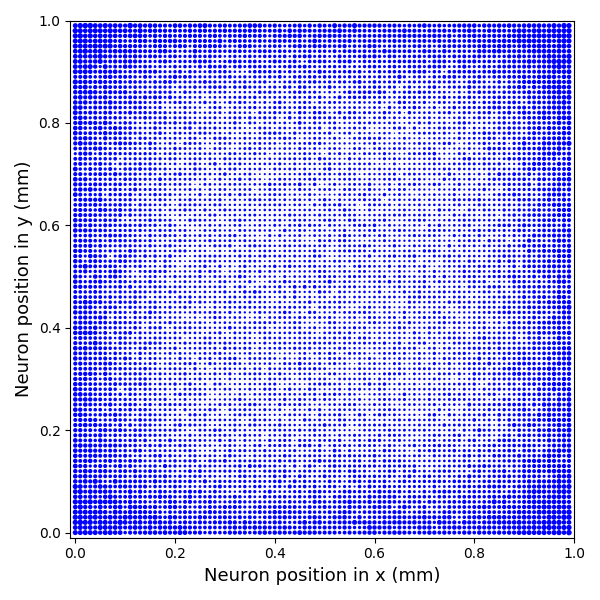}}
  
\subfloat
  []
  {\label{fig:boundary_graf_01B}\includegraphics[width=\textwidth,height=2cm]{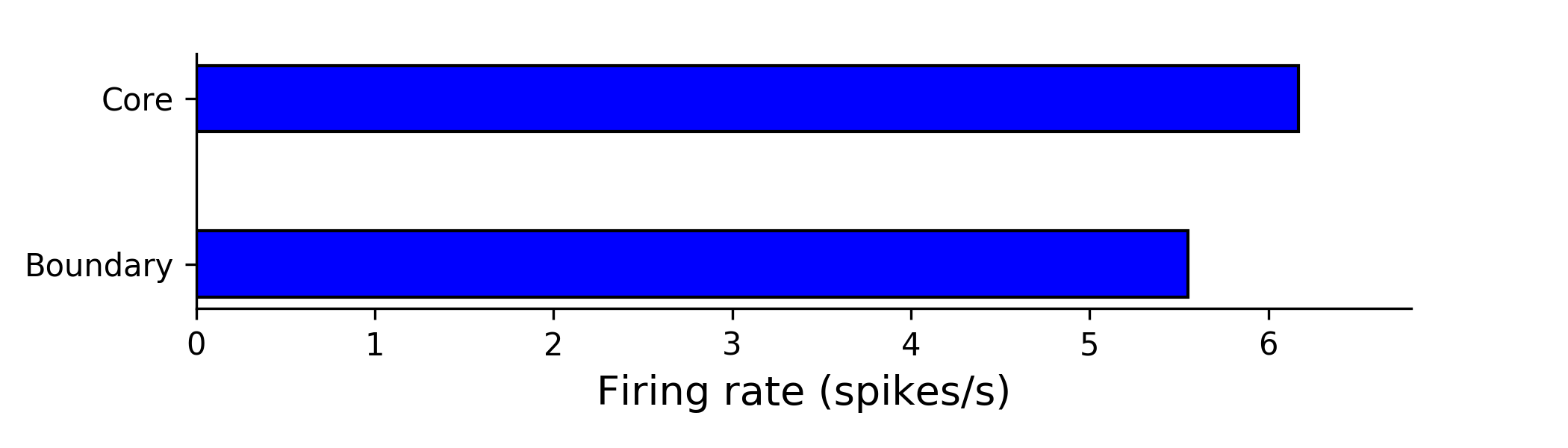}}
  
\subfloat
  []
  {\label{fig:boundary_graf_01C}\includegraphics[width=\textwidth,height=1cm]{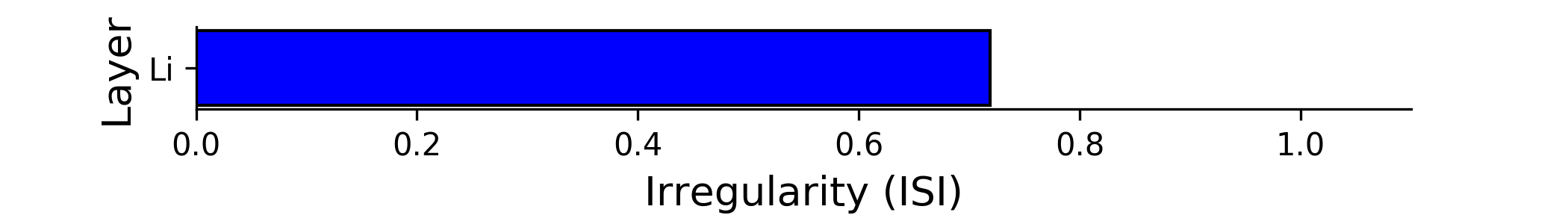}}
  
\end{minipage}
\begin{minipage}[b]{.45\textwidth}
\subfloat
  []
  {\label{fig:boundary_graf_01D}\includegraphics[width=\textwidth,height=5cm]{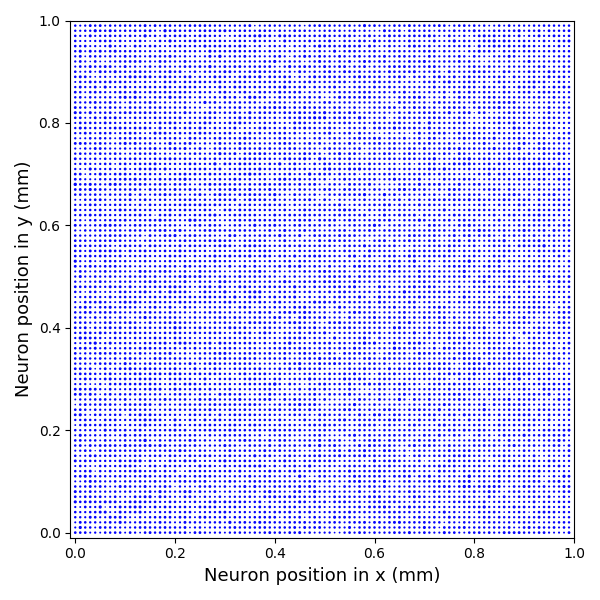}}
  
\subfloat
  []
  {\label{fig:boundary_graf_01E}\includegraphics[width=\textwidth,height=2cm]{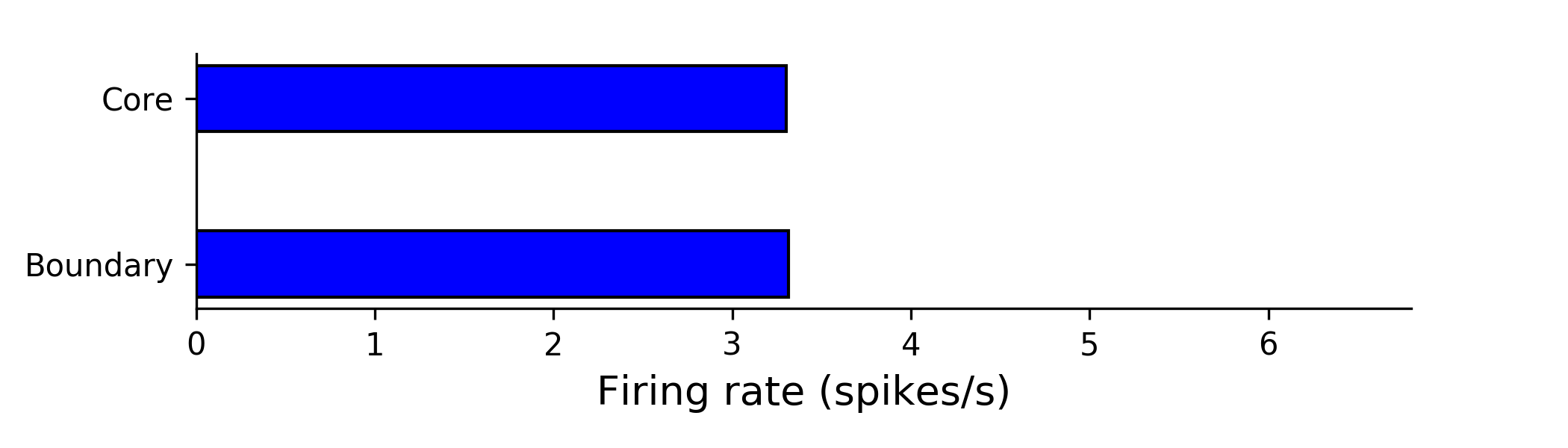}}
  
\subfloat
  []
  {\label{fig:boundary_graf_01F}\includegraphics[width=\textwidth,height=1cm]{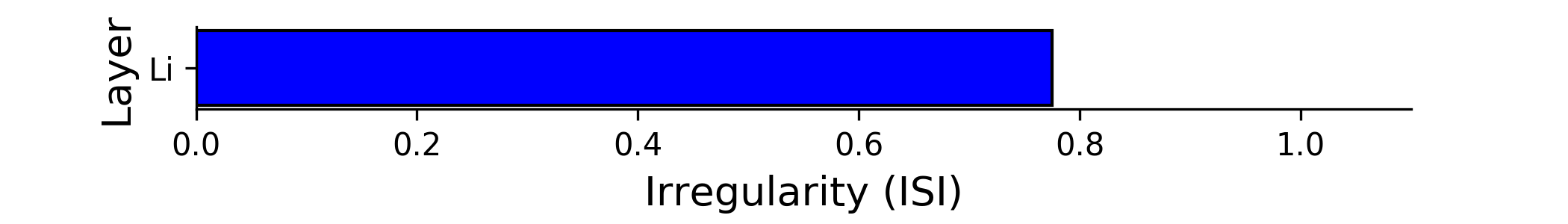}}
  
\end{minipage}
\caption{ Topographic sparse random connected inhibitory neurons network model with $\sigma_{g}=0.25mm$. Average firing rate per neuron (A) without boundary correction (D) with boundary correction. (B) Core and boundary average firing
rate and (C) Irregularity without boundary correction and (E) average firing rate and (F) irregularity without boundary correction.}
\label{fig:boundary_graf_01}
\end{figure}

All neurons from the model presented in \ref{sec:Sparse random connected network} were homogeneous distributed on $1 mm^2$ and a $\sigma_g$ = 0.25 mm was utilized. Figure \ref{fig:boundary_graf_01} presents the average firing-rate per neurons before and after the boundary correction, the mean fire rate of neurons in the core (around 50 \% of all neurons) and on the boundary (the complementary 50 \% of all neurons) and the network irregularity. Visually the boundary neurons spike more than core neurons in the model without boundary correction due to the leak of inhibition connections.

\vspace{0.4 cm}

\subsubsection{Somatosensory S1 network: Eight layers excitatory-inhibitory interconnected Network}
\vspace{0.4 cm}

All neurons from the full version of the model presented on \ref{sec:PD network: Eight layers excitatory-inhibitory interconnected Network} were homogeneously distributed on $1 mm^2$ and a $\sigma_g = 0.275 mm$ was utilised. The same was done for the rescaling to 50\%.
\vspace{0.4 cm}

\begin{figure}[H]
\centering

\begin{minipage}[b]{.24\textwidth}
\subfloat
  []
  {\label{fig:boundary_graf_02bA}\includegraphics[width=\textwidth,height=3cm]{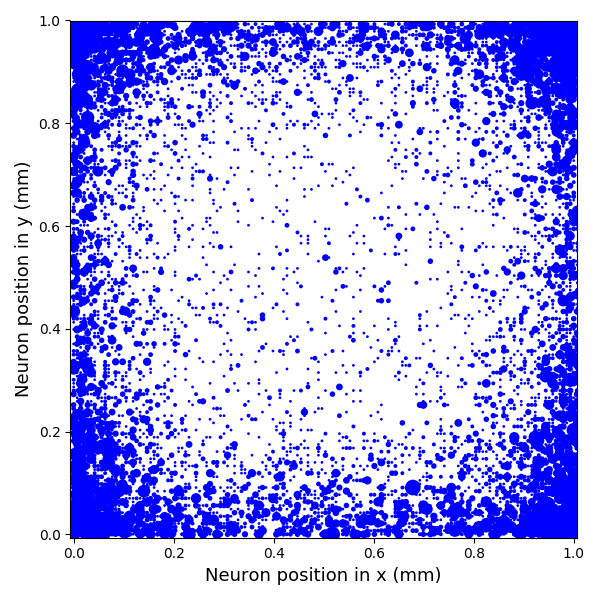}}
  
  \subfloat
  []
  {\label{fig:boundary_graf_02bB}\includegraphics[width=\textwidth,height=3cm]{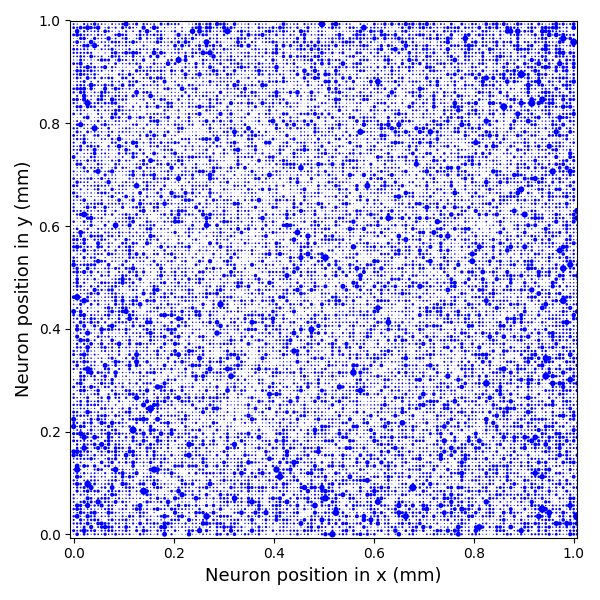}}
  \end{minipage}
\begin{minipage}[b]{.24\textwidth}
\subfloat
  []
  {\label{fig:boundary_graf_02bC}\includegraphics[width=\textwidth,height=3cm]{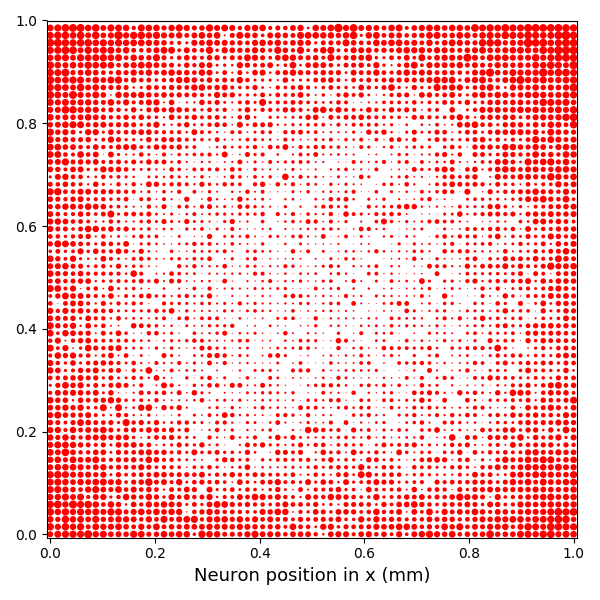}}
  
\subfloat
  []
  {\label{fig:boundary_graf_02bD}\includegraphics[width=\textwidth,height=3cm]{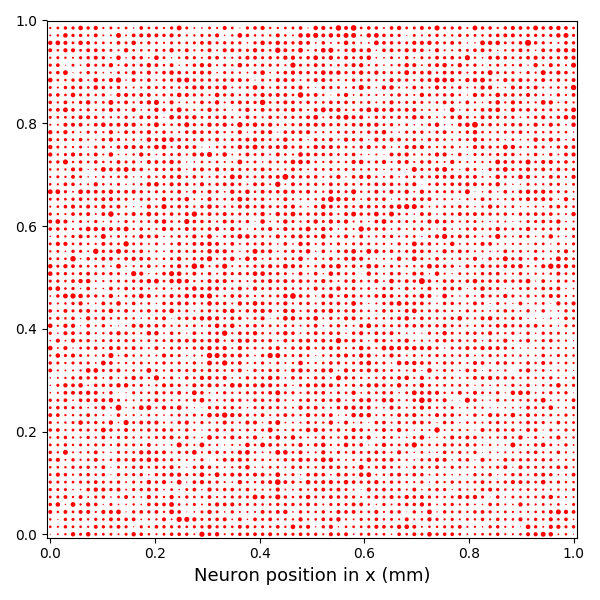}}

\end{minipage}
\begin{minipage}[b]{.24\textwidth}
  \subfloat
    []
    {\label{fig:boundary_graf_02bE}\includegraphics[width=\textwidth,height=3cm]{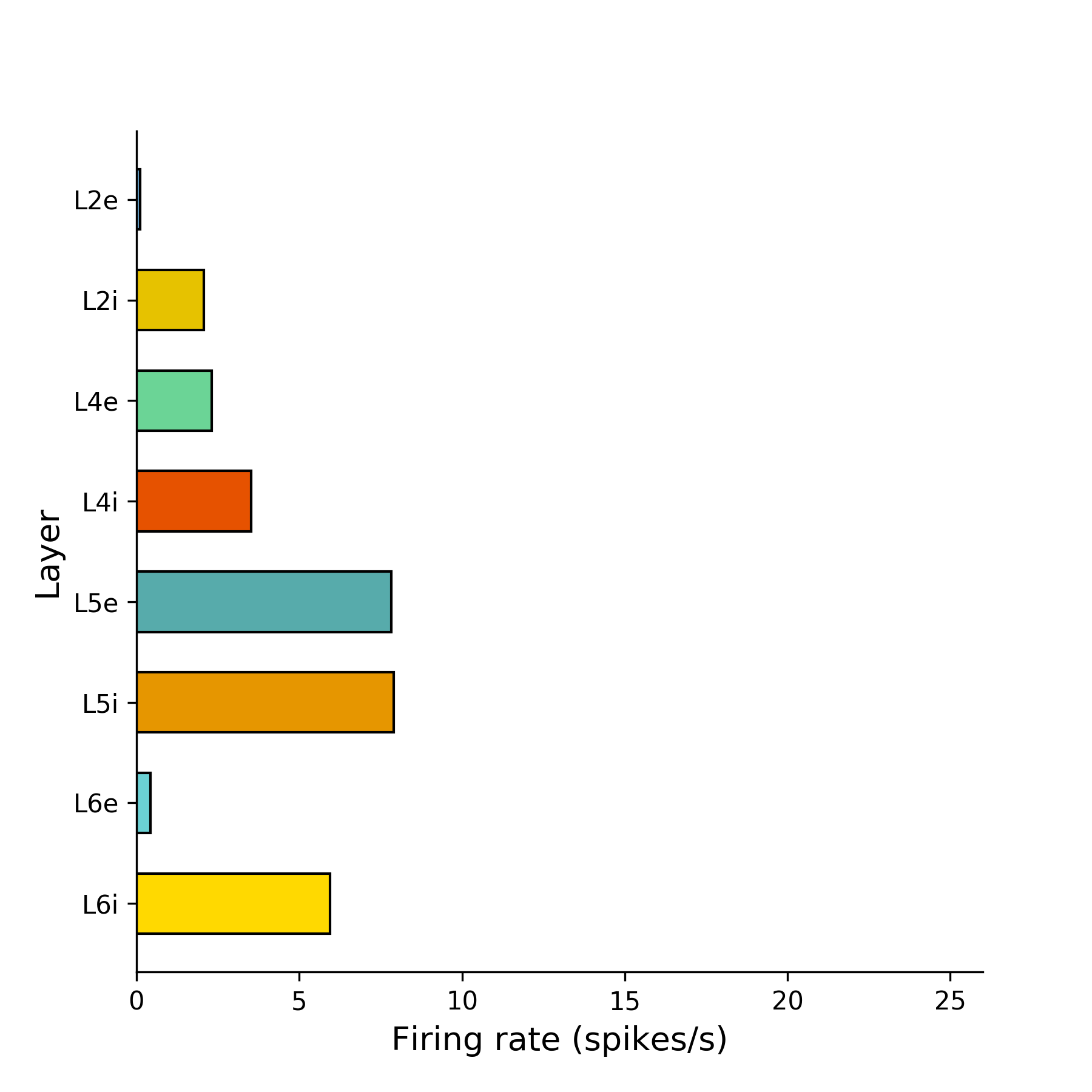}}
    
    \subfloat
  []
  {\label{fig:boundary_graf_02bF}\includegraphics[width=\textwidth,height=3cm]{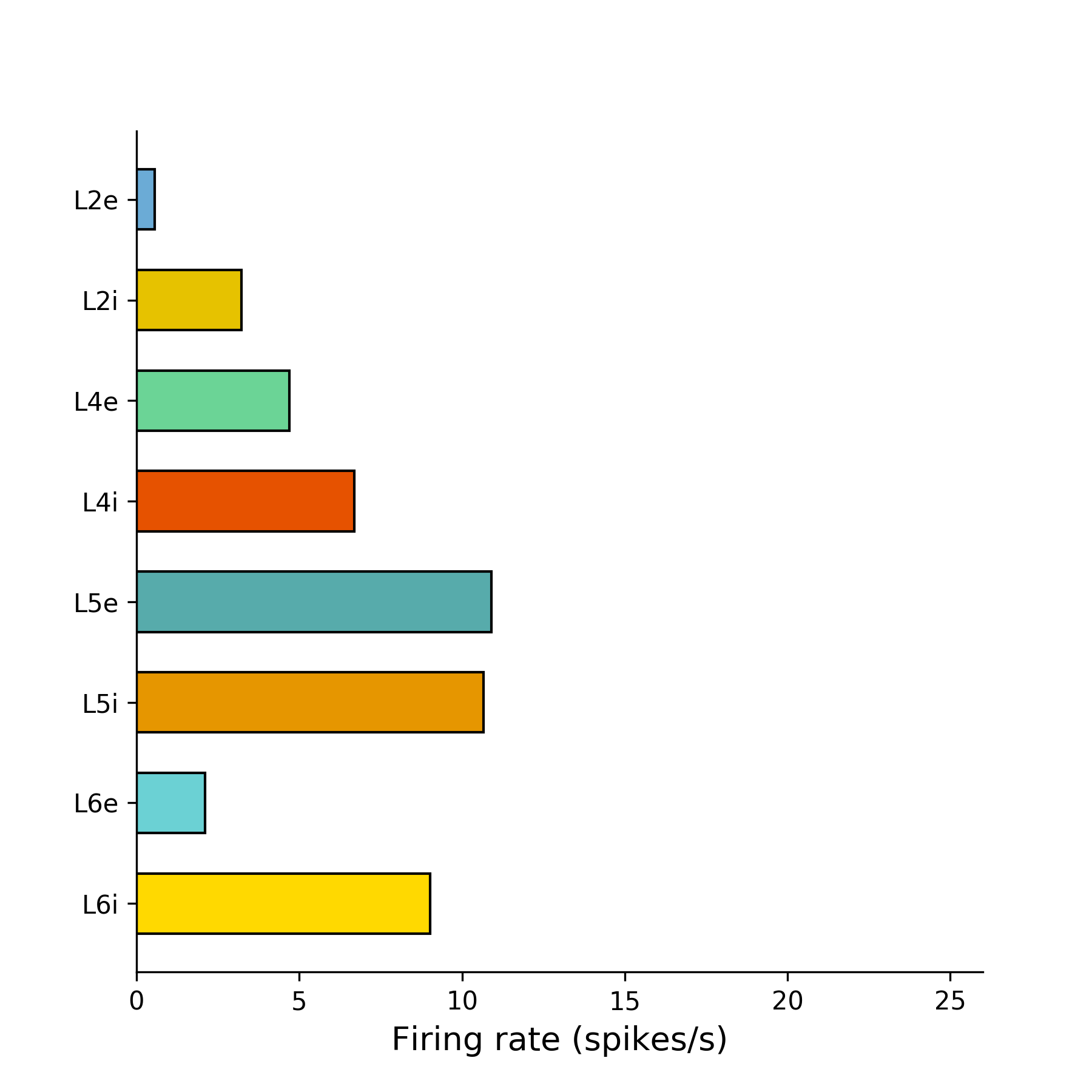}}
  \end{minipage}
\begin{minipage}[b][\ht\measurebox]{.24\textwidth}
\centering
\vfill

\subfloat
  []
  {\label{fig:boundary_graf_02bG}\includegraphics[width=\textwidth,height=3cm]{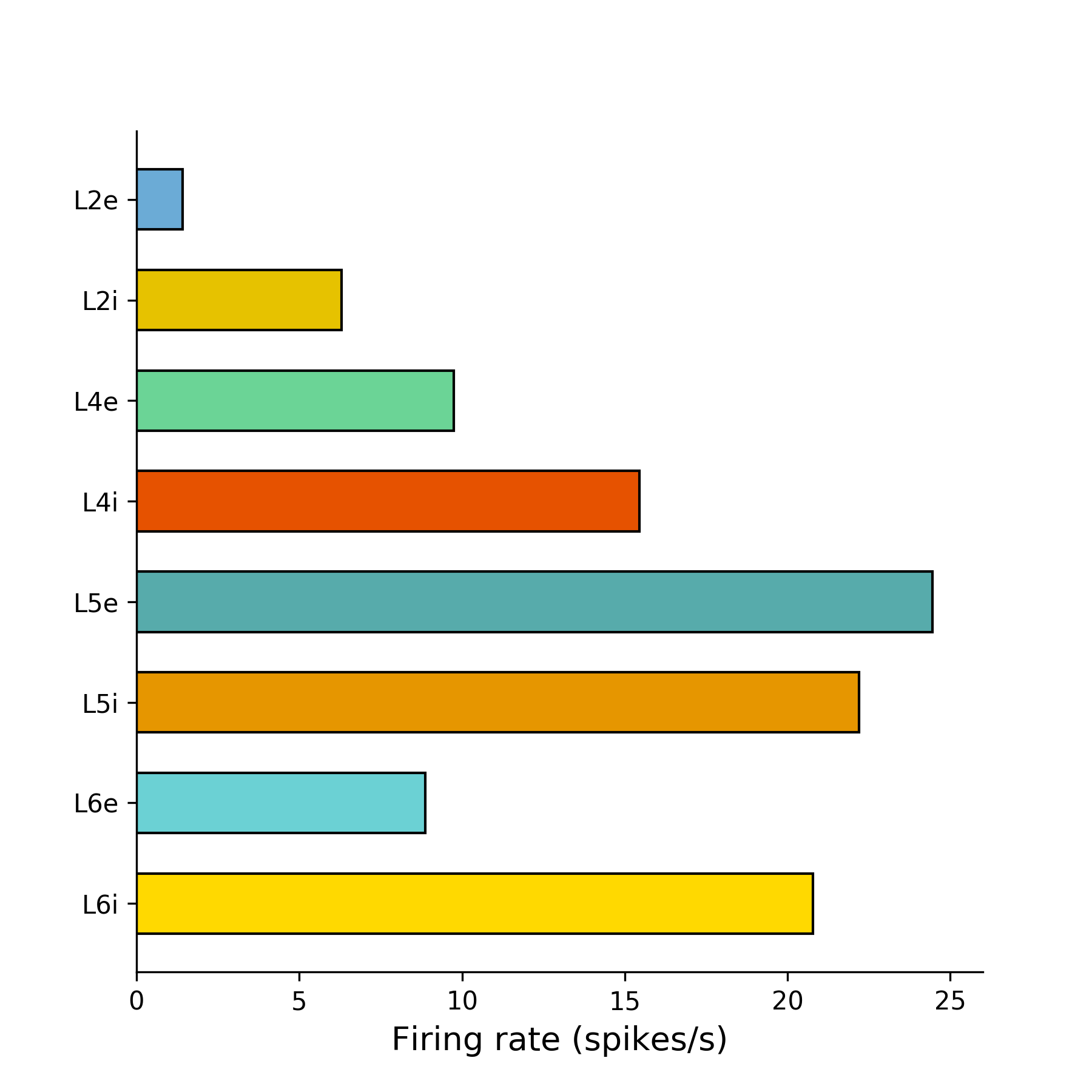}}
  
\subfloat
  []
  {\label{fig:boundary_graf_02bH}\includegraphics[width=\textwidth,height=3cm]{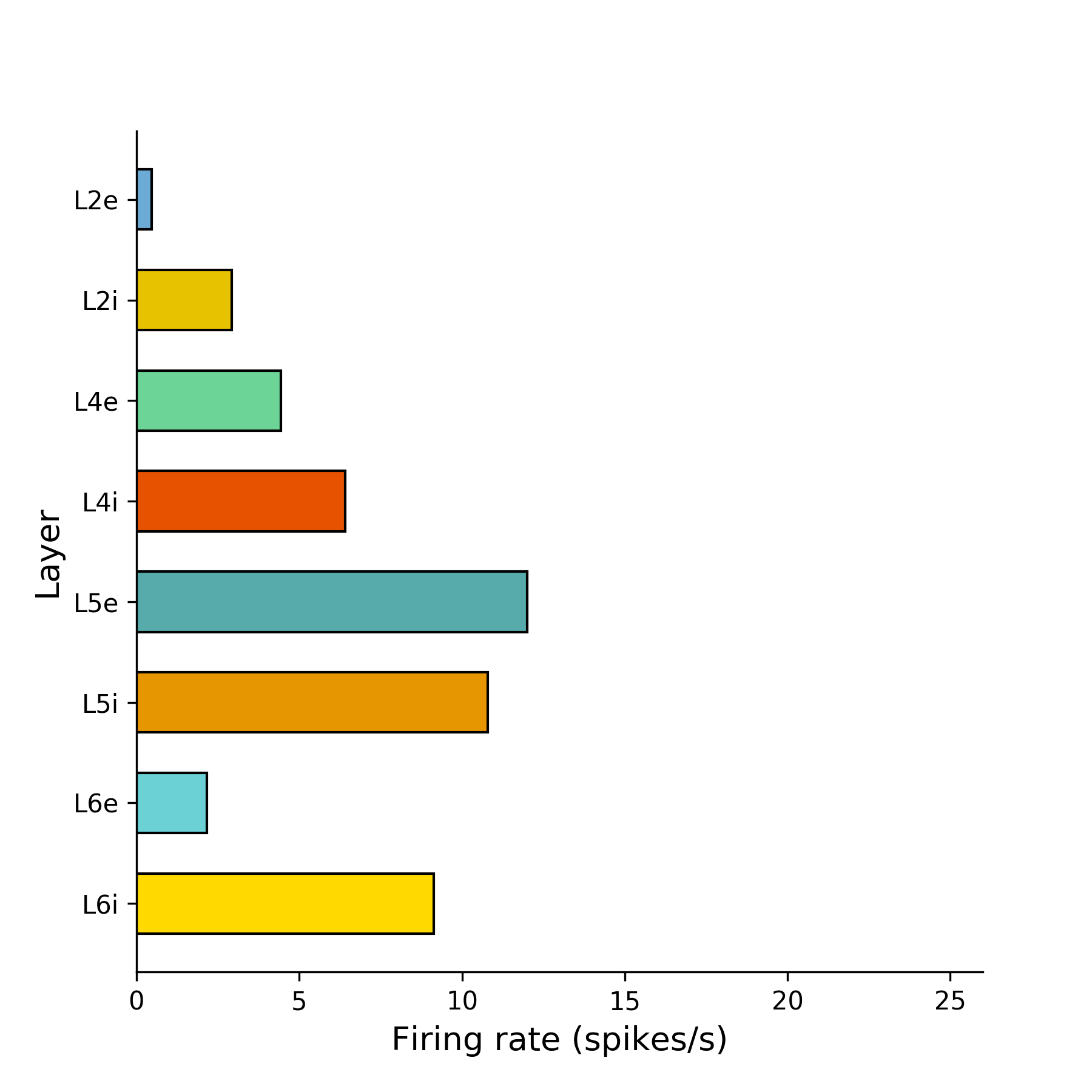}}

\end{minipage} 
\caption{The PD full scale average firing-rate per neuron and per layer. Neurons from L2 excitatory (A)without and (B) with boundary correction. Neurons in L5 excitatory (C) without and (D) with boundary correction.
Each dot represents the position of neuron and the size of the dot is proportional to the average firing rate of that neuron.  The core (around 50 \% of all neurons) layers average firing rate (E) without boundary correction and (F) with boundary correction. The boundary (the complementary 50 \% of all neurons) layers average firing rate (G) without boundary correction and (H) with boundary correction. }
\label{fig:boundary_graf_02b}
\end{figure}

\begin{figure}[H]
\centering

\begin{minipage}[b]{.24\textwidth}
\subfloat
  []
  {\label{fig:boundary_graf_02aA}\includegraphics[width=\textwidth,height=3cm]{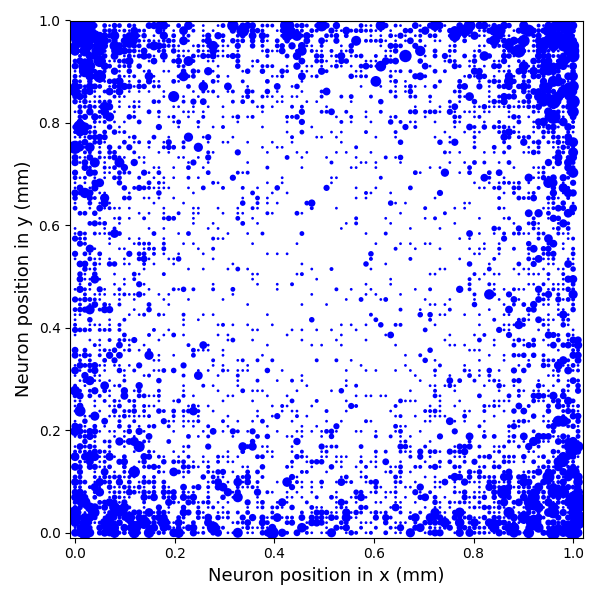}}
  
  \subfloat
  []
  {\label{fig:boundary_graf_02aB}\includegraphics[width=\textwidth,height=3cm]{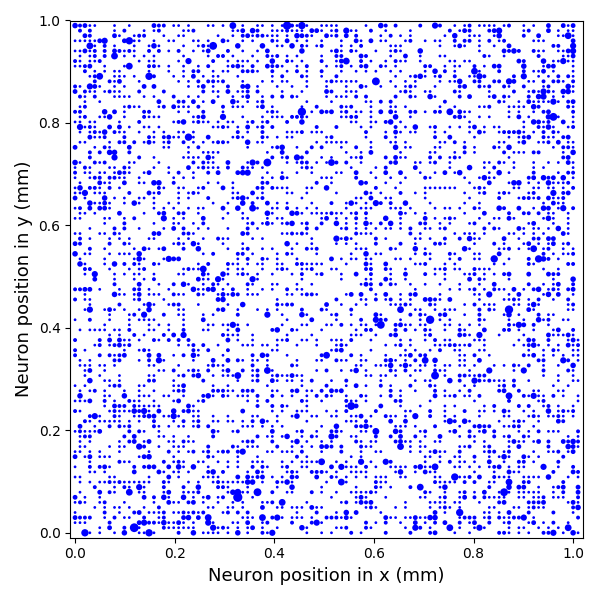}}
  \end{minipage}
\begin{minipage}[b]{.24\textwidth}
\subfloat
  []
  {\label{fig:boundary_graf_02aC}\includegraphics[width=\textwidth,height=3cm]{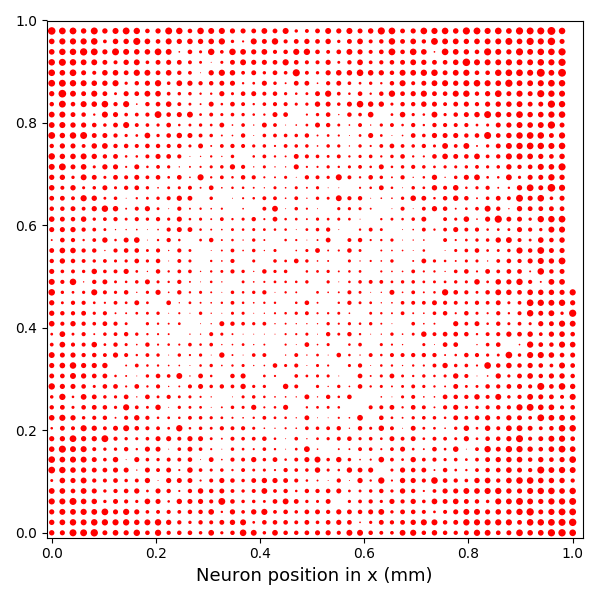}}
  
\subfloat
  []
  {\label{fig:boundary_graf_02aD}\includegraphics[width=\textwidth,height=3cm]{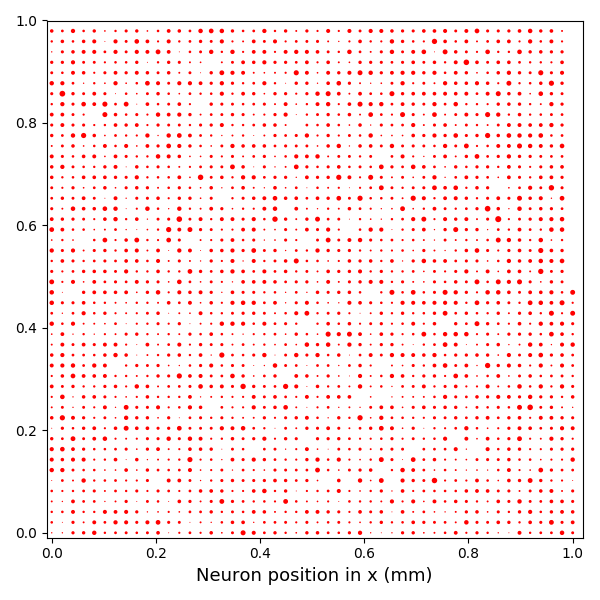}}

\end{minipage}
\begin{minipage}[b]{.24\textwidth}
  \subfloat
    []
    {\label{fig:boundary_graf_02aE}\includegraphics[width=\textwidth,height=3cm]{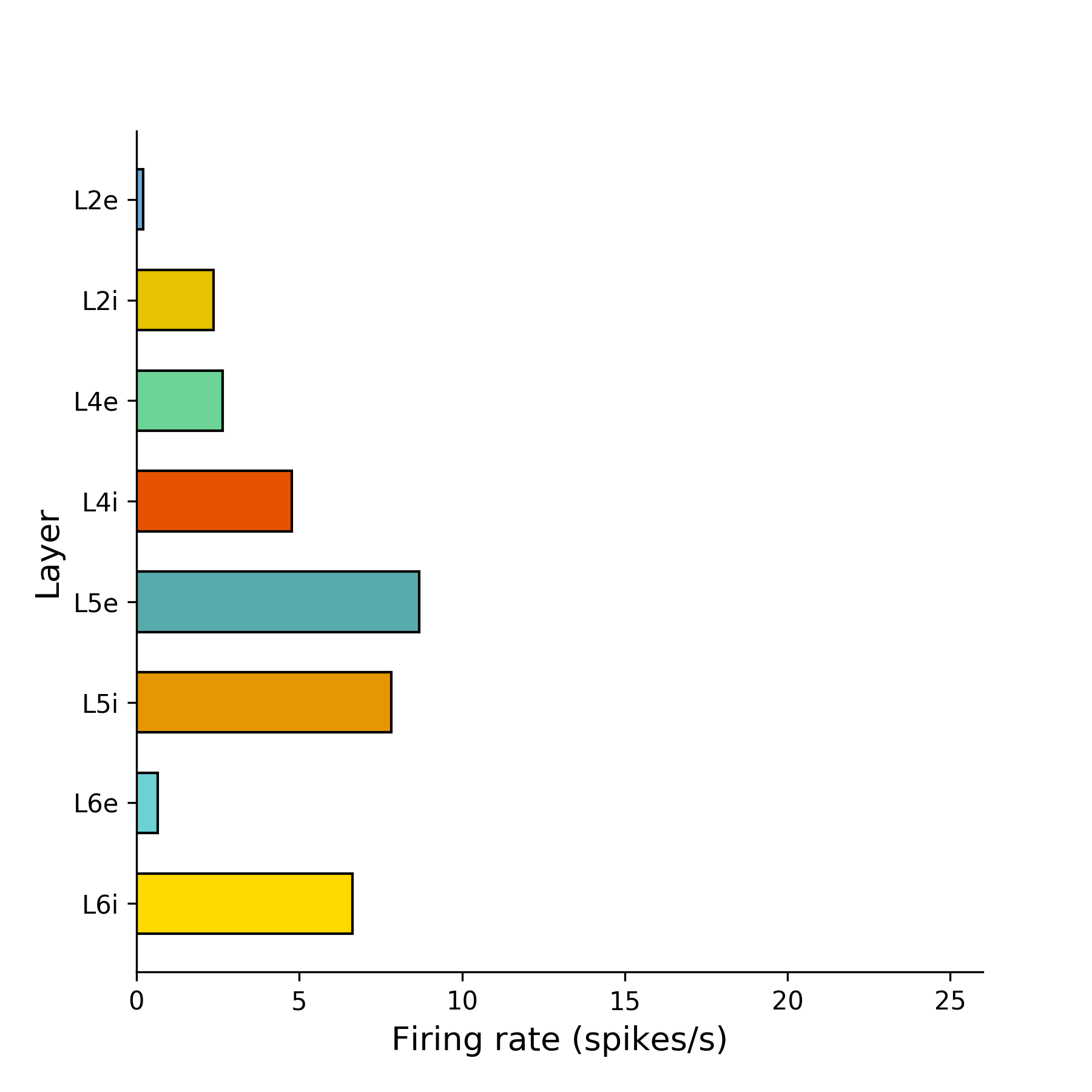}}
    
    \subfloat
  []
  {\label{fig:boundary_graf_02aF}\includegraphics[width=\textwidth,height=3cm]{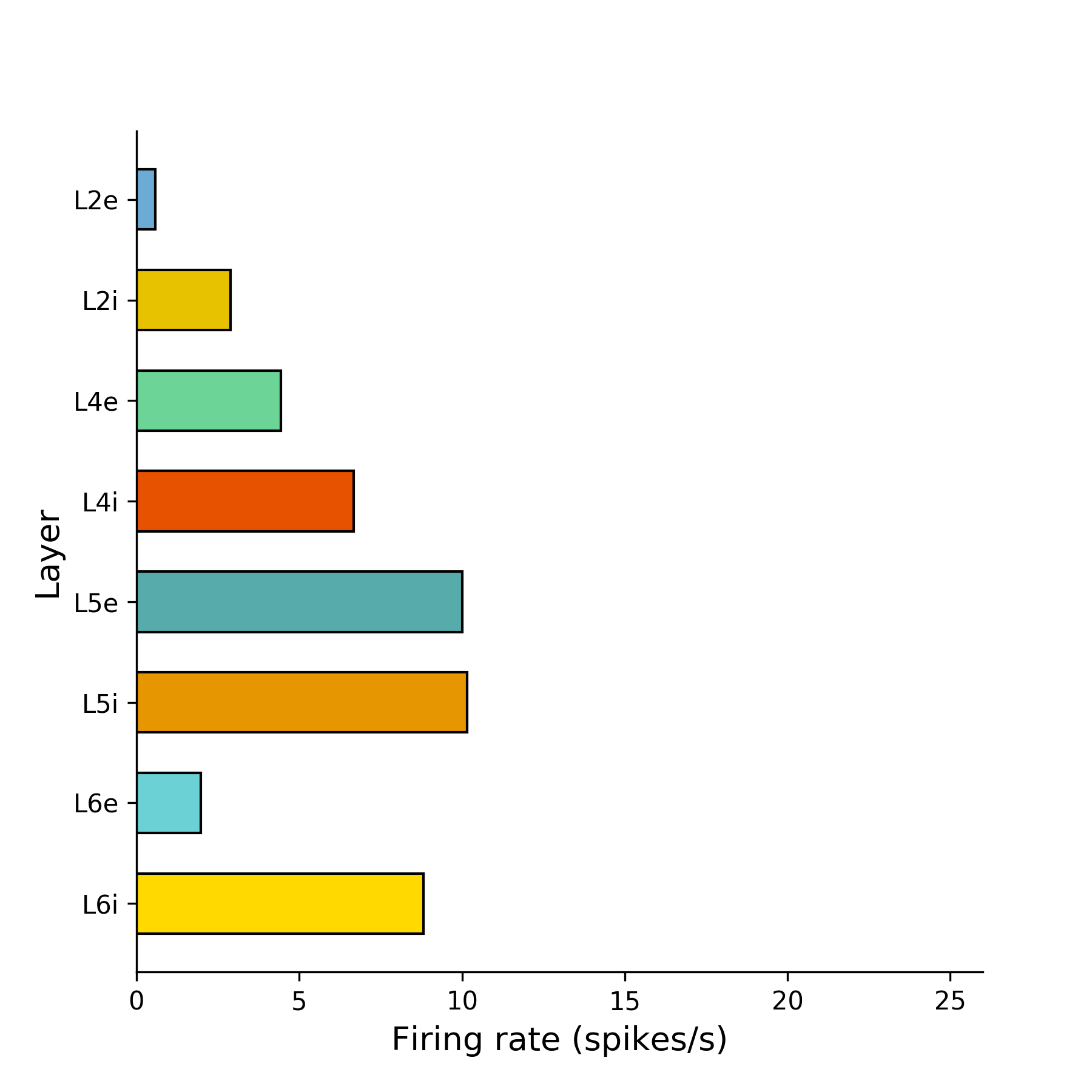}}
  \end{minipage}
\begin{minipage}[b][\ht\measurebox]{.24\textwidth}
\centering
\vfill

\subfloat
  []
  {\label{fig:boundary_graf_02aG}\includegraphics[width=\textwidth,height=3cm]{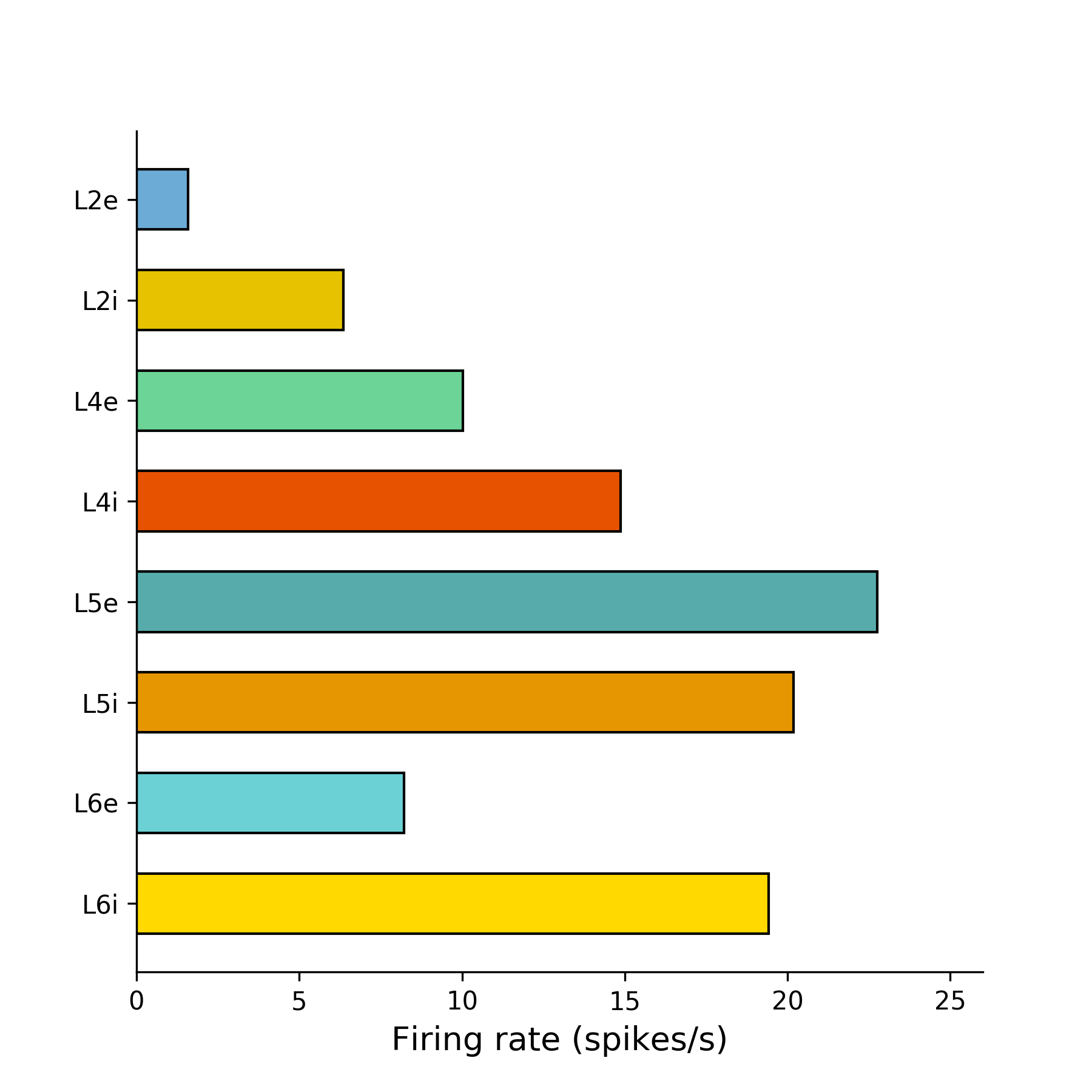}}
  
\subfloat
  []
  {\label{fig:boundary_graf_02aH}\includegraphics[width=\textwidth,height=3cm]{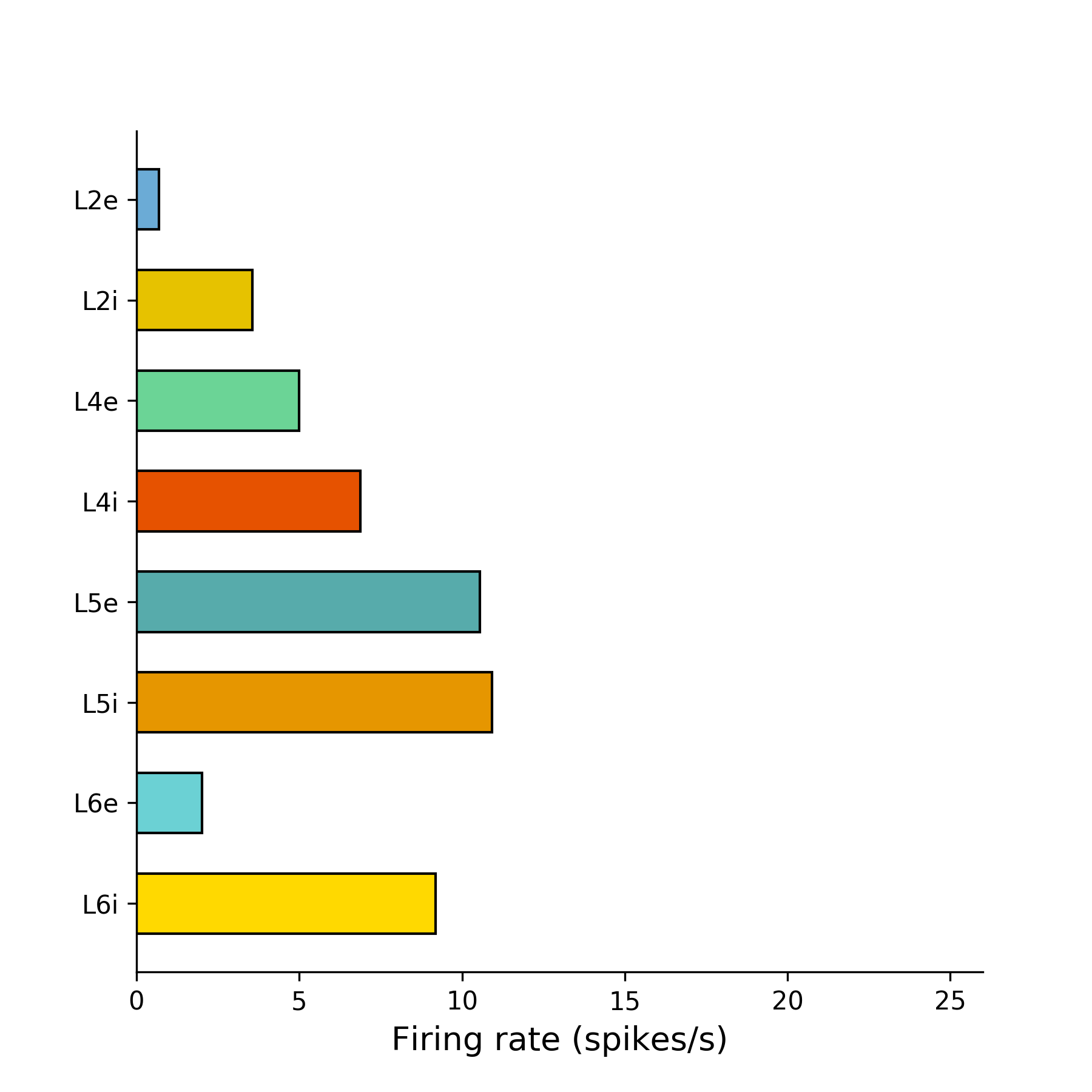}}

\end{minipage} 
\caption{The PD rescaled to 50 \% average firing-rate per neuron and per layer. Neurons from L2 excitatory (A)without and (B) with boundary correction. Neurons in L5 excitatory (C) without and (D) with boundary correction.
Each dot represents the position of neuron and the size of the dot is proportional to the average firing rate of that neuron.  The core (around 50 \% of all neurons) layers average firing rate (E) without boundary correction and (F) with boundary correction. The boundary (the complementary 50 \% of all neurons) layers average firing rate (G) without boundary correction and (H) with boundary correction. }
\label{fig:boundary_graf_02a}
\end{figure}

Figures \ref{fig:boundary_graf_02bA} to \ref{fig:boundary_graf_02bD} present the average firing-rate per neurons. Each dot represents the position of the neuron and the size of the dot is proportional to the average firing rate of that neuron. Figures \ref{fig:boundary_graf_02bA} and \ref{fig:boundary_graf_02bB} correspond to excitatory layer L2 without and with boundary correction respectively. Figures \ref{fig:boundary_graf_02bC} and \ref{fig:boundary_graf_02bD} present excitatory L5 without and with boundary correction respectively. Figures \ref{fig:boundary_graf_02bE} and \ref{fig:boundary_graf_02bG} present the core (around 50 \% of all neurons)- boundary (the complementary 50 \% of all neurons) layers average firing rate respectively without boundary correction and Figures \ref{fig:boundary_graf_02bF} and \ref{fig:boundary_graf_02bH} are with boundary correction. Figure \ref{fig:boundary_graf_02a} presents the same of the Figure \ref{fig:boundary_graf_02b} for the network rescaled in 50\% of original size. This show that it is possible to combine both methods. In all cases, the boundary neurons spikes visually more than core on the model without boundary correction due to the leak of inhibition connections.
\vspace{0.4 cm}

\vspace{0.4 cm}

\vspace{0.6 cm}

\subsubsection{Brunel \cite{brunel2000dynamics} network: Excitatory-inhibitory interconnected}
\vspace{0.4 cm}

\begin{figure}[H]
\centering

\begin{minipage}[b]{.24\textwidth}
\subfloat
  []
  {\label{fig:boundary_graf_02cA}\includegraphics[width=\textwidth,height=3.5cm]{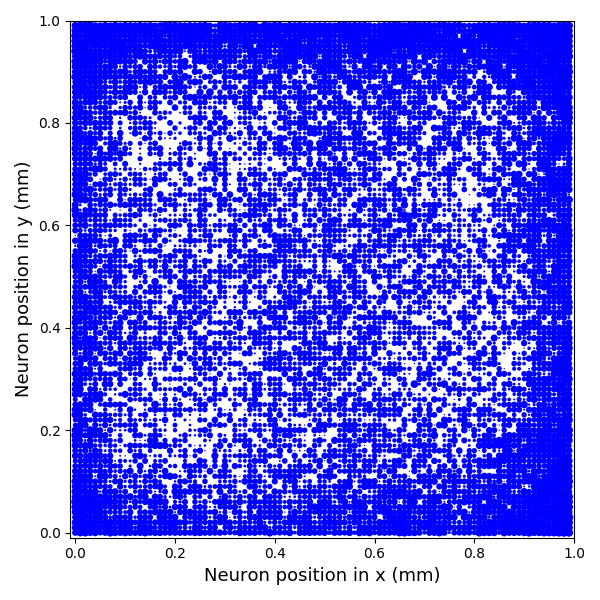}}
  
  \subfloat
  []
  {\label{fig:boundary_graf_02cB}\includegraphics[width=\textwidth,height=3.5cm]{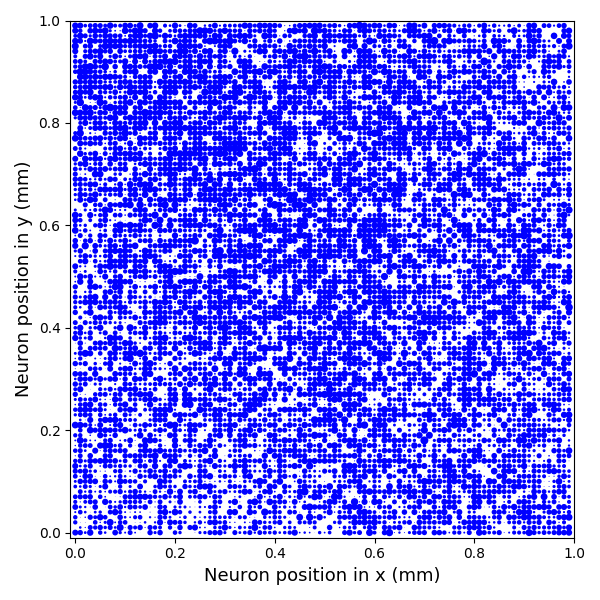}}
  \end{minipage}
\begin{minipage}[b]{.24\textwidth}
\subfloat
  []
  {\label{fig:boundary_graf_02cC}\includegraphics[width=\textwidth,height=3.5cm]{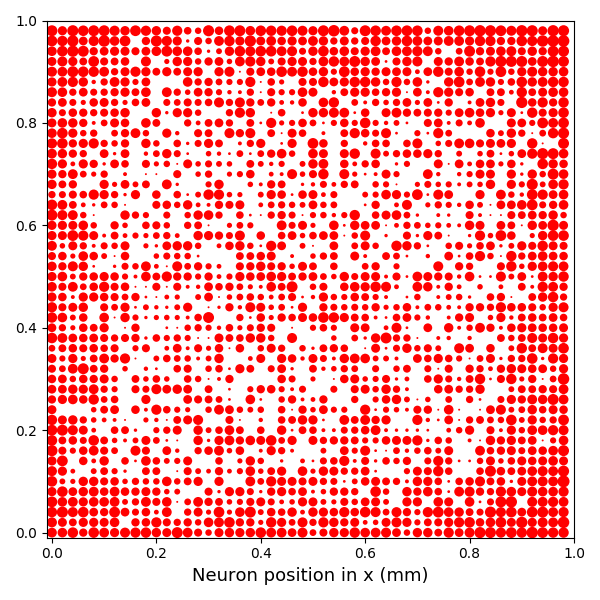}}
  
\subfloat
  []
  {\label{fig:boundary_graf_02cD}\includegraphics[width=\textwidth,height=3.5cm]{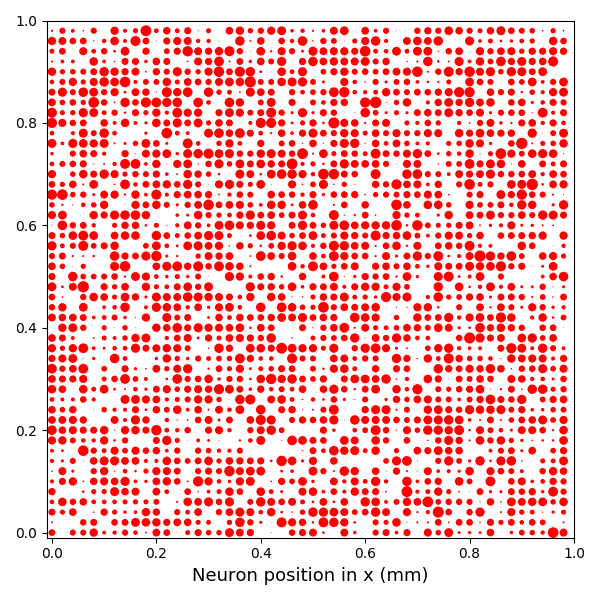}}

\end{minipage}
\begin{minipage}[b][\ht\measurebox]{.20\textwidth}
\centering
\vfill

\subfloat
  []
  {\label{fig:boundary_graf_02cE}\includegraphics[width=\textwidth,height=1.0cm]{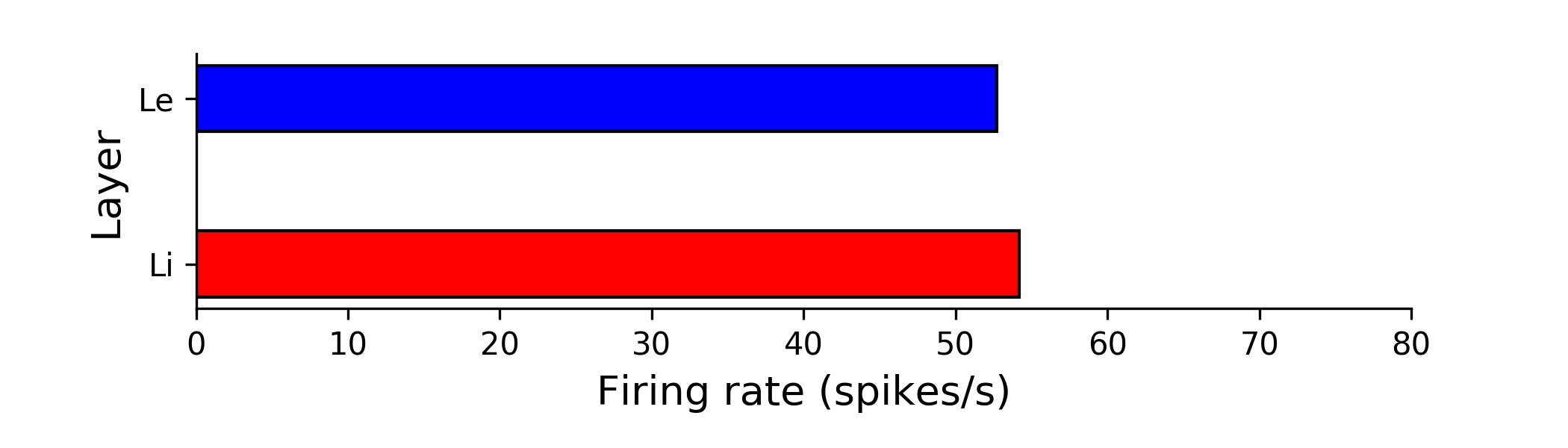}}

\subfloat
  []
  {\label{fig:boundary_graf_02cF}\includegraphics[width=\textwidth,height=1.0cm]{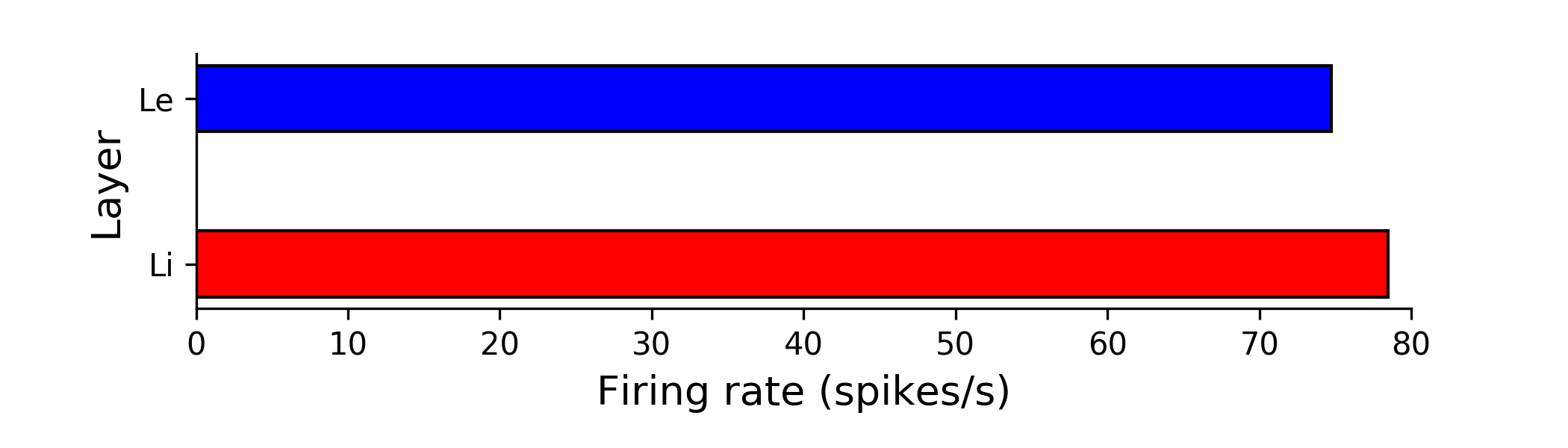}}
  
 \subfloat
  []
  {\label{fig:boundary_graf_02cF2}\includegraphics[width=\textwidth,height=0.6cm]{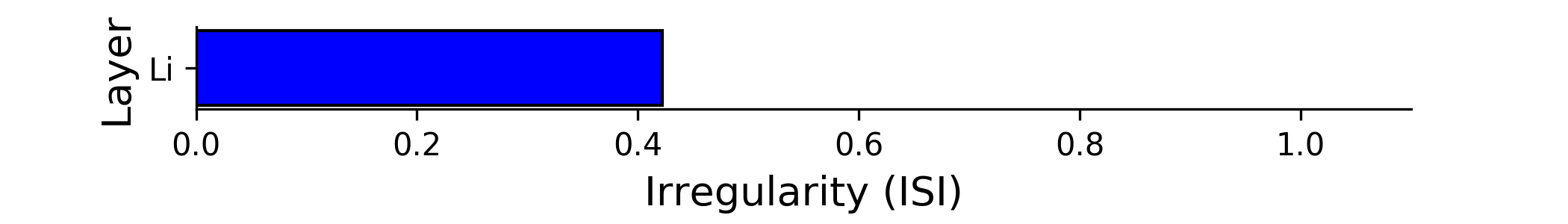}}
  
\subfloat
  []
  {\label{fig:boundary_graf_02cG}\includegraphics[width=\textwidth,height=1.0cm]{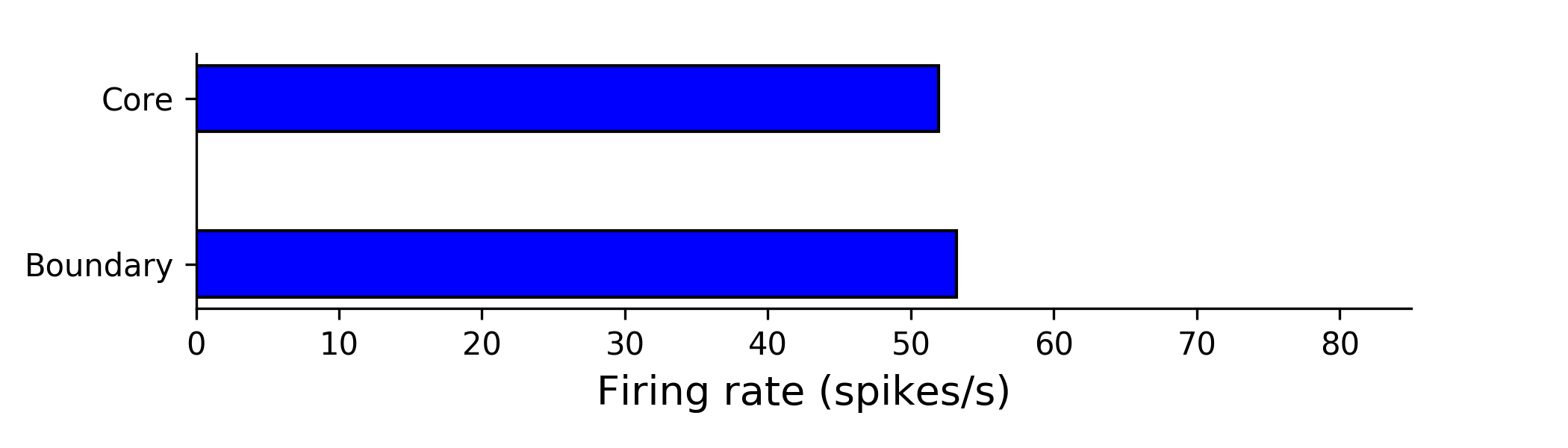}}
  
 \subfloat
  []
  {\label{fig:boundary_graf_02cH}\includegraphics[width=\textwidth,height=0.6cm]{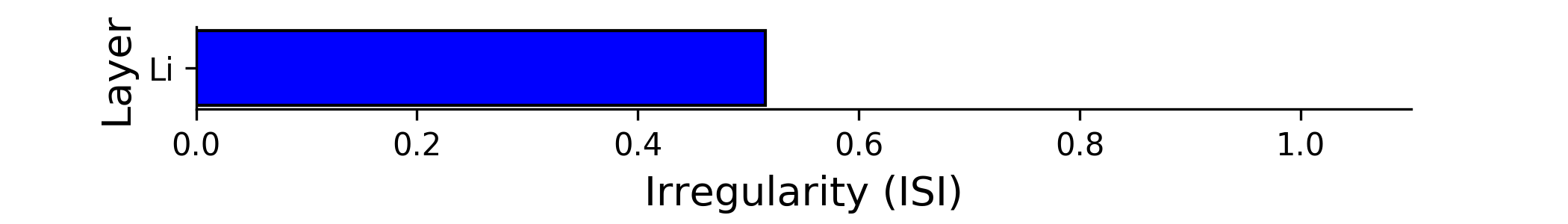}}

\end{minipage} 
\begin{minipage}[b]{.28\textwidth}
  \subfloat
    []
    {\label{fig:boundary_graf_02cI}\includegraphics[width=\textwidth,height=6cm]{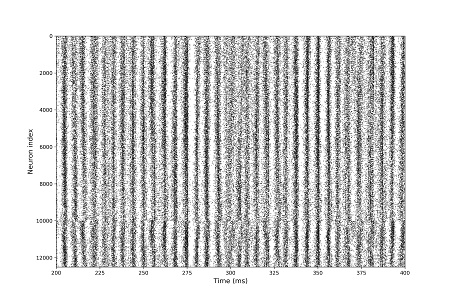}}
    
    \subfloat
  []
  {\label{fig:boundary_graf_02cJ}\includegraphics[width=\textwidth,height=1.5cm]{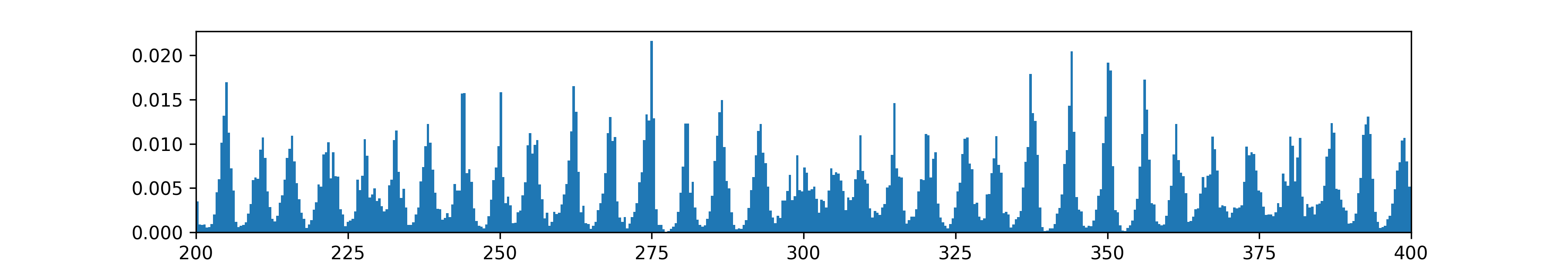}}
  \end{minipage}
\caption{The topographic Brunel \cite{brunel2000dynamics} model average firing-rate per neuron and per layer for $g=6$ and $\Theta = 4.V_{th}$ configuration. Neurons from layer excitatory (A) without and (B) with boundary correction. Neurons in layer inhibitory (C) without and (D) with boundary correction.
Each dot represents the position of neuron and the size of the dot is proportional to the average firing rate of that neuron.  (E) The core (around 50 \% of all neurons) layers average firing rate and (F) the boundary (the complementary 50 \% of all neurons) layers average firing rate without boundary correction  with boundary correction. The average of the single-unit irregularity calculated by the coefficient of variance of the interspike intervals (ISI) without boundary correction (G) and with boundary correction (I). (H) The average firing rate of all neurons with boundary correction. (J) Raster plot and (K) histogram of the network for 200ms run. All others graphics showed results for 5s run.}
\label{fig:boundary_graf_02c}
\end{figure}

\begin{figure}[H]
\centering

\begin{minipage}[b]{.24\textwidth}
\subfloat
  []
  {\label{fig:boundary_graf_03bA}\includegraphics[width=\textwidth,height=3.5cm]{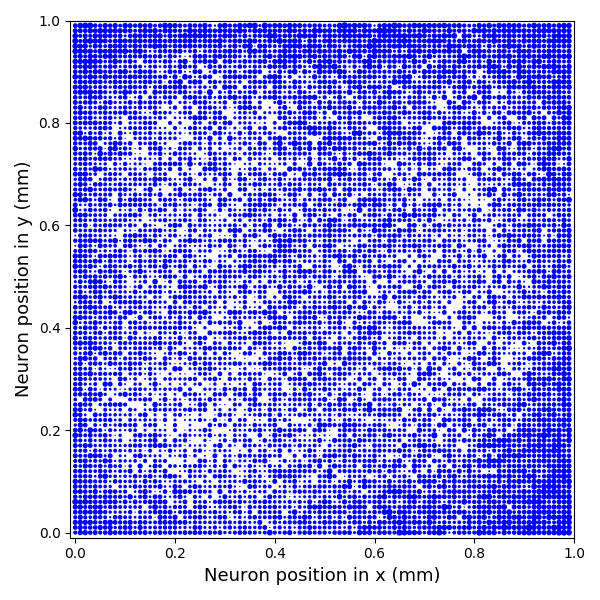}}
  
  \subfloat
  []
  {\label{fig:boundary_graf_03bB}\includegraphics[width=\textwidth,height=3.5cm]{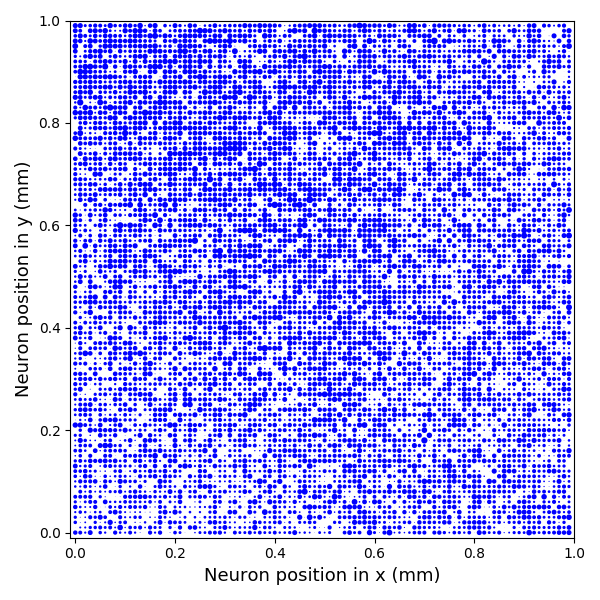}}
  \end{minipage}
\begin{minipage}[b]{.24\textwidth}
\subfloat
  []
  {\label{fig:boundary_graf_03bC}\includegraphics[width=\textwidth,height=3.5cm]{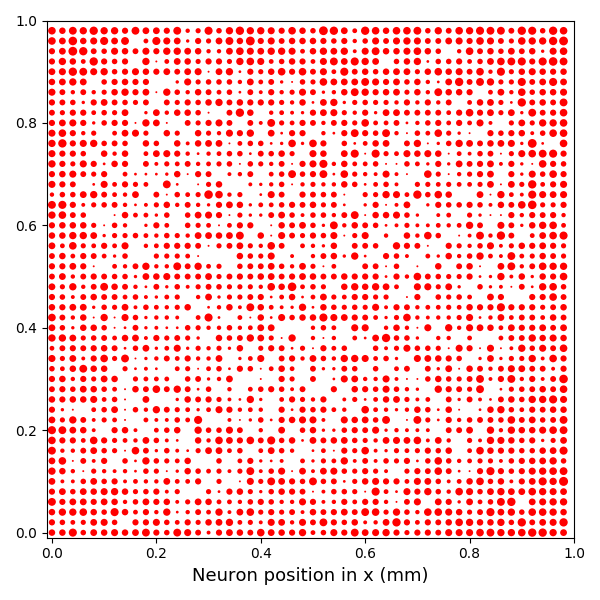}}
  
\subfloat
  []
  {\label{ffig:boundary_graf_03bD}\includegraphics[width=\textwidth,height=3.5cm]{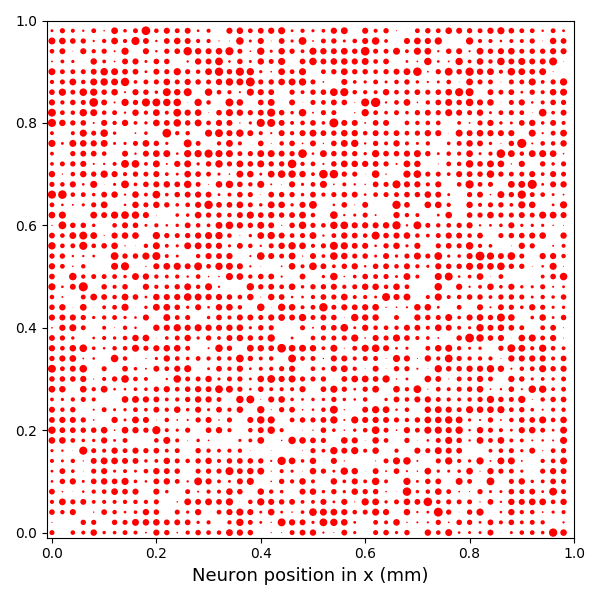}}

\end{minipage}
\begin{minipage}[b][\ht\measurebox]{.20\textwidth}
\centering
\vfill

\subfloat
  []
  {\label{fig:boundary_graf_03bE}\includegraphics[width=\textwidth,height=1.0cm]{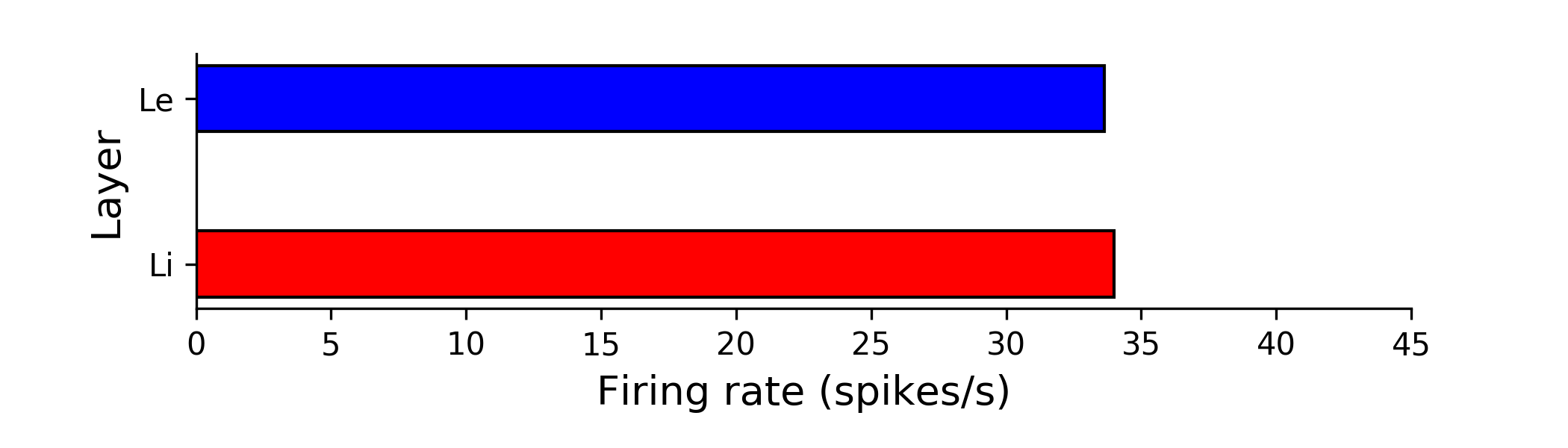}}

 \subfloat
  []
  {\label{fig:boundary_graf_03bF}\includegraphics[width=\textwidth,height=1.0cm]{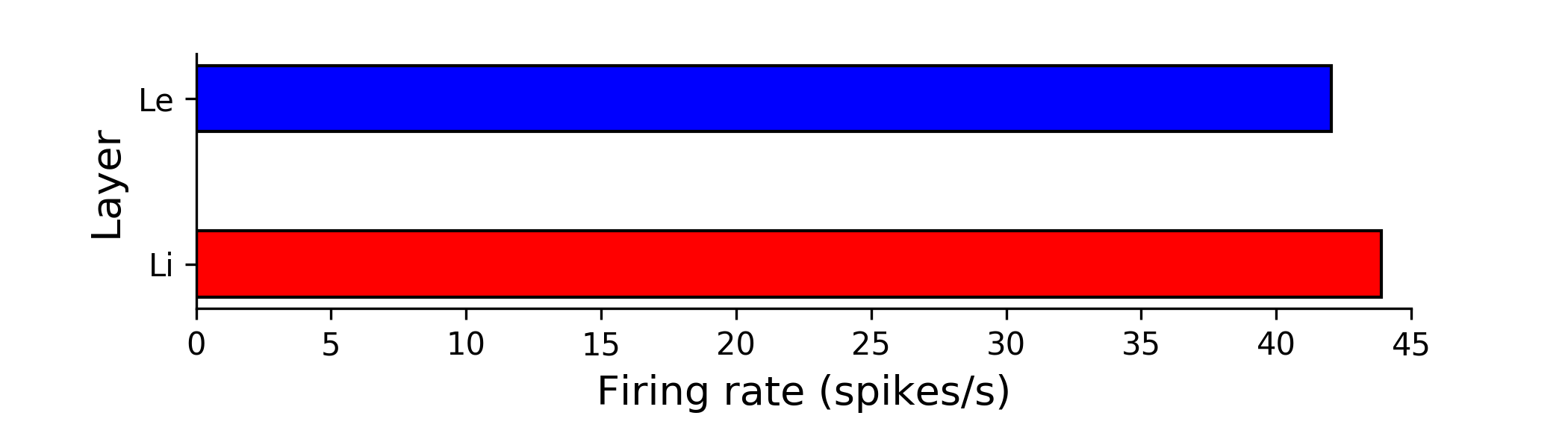}}
  
 \subfloat
  []
  {\label{fig:boundary_graf_03bF2}\includegraphics[width=\textwidth,height=0.6cm]{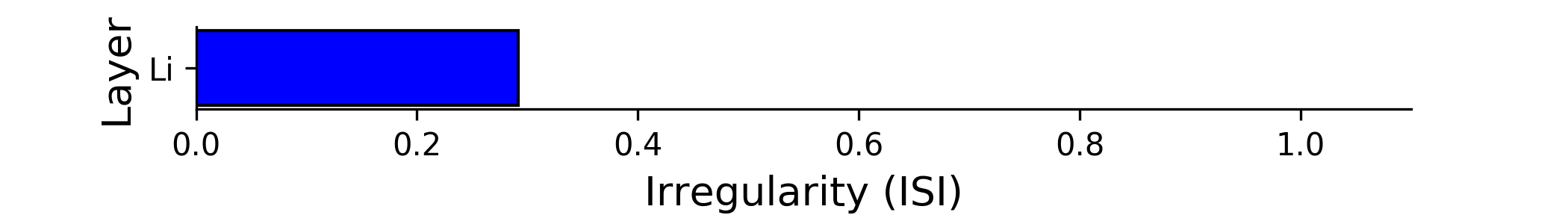}}
  
\subfloat
  []
  {\label{fig:boundary_graf_03bG}\includegraphics[width=\textwidth,height=1.0cm]{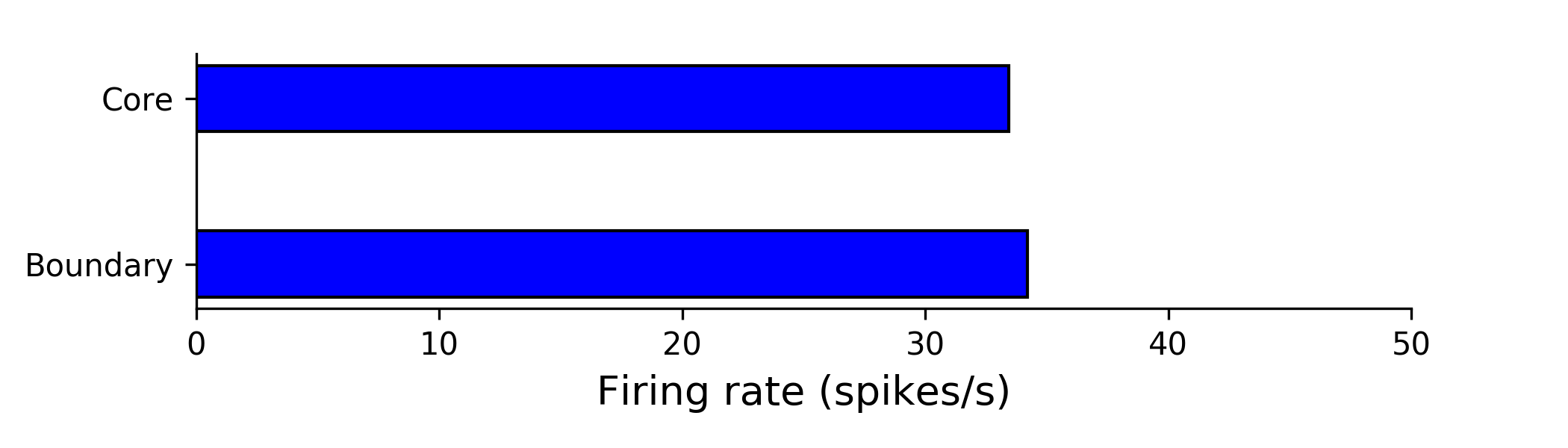}}
  
 \subfloat
  []
  {\label{fig:boundary_graf_03bH}\includegraphics[width=\textwidth,height=0.6cm]{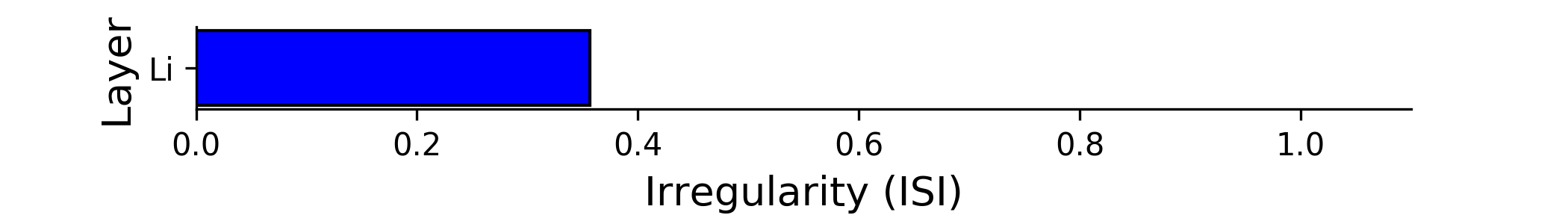}}

\end{minipage} 
\begin{minipage}[b]{.28\textwidth}
  \subfloat
    []
    {\label{fig:boundary_graf_03bI}\includegraphics[width=\textwidth,height=6cm]{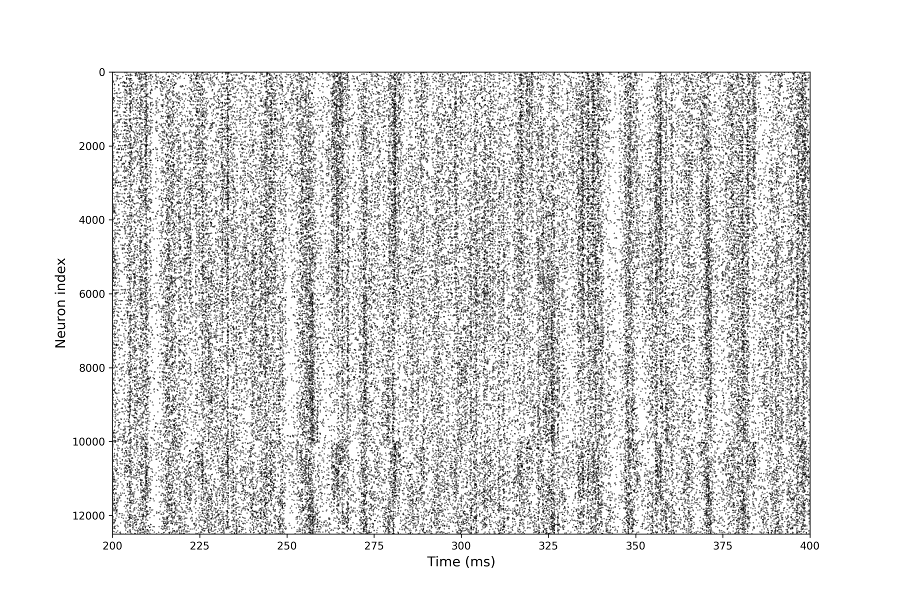}}
    
    \subfloat
  []
  {\label{fig:boundary_graf_03bK}\includegraphics[width=\textwidth,height=1.5cm]{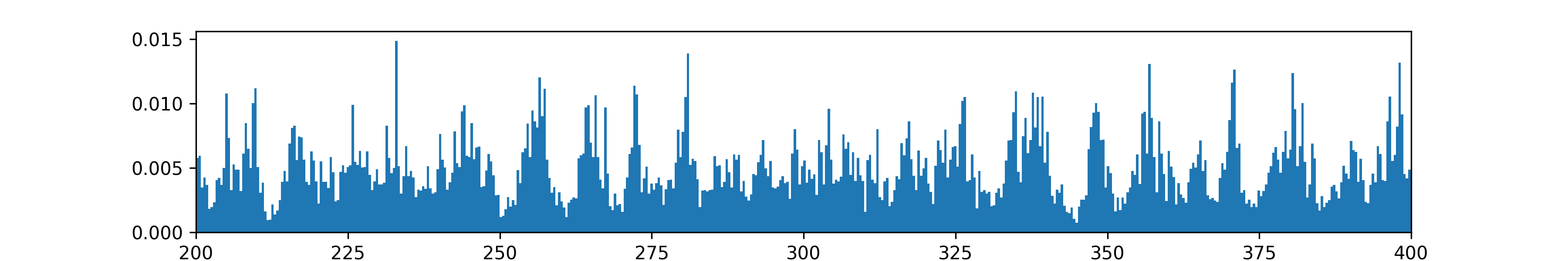}}
  \end{minipage}
\caption{The topographic Brunel \cite{brunel2000dynamics} model average firing-rate per neuron and per layer for $g=5$ and $\Theta = 2.V_{th}$ configuration. Neurons from layer excitatory (A) without and (B) with boundary correction. Neurons in layer inhibitory (C) without and (D) with boundary correction.
Each dot represents the position of neuron and the size of the dot is proportional to the average firing rate of that neuron.  (E) The core (around 50 \% of all neurons) layers average firing rate and (F) the boundary (the complementary 50 \% of all neurons) layers average firing rate without boundary correction  with boundary correction. The average of the single-unit irregularity calculated by the coefficient of variance of the interspike intervals (ISI) without boundary correction (G) and with boundary correction (I). (H) The average firing rate of all neurons with boundary correction. (J) Raster plot and (K) histogram of the network for 200ms run. All others graphics showed results for 5s run.
}
\label{fig:boundary_graf_03b}
\end{figure}

\begin{figure}[H]
\centering

\begin{minipage}[b]{.45\textwidth}
\subfloat
  []
  {\label{fig:boundary_graf_03cA}\includegraphics[width=\textwidth,height=2cm]{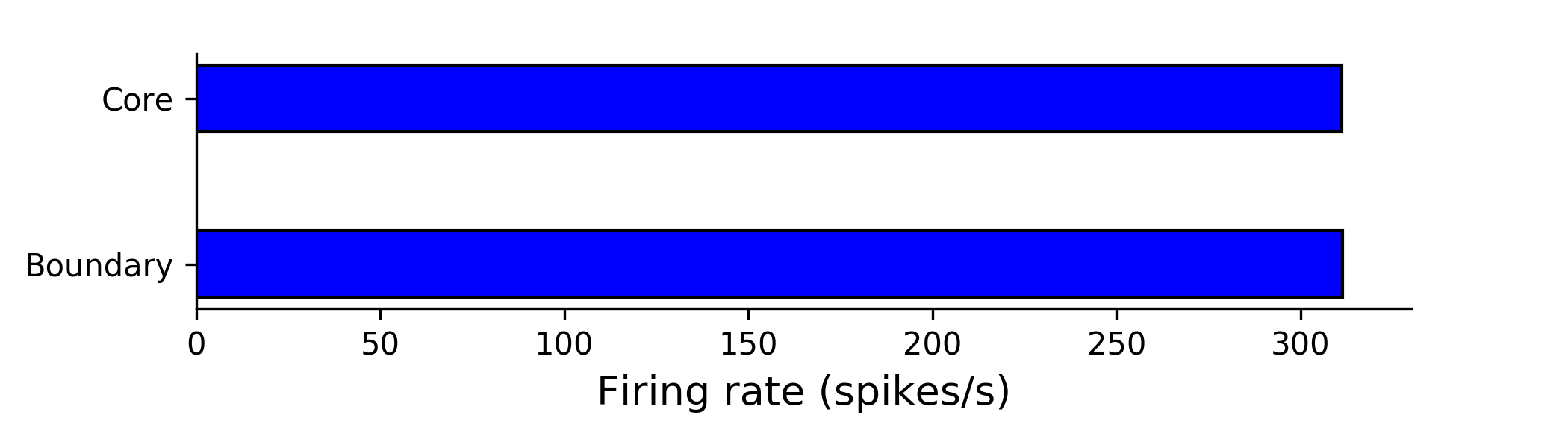}}
  
  \subfloat
  []
  {\label{fig:boundary_graf_03cB}\includegraphics[width=\textwidth,height=0.7cm]{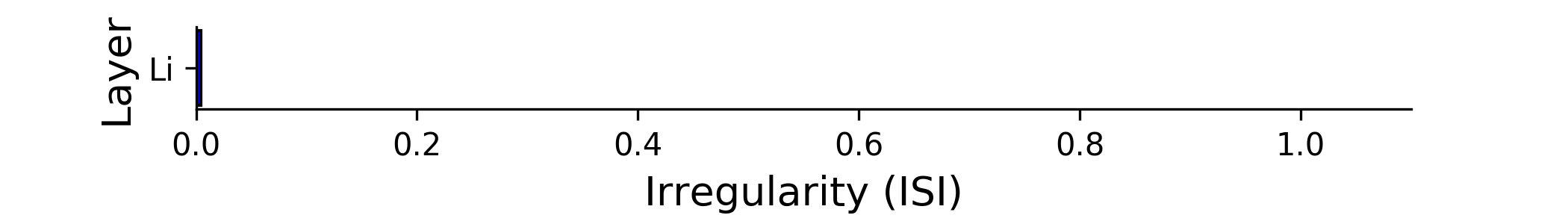}}
  
  \subfloat
  []
  {\label{fig:boundary_graf_03cC}\includegraphics[width=\textwidth,height=2cm]{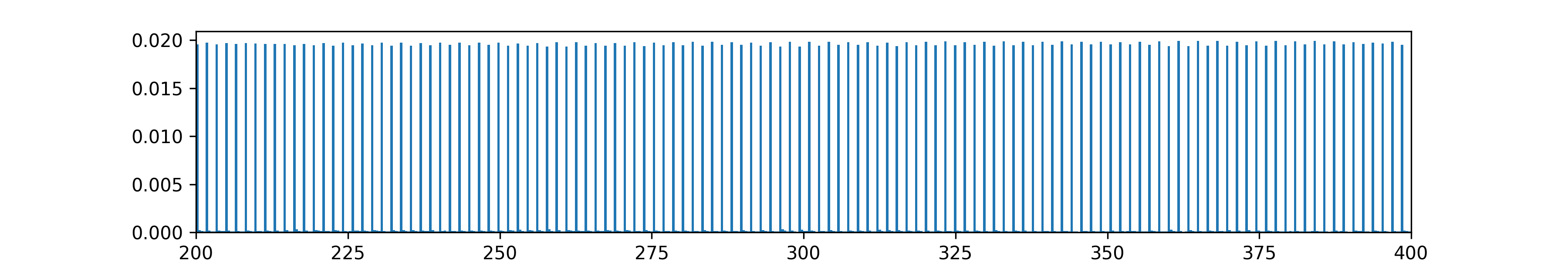}}
  \end{minipage}
\begin{minipage}[b]{.45\textwidth}
\subfloat
  []
  {\label{fig:boundary_graf_03cD}\includegraphics[width=\textwidth,height=2cm]{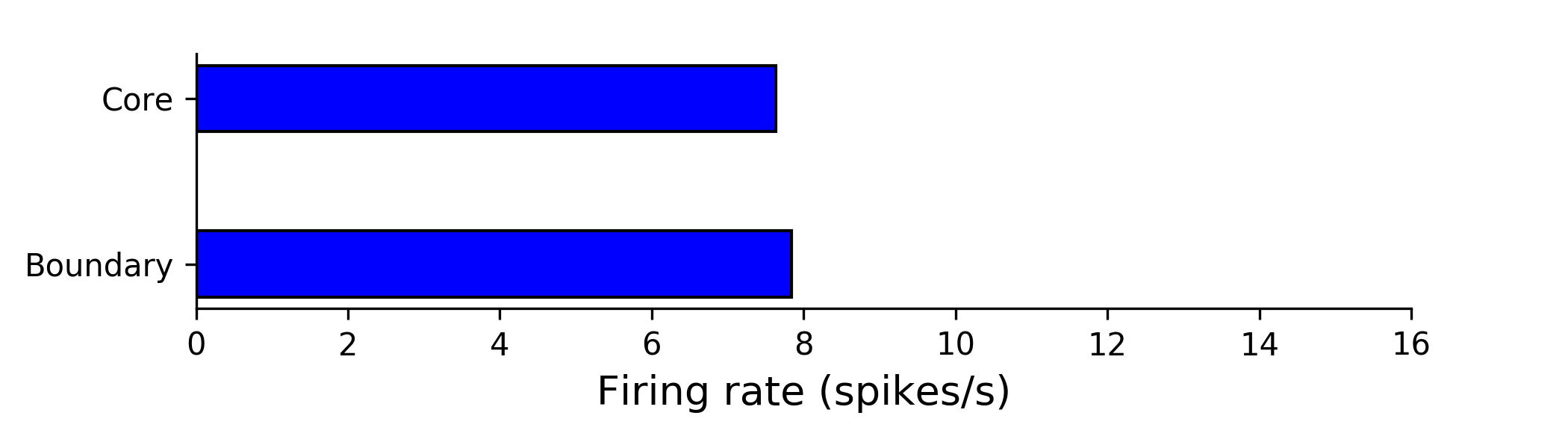}}
  
\subfloat
  []
  {\label{fig:boundary_graf_03cE}\includegraphics[width=\textwidth,height=0.7cm]{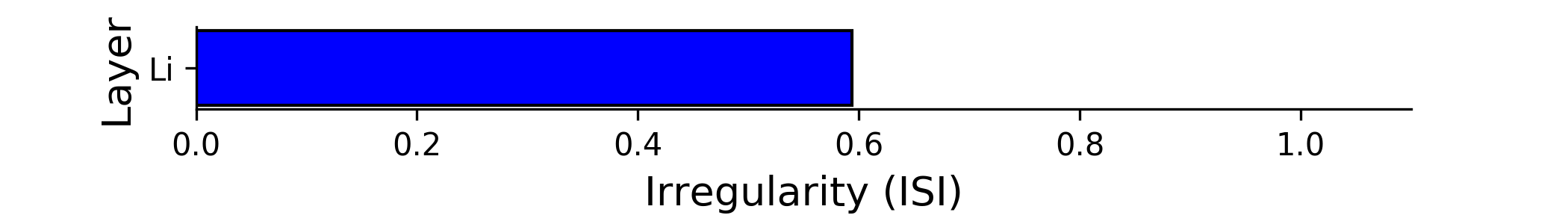}}

\subfloat
  []
  {\label{fig:boundary_graf_03cF}\includegraphics[width=\textwidth,height=2cm]{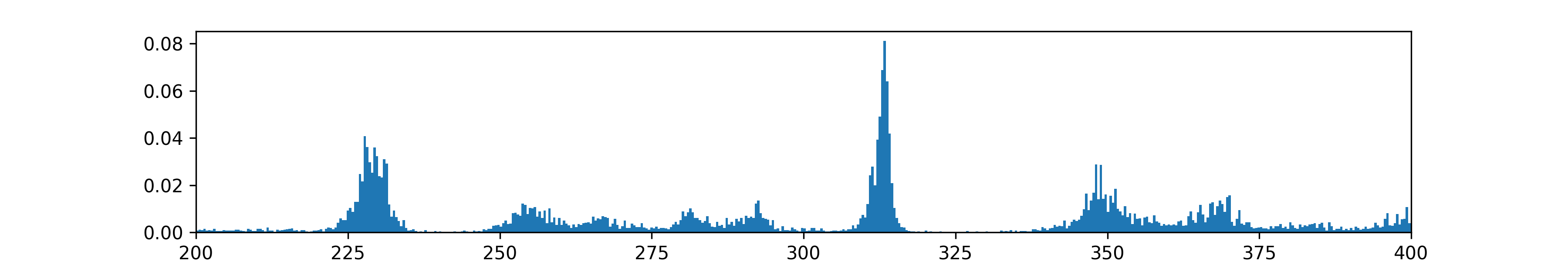}}
  
\end{minipage}
\caption{The topographic Brunel \cite{brunel2000dynamics} model average firing-rate, irregularity and spike histogram (A, B, C) for $g=3$ and $\Theta = 2.V_{th}$ and (D, E, F) for $g=4$ and $\Theta = 1.001.V_{th}$ configurations. Average firing rate (A, D), irregularity  (B, E) and spike histogram (C, F) of neurons from layers excitatory and inhibitory with boundary correction. Average firing rate and Irregularity for 5s run and histogram of the network for 200ms run.}
\label{fig:boundary_graf_03c}
\end{figure}

All neurons from the full version of the model presented on \ref{sec:Brunel network: Excitatory-inhibitory interconnected} were homogeneous distributed on $1 mm^2$ and a $\sigma_g = 0.150 mm$ was utilized. We ran the model for $g=6$ and $\Theta = 4.V_{th}$  configuration (Figure \ref{fig:boundary_graf_02c}), for $g=5$, $\Theta = 2.V_{th}$  configuration  (Figure \ref{fig:boundary_graf_03b}), for $g=3$ and $\Theta = 2.V_{th}$ and for $g=4$ and $\Theta = 1.001.V_{th}$ configurations (Figure \ref{fig:boundary_graf_03c}), respectively Figure 8 B, C, A and D configuration network of \cite{brunel2000dynamics}. .
\vspace{0.4 cm}

Figures \ref{fig:boundary_graf_02cA} to \ref{fig:boundary_graf_02cD} present the average firing-rate per neurons. Each dot represents the neuron's position and the size of the dot is proportional to the average firing rate of that neuron. The  Figures \ref{fig:boundary_graf_02cA} and \ref{fig:boundary_graf_02cB} correspond to excitatory layer without and with boundary correction respectively. The Figures \ref{fig:boundary_graf_02cC} and \ref{fig:boundary_graf_02cD} to inhibitory layer without and with boundary correction respectively. The Figures \ref{fig:boundary_graf_02cE} to \ref{fig:boundary_graf_02cG} present the core (around 50 \% of all neurons)- boundary (the complementary 50 \% of all neurons) layers average firing rate  and network irregularity respectively without boundary correction. Figures \ref{fig:boundary_graf_02cH} to \ref{fig:boundary_graf_02cI} are network average firing rate and network irregularity with boundary correction. The Figure \ref{fig:boundary_graf_02cI} presents the raster plot and Figure \ref{fig:boundary_graf_02cJ} presents the spikes histogram of the network. All those for $g=6$ and $\Theta = 4.V_{th}$  configuration. For $g=5$ and $\Theta = 2.V_{th}$ configuration the same can be found in the for the configuration $g=6$ and $\Theta = 4.V_{th}$  (Figure \ref{fig:boundary_graf_03b}).
\vspace{0.4 cm}

Note that the oscillation presents in Figure \ref{fig:rescaling_graf_04bB}, vanished in \ref{fig:rescaling_graf_04bD} is visually back on Figure \ref{fig:boundary_graf_02cJ}. This phenomenon will be explained in the Section \ref{Boundary:Model requirements, mathematical explication and method limitations} - Model requirements, mathematical explications and method limitations.
\vspace{0.4 cm}

Figure \ref{fig:boundary_graf_03c} presents the results of the reproduction the Figure 8-A ($g=3$ and $\Theta = 2.V_{th}$) configuration network and Figure 8 D ($g=4$ and $\Theta = 1.001.V_{th}$) configuration network of \cite{brunel2000dynamics}. Figures \ref{fig:boundary_graf_03cA} and \ref{fig:boundary_graf_03cB} present the average firing-rate and irregularity with boundary correction for 5s run simulation. Figure \ref{fig:boundary_graf_03cC} presents the spikes histogram of the network. All those for $g=3$ and $\Theta = 2.V_{th}$  configuration. For $g=4$ and $\Theta = 1.001.V_{th}$ configuration the same can be found in Figure \ref{fig:boundary_graf_03cD} to \ref{fig:boundary_graf_03cF} but to the $\Theta$ replaced for an equivalent Poisson input.
\vspace{0.4 cm}

\vspace{0.4 cm}

\vspace{0.6 cm}

\vspace{0.6 cm}

\subsection{Model requirements, mathematical explication and method limitations}
\label{Boundary:Model requirements, mathematical explication and method limitations}
\vspace{0.4 cm}

This method works in any network that satisfies the rescaling condition (Section \ref{Model requirements, math explication and method limitations}) for a reduction to 25\%: the minimum of rescaling to which a corner neuron can be submitted. This is a sufficient condition even though it is not a necessary condition. 
\vspace{0.4 cm}

Note that this boundary solution was able to retrieve the firing rate of:
\vspace{0.1 cm}

- Figure \ref{fig:boundary_graf_01E} back to Figure \ref{fig:rescaling_graf_01c} lost in Figure \ref{fig:boundary_graf_01B}; 
\vspace{0.1 cm}

- Figures \ref{fig:boundary_graf_02bF} to  \ref{fig:boundary_graf_02bH} and Figures \ref{fig:boundary_graf_02aF} to  \ref{fig:boundary_graf_02aH} back to Figure \ref{fig:rescaling_graf_03B} lost in Figures \ref{fig:boundary_graf_02bE} to  \ref{fig:boundary_graf_02bG} and Figures \ref{fig:boundary_graf_02aE} to  \ref{fig:boundary_graf_02aG};
\vspace{0.1 cm}

- Figures \ref{fig:boundary_graf_02cG} back to Figure \ref{fig:Ex04B1} lost in  Figures \ref{fig:boundary_graf_02cE} and  \ref{fig:boundary_graf_02cF};
\vspace{0.1 cm}

- Figures \ref{fig:boundary_graf_03bG} back to Figure \ref{fig:Ex04C1} lost in  Figures \ref{fig:boundary_graf_03bE} and  \ref{fig:boundary_graf_03bF};
\vspace{0.4 cm}

To retain the firing rate of:
\vspace{0.1 cm}

- Figure \ref{fig:Ex04A1} in Figure \ref{fig:boundary_graf_03cA};
\vspace{0.1 cm}

- Figure \ref{fig:Ex04D1} in Figure \ref{fig:boundary_graf_03cD};
\vspace{0.4 cm}

To retrieve the irregularity of 
\vspace{0.1 cm}

- Figure \ref{fig:boundary_graf_01F} back to Figure \ref{fig:rescaling_graf_01d} lost in Figure \ref{fig:boundary_graf_01C};
\vspace{0.1 cm}

- Figures \ref{fig:boundary_graf_03bH} back to Figure \ref{fig:Ex04bC1} lost in  Figure \ref{fig:boundary_graf_03bF2} 
\vspace{0.4 cm}

and to retain the irregularity of
\vspace{0.1 cm}

- Figure \ref{fig:Ex04bA1} in Figure \ref{fig:boundary_graf_03cB}
\vspace{0.1 cm}

- Figure \ref{fig:Ex04bD1} in Figure \ref{fig:boundary_graf_03cE}
\vspace{0.4 cm}

It is possible that even networks that could lose some synchrony with the application of the rescaling to 25\% (see Figure \ref{fig:rescaling_graf_04bD}), could remain the synchrony with the application of boundary correction (see Figure \ref{fig:boundary_graf_03bK} and \ref{fig:rescaling_graf_04bB} ). This is due to the weighting of neurons on network that had the $\mu_{int}$ reduced. If that is low enough to do not perturb the $G$ of the system (see Equation: \ref{eq:BrunelG}), the network proprieties to oscillation remain valid.
\vspace{0.6 cm}

\section{Acknowledgments}
This work was produced as part of the activities of FAPESP Research, Disseminations and Innovation Center for Neuromathematics (Grant 2013/07699-0, S. Paulo Research Foundation). The author is the recipient of PhD scholarships from the Brazilian Coordenação de Aperfeiçoamento de Pessoal de Nível Superior (CAPES). The author is thankful to Mauricio Girardi-Schappo, who encourage her to explain the method in a paper and author's little sister, Cinthia Romaro, who found time to comment on the manuscript even during her Harvard MBA. 
\vspace{0.6 cm}

\bibliographystyle{acm}
\bibliography{Bibliografia}
\end{document}